\documentclass[times,final,preprint]{elsarticle}

\usepackage{jcomp}
\usepackage{framed,multirow}

\usepackage{amssymb}
\usepackage{latexsym}

\usepackage{mathrsfs}
\usepackage{amsmath,amssymb,amsfonts,bm} 
\usepackage{subfigure}

\newtheorem{thm}{Theorem}
\newtheorem{lem}[thm]{Lemma}
\newdefinition{rmk}{Remark}
\newproof{pf}{Proof}
\newproof{pot}{Proof of Theorem \ref{thm2}}

\usepackage{tikz}
\newcommand*{\circled}[1]{\lower.7ex\hbox{\tikz\draw (0pt, 0pt)%
		circle (.5em) node {\makebox[1em][c]{\small #1}};}}

\usepackage[breaklinks,colorlinks]{hyperref} 
\newtheorem{example}{Example}[section]

\newcommand{\reffig}[1]{Fig.\,\ref{#1}}
\newcommand{\reffigs}[1]{Figs.\,\ref{#1}}
\newcommand{\reftab}[1]{Tab.\,\ref{#1}}
\newcommand{\reftabs}[1]{Tabs.\,\ref{#1}}

\usepackage{url}
\usepackage{xcolor}
\definecolor{newcolor}{rgb}{.8,.349,.1}
\biboptions{longnamesfirst,sort&compress} 

\journal{Journal of Computational Physics}

\begin{document}

\verso{Li Zhao \textit{et al.}}


\begin{frontmatter}

\title{Matter flow method for alleviating checkerboard oscillations in triangular mesh SGH Lagrangian simulation }%

\author[1]{Li Zhao}
\author[2]{Bo Xiao \corref{cor1}}
\cortext[cor1]{e-mail: homenature@163.com;}
\author[2]{Ganghua Wang}
\author[3]{Haibo Zhao}
\author[2]{Jinsong Bai}
\author[1]{Chunsheng Feng}
\author[1]{Shi Shu}

\address[1]{School of Mathematics and Computational Science, Xiangtan University, Xiangtan 411105, China}
\address[2]{Institute of Fluid Physics, CAEP, Mianyang 621999, China}
\address[3]{Department of Mechanics and Aerospace Engineering, Southern University of Science and Technology, Shenzhen 518055, China}

\received{}
\finalform{}
\accepted{}
\availableonline{}
\communicated{}

\begin{abstract}
When the SGH Lagrangian based on triangle mesh is used to simulate compressible hydrodynamics, because of the stiffness of triangular mesh, 
the problem of physical quantity cell-to-cell spatial oscillation (also called "checkerboard oscillation") is easy to occur. 
A matter flow method is proposed to alleviate the oscillation of physical quantities caused by triangular stiffness.
The basic idea of this method is to attribute the stiffness of triangle to the fact that the edges of triangle mesh can not do bending motion, 
and to compensate the effect of triangle edge bending motion by means of matter flow. 
Three effects are considered in our matter flow method: (1) transport of the mass, momentum and energy carried by the moving matter; 
(2) the work done on the element, since the flow of matter changes the specific volume of the grid element; 
(3) the effect of matter flow on the strain rate in the element.
Numerical experiments show that the proposed matter flow method can effectively alleviate the spatial oscillation of physical quantities.
\end{abstract}


\end{frontmatter}


\section{Introduction}
The motion of compressible multi-material large deformation fluid is a common hydrodynamic process in the fields of high energy density such as detonation, inertial confinement fusion, superhigh velocity collision, astrophysics. It is also a difficult point in hydrodynamic numerical simulation. Currently, among the main technical schemes to simulate the motion of compressible multi-material large deformation fluid are Euler method \cite{Banks-Godunov,Niem-volume-of-fluid,Zheng-Richtmyer-Meshkov-instability,Sambasivan-Local-Mesh-Refinement,Zheng-adaptive-solution,Movahed-Richtmyer-Meshkov-instability,Chen-real-ghost-fluid,Zhang-Hydrodynamic-instabilities,Mainc-FVM,Kapahi-multi-material-flow,Diot-higher-order,Sijoy-Eulerian-multi-material,Dimitrios-multi-component,He-Characteristic-based,Wang-multiphase-flows,Man-nonlinear,Liu-Adaptive-THINC-GFM}~and Arbitrary Lagrange-Euler (ALE) method \cite{Barlow-ALE,Anbarlooei-MMALE,Milan-MMALE,Tian-global-ALE,Galera-CCALE,Jia-MOF,Zeng-MMALE}. The ALE method here usually refers to the ALE method that allows the interface mesh to move across different matters, also known as MMALE (Multi-Material ALE). Because of the cross-matter motion of the grid in the ALE method, there will be mixed matter grid elements similar to the Eulerian method, and the mixing can cause the dispersion of the material interface. In order to control the interface dispersion, it is necessary to introduce interface reconstruction (such as VOF \cite{Niem-volume-of-fluid,Sijoy-Eulerian-multi-material,Milan-MMALE} and MOF \cite{Anbarlooei-MMALE,Milan-MMALE,Galera-CCALE,Jia-MOF}) into Euler method and ALE method.

At present, Lagrangian method \cite{Cheng-Positivity-preserving,WickeDynamic,wangruili,WangRuiLi2014AdaptiveLagrange,Liu2016,Cheng-Second-order-symmetry-preserving,Georges-3DGCL,zhaohaibobiyelunwen,zhaohaibo2018paper,Waltz-nodal-Godunov,Scovazzi-tetrahedral-meshes,Morgan-Godunov-like}, despite its advantage in edge-tracking for multi-material fluids, has not become the mainstream method of compressible multi-material large deformation hydrodynamic simulation. Its main reasons include mesh distortion, physical quantity oscillation (mainly in two-dimensional triangular mesh and three-dimensional tetrahedral mesh), and it is not easy to deal with the interface topology changes caused by material collision or fracture. The oscillation of physical quantities is the focus of this paper.

In Lagrangian hydrodynamic simulation, when triangular or tetrahedral meshes are used, it is easy to appear the cell-to-cell oscillation distribution of physical quantities between mesh elements, that is, the problem of checkerboard oscillation of physical quantities. The reason can be attributed to the mesh stiffness of triangular or tetrahedral meshes. For the nonphysical checkerboard oscillation problem, some research work  \cite{Scovazzi-tetrahedral-meshes,Morgan-Godunov-like} has been done at present. In 2012, G.Scovazzi \cite{Scovazzi-tetrahedral-meshes} took the lead in discussing the use of "Flux" to alleviate stiffness. Starting from variable multiscale anaysis (VMS), scovazzi makes a linear approximation of the mesh motion on a finer scale and transforms it into flux on the edge of a large-scale mesh. After a series of approximations, Scovazzi discarded many complex terms, and finally retained an energy diffusion term proportional to the pressure gradient in flux. Scovazzi's "Flux" method can alleviate the oscillation in some typical shock wave problems. However, this method also causes some other non physical effects, such as the density increases instead of decreasing at the wall heating. This may be due to the over simplification of Scovazzi's "Flux" term. In 2015, N. R. Morgan \cite{Morgan-Godunov-like} also discussed a method of using "Flux" to alleviate the stiffness. The author thinks that in the Point-Centered Lagrangian hydrodynamics (PCH) discretization, stiffness originates from the volume error, and then proposes a matter flow term to correct the error. The so-called matter flow means that matter is allowed to be transported from one control body to another. In addition to mass, the energy and momentum carried by matter will be transported along with the matter.

Similar to the method in the literature \cite{Scovazzi-tetrahedral-meshes,Morgan-Godunov-like}, this paper also constructs a "Flux" method to alleviate the physical quantity oscillation caused by triangle stiffness in two-dimensional SGH Lagrange simulation. The core idea of this method is to attribute the stiffness of triangular mesh to the fact that there is no proper bending on the edge of the mesh element, and a "Flux" is constructed to replace the edge bending effect. "Flux" method in this paper considers a variety of effects, including the mass, momentum, and energy transport caused by the transport of matter between elements (similar to that done in the Morgan), the energy transport between elements due to the work done by the specific volume change of the elements (this is a bit like that done in the Scovazzi), and the influence of material transport on the strain rate of the element. To facilitate the application of matter flow method in parallel hydrodynamic program, this paper also designs a parallel implementation scheme of matter flow SGH Lagrange method based on OpenMP \cite{Chapman-Parallel}.

The following chapters are arranged as follows: Section \ref{sec:two} introduces the compressible hydrodynamic equations and discrete format. Section \ref{sec:three} presents a matter compensation flow method. Section \ref{sec:four} discusses the parallel implementation algorithm based on shared memory. Section \ref{sec:five} gives numerical examples and analysis. Section \ref{sec:six} summarizes and discusses the research work of this paper.

\section{Compressible hydrodynamic equations and discrete format}\label{sec:two}
\subsection{Two dimensional compressible hydrodynamic equations}
Consider the following two dimensional compressible hydrodynamic equations in this article. Equations \eqref{eq:mass-equation} - \eqref{eq:state-equation} are mass equation, momentum equation, internal energy equation and equation of state,respectively.

\begin{equation}\label{eq:mass-equation}
\frac{1}{\rho }\frac{{d\rho }}{{dt}} + \nabla  \cdot {\bm{u}} = 0,
\end{equation}
\begin{equation}\label{eq:momentum-equation}
\rho \frac{{d{\bm{u}}}}{{dt}} =  - \nabla (p + q),
\end{equation}
\begin{equation}\label{eq:inner-energy-equation}
\rho \frac{{de}}{{dt}} =  - (p + q) \nabla  \cdot {\bm{u}},
\end{equation}
\begin{equation}\label{eq:state-equation}
p =  p(\rho,e).
\end{equation}
where $\rho$, $e$, $\bm{u}$ are the density, specific internal energy and velocity of the fluid, $p$ is the pressure and $q $ is the artificial viscosity. The specific forms of differential operators $ \nabla \cdot $ and $ \nabla $ are respectively $ \frac{\partial}{\partial x}$+$ \frac{\partial}{\partial y} $ and $ (\frac{\partial}{\partial x},\frac{\partial}{\partial y})^T $, and $ \frac{d}{dt} $ is the Lagrangian time derivative.

In the example of this paper, the equation of state \eqref{eq:state-equation} is taken as the ideal gas equation of state, which is expressed as follows:
\begin{equation}\label{eq:gas-state2}
p(\rho,e) = (\gamma - 1)\rho e
\end{equation}
where $ \gamma$ is the gas adiabatic index.

\subsection{Finite volume discretization scheme based on SGH}\label{sec:sgh}
\begin{figure}[h]
	\subfigure[triangular element $ c $]{\label{fig:11}
		\begin{minipage}[t]{0.5\linewidth}
			\centering
			\includegraphics[width=4cm]{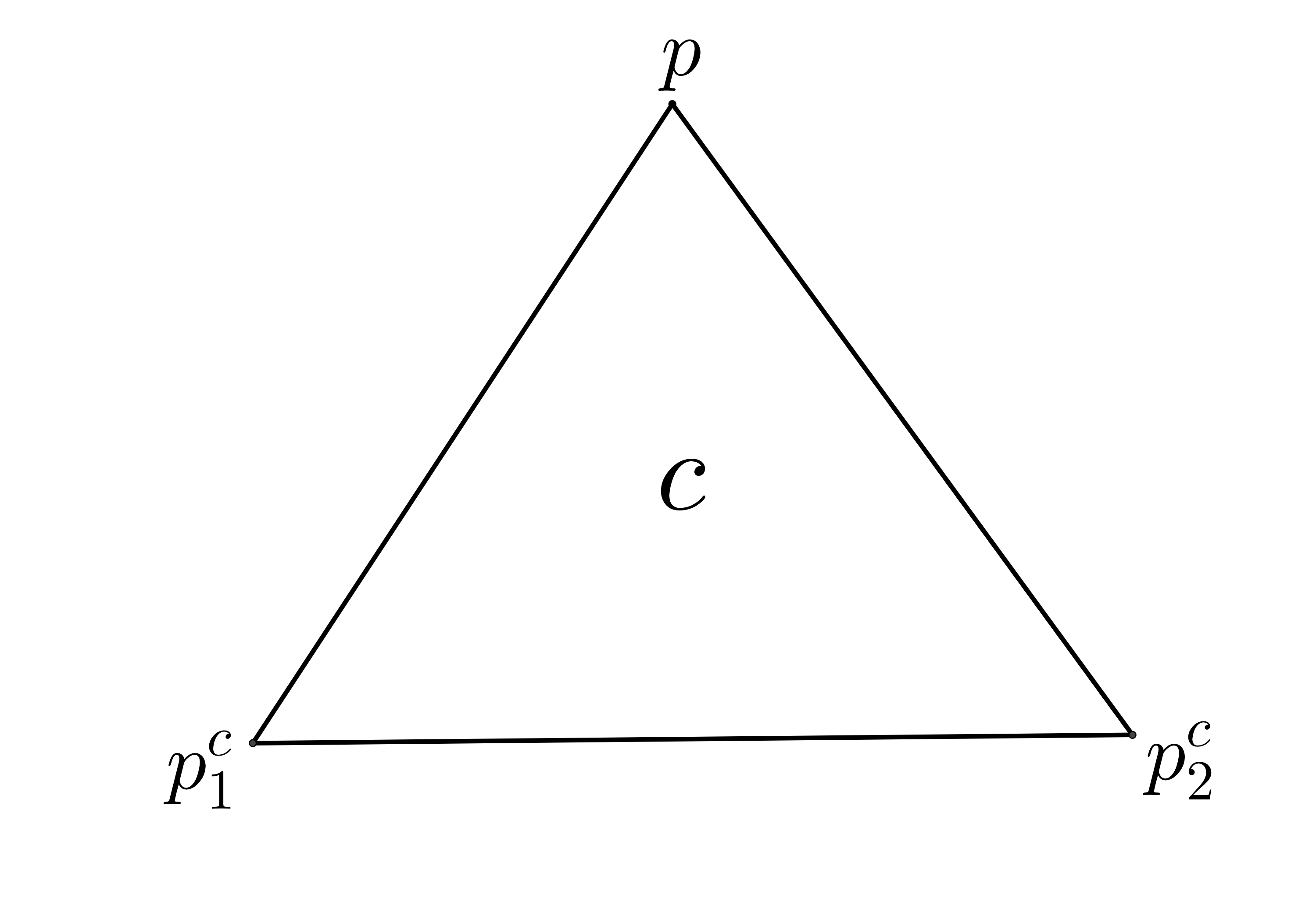}
		\end{minipage}
	}%
	\subfigure[node~$p$]{\label{fig:2.2-1:b}
		\begin{minipage}[t]{0.5\linewidth}
			\centering
			\includegraphics[width=4cm]{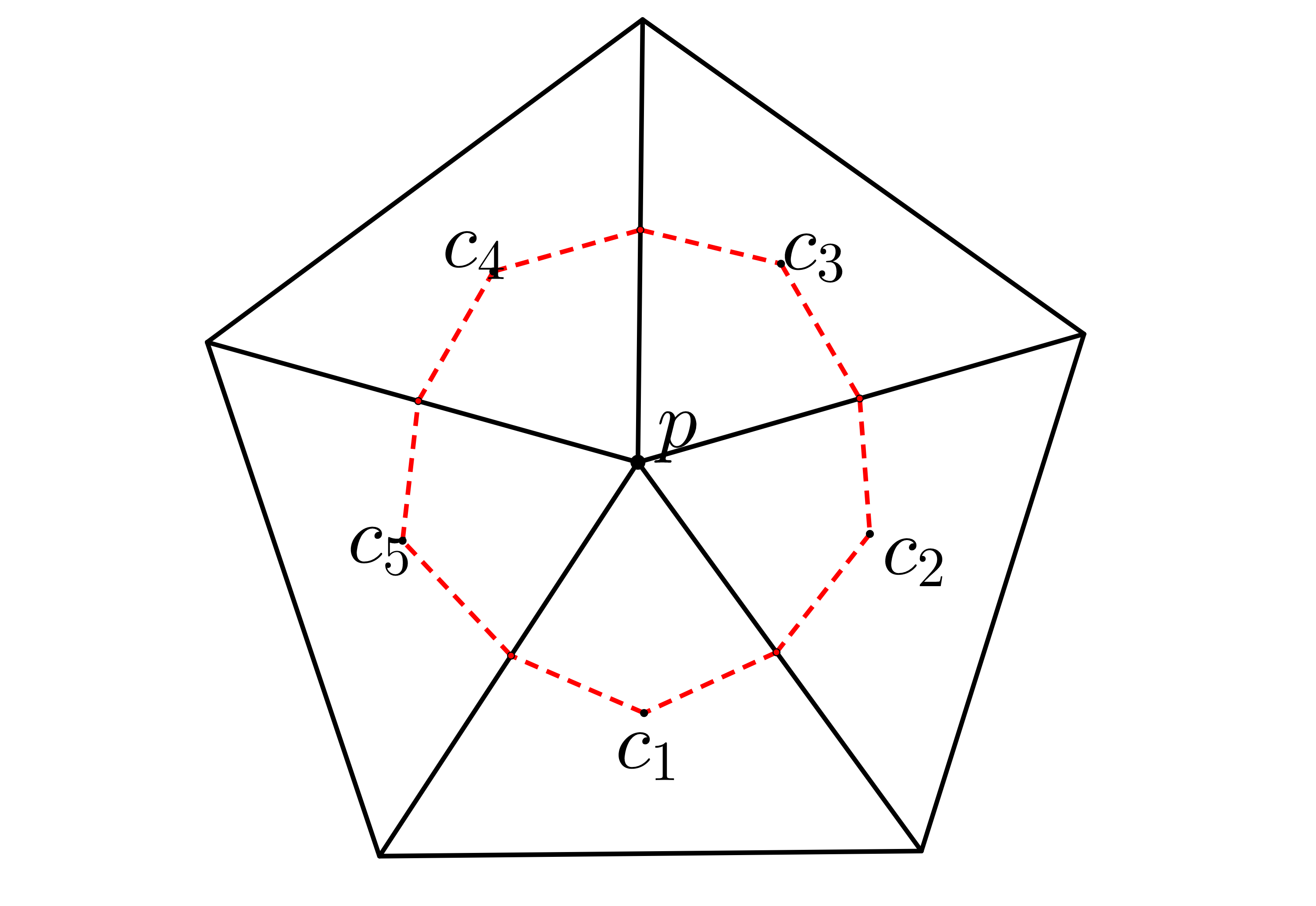}
		\end{minipage}
	}%
	\caption{Control volume diagram}
	\label{fig:control-volume}
\end{figure}

In computational fluid dynamics, according to locations (in elements or on grid points) on which physical quantities in the Lagrangian method are defined, it can usually be divided into Staggered-Grid Hydrodynamics (SGH) method, Cell-Centered Hydrodynamics (CCH) method and Point-Centered Hydrodynamics (PCH) method. In the SGH method, the pressure, density and internal energy are defined in the center of the element, and the velocity and kinetic energy are defined on the nodes. In this paper, the SGH Lagrangian finite volume method is used to discretize the control equations, and the physical quantities defined in cells are treated as piece-wise constant. The discrete control volume is shown in the \reffig{fig:control-volume}.

For the SGH Lagrangian finite volume method, the discrete form of compressible hydrodynamics equations~\eqref{eq:mass-equation}-~\eqref{eq:inner-energy-equation}~ are expressed as follows:
\begin{equation}\label{eq:cell-mass-discrete}
m_c = {\rm constant},
\end{equation}
\begin{equation}\label{eq:point-mass-discrete}
m_p = \frac{1}{3} \sum_{c \in T(p)} m_c,
\end{equation}
\begin{equation}\label{eq:velocity-discrete}
{\bm{u}}^{n+1}_p = {\bm{u}}^{n}_p + \frac{ \sum_{c \in T(p)} {  \bm{f}}^{n}_{pc} }{m_p} \Delta t,
\end{equation}
\begin{equation}\label{eq:position-discrete}
{\bm{x}}^{n+1}_p = {\bm{x}}^{n}_p + {\bm{u}}^{n}_p \Delta t + \frac{\sum_{c \in T(p)} {\bm{f}}^{n}_{pc}  }{2 m_p} (\Delta t)^2,
\end{equation}
\begin{equation}\label{eq:inner-energy-discrete}
E_c^{n+1} = E_c^{n} - \sum_{p \in P(c)} {\bm{f}}^{n}_{pc} \cdot ({\bm{x}}^{n+1}_{p} - {\bm{x}}^{n}_{p}).
\end{equation}

\noindent where $m_c$ represents the mass of the element $c$, $m_p$ denotes the mass of the node $p$, $ {\bm{u}}^{n}_p$ and $ {\bm{u}}^{n+1}_p$ respectively represent the velocity of the node $p$ at the moments $t^n$ and $t^{n+1}$, $ {\bm{x}}^{n}_p$ and ${\bm{x}}^{n+1}_p $ respectively represent the position of the nodes $ p$ at $ t^n$ and $ t^{n+1}$ moments, $E^n_c $ and $ E^{n+1}_c$ respectively indicate the internal energy of element $c $ at $ t^n$ and $t^{n+1}$, $ T(p)$ represents all the element sets containing nodes $p$, $P(c)$ represents all the node sets in element $ c$, $ {\bm{f}}^{n}_{pc}$ represents the force of the element $ c$ on the node $ p$, its expression is as follows:
\begin{equation*}
{\bm{f}}_{pc} =  (p_c + q_c)
\left(
\begin{array}{c}
- \frac{ y_{p_2^c} -y_{p_1^c}  }{2} \\
\frac{ x_{p_2^c} - x_{p_1^c} }{2} \\
\end{array}
\right),
\end{equation*}
where~$p_c$, $q_c$ denotes the pressure and viscous in the element $c$. The expression of $q_c$ is as follows:
\begin{equation}\label{eq:qc}
q_c =  - c_{visc} \rho_c \dot{v},
\end{equation}
where $c_{visc}$ is the viscosity coefficient, $\rho_c$ is the density in the grid element $c$, and $\dot{v}$ is the relative change rate of volume. The viscosity coefficient is selected as follows:
\begin{equation}\label{eq:visc}
c_{visc} = k \max \big\{ -2\dot{v} h^2, v_s h \big\}. 
\end{equation}
where $h$ is the maximum value of the three sides of the triangular element, and $v_s$ is the sound velocity and $k$ is an adjustable factor, which is taken as 2.0 in this paper.

\subsection{Time step control}
To retain time stability, the time step $ \Delta t $ in the evolution of Lagrange hydrodynamics needs to satisfy the stability condition. In this paper, the time step is selected according to the following conditions.
\begin{equation*}
\Delta t = \min\limits_{c \in \mathscr{C}} \Big\{ \Delta t^c_{s},~ \Delta t^c_{v},~\Delta t^c_{fv},~\Delta t^c_{fa}   \Big\}.
\end{equation*}
where $ \mathscr{C} $ is the set of grid elements, $ \Delta t^c_{s},~ \Delta t^c_{v},~\Delta t^c_{fv},~\Delta t^c_{fa}$ for each element are as follows.

\begin{enumerate}[(1)]
	\item Time step determined by sound velocity:
	\begin{equation*}
	\Delta t^c_{s}  =  C_{safe} \cdot \frac{h_{min}}{v_{s}}.
	\end{equation*}
	where $C_{safe}=0.05$ is an adjustable safe factor, which is taken to be 0.05 all through the following, $h_{min}$ is the minimum height of the triangle, $v_{s} $ is the sound velocity of the triangle.

	\item Time step determined by viscosity:
	\begin{equation*}
	\Delta t^c_{v}  = C_{safe} \cdot \frac{h_{min}^{2}}{c_{visc}}.
	\end{equation*}
	where $c_{visc}$ is the viscosity coefficient of the triangle.
	
	\item  The time step determined by the velocity of matter flow in a triangle:
	\begin{equation*}
	\Delta t^c_{fv}  = C_{safe} \cdot \min \Big\{ \frac{h_{1}}{3|u_{flow}^{1}|},~\frac{h_{2}}{3|u_{flow}^{2}|},~\frac{h_{3}}{3|u_{flow}^{3}|} \Big\}.
	\end{equation*}
	where $h_{1},~h_{2},~h_{3}$ is the three heights of the triangle, $u_{flow}^{1},~u_{flow}^{2},~u_{flow}^{3}$ are the matter flow velocities on the three sides of the triangle (see section \ref{sec:three}).

	\item The time step determined by the acceleration of triangular matter flow:
	\begin{equation*}
	\Delta t^c_{fa}  = C_{safe} \cdot \min \Big\{\sqrt{ \frac{2h_1}{3|a^1_{flow}|} },~\sqrt{ \frac{2h_2}{3|a^2_{flow}|} },~\sqrt{ \frac{2h_3}{3|a^3_{flow}|} } \Big\}.
	\end{equation*}
	where $a^1_{flow},~a^2_{flow},~a^3_{flow}$ is the acceleration of matter flow on the three sides of the triangle (see section \ref{sec:three}).
	
\end{enumerate}

\section{Matter compensation flow method}\label{sec:three}

\begin{figure}[h]
	\subfigure[Quadrilateral computing grid]{\label{fig:physics-shock-1}
		\begin{minipage}[t]{0.3\linewidth}
			\centering
			\includegraphics[width=3cm]{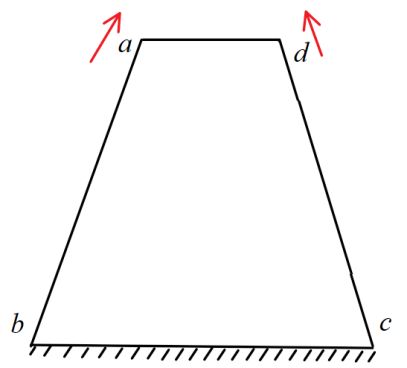}		
		\end{minipage}
	}%
	\subfigure[Triangular computing grid]{\label{fig:physics-shock-2}
		\begin{minipage}[t]{0.3\linewidth}
			\centering
			\includegraphics[width=3cm]{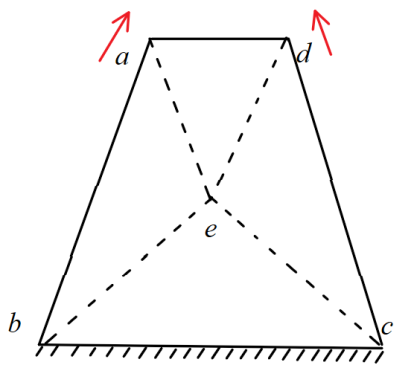}
		\end{minipage}
	}%
	\subfigure[Non-physical oscillation phenomena]{\label{fig:physics-shock-3}
		\begin{minipage}[t]{0.3\linewidth}
			\centering
			\includegraphics[width=3cm]{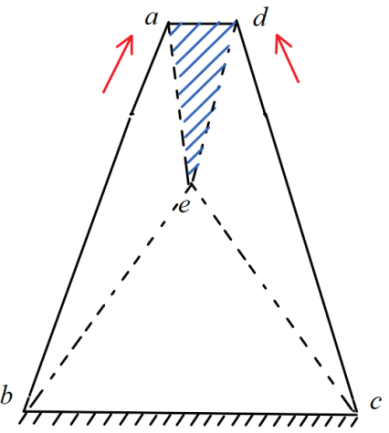}
		\end{minipage}
	}%
	\caption{Oscillation of physical quantities in triangular meshes \cite{zhaohaibobiyelunwen}}
	\label{fig:physics-shock}
\end{figure}

When the SGH Lagrangian method based on section \ref{sec:sgh} is used to simulate the motion of compressible hydrodynamics, it is easy to appear the phenomenon of physical quantity cell-to-cell oscillation caused by the stiffness of triangular mesh. \reffig{fig:physics-shock} gives an intuitive description of the physical quantity spatial oscillation caused by the stiffness of the triangular mesh: suppose \reffig{fig:physics-shock-1} the quadrilateral mesh $ abcd $ be filled with fluid, nodes $ b $ and $ c $ are fixed, nodes $ a $ and $ d $ move in the direction of the arrow. Under these conditions, fluid density in the quadrilateral $ abcd $ will decrease as the mesh area increases. While in \reffig{fig:physics-shock-2}, a quadrilateral grid $ abcd $ is divided into four triangular grids. The rest of the conditions do not change. With the movement of nodes $ a $ and $ d $, the length of the edge $ ad $ decreases, and the area of the triangle $ ade $ becomes smaller. Finally, the density and pressure in the triangle $ ade $ increase significantly higher than that in the adjacent triangle. As a result, the cell-to-cell oscillation phenomenon of physical quantities appears, as shown in \reffig{fig:physics-shock-3}.

\begin{figure}[h]
	\centering
	\subfigure[Triangle mesh edge bending]{\label{fig:triangle-edge-bend-1}
		\begin{minipage}[t]{0.4\linewidth}
			\centering
			\includegraphics[width=3cm]{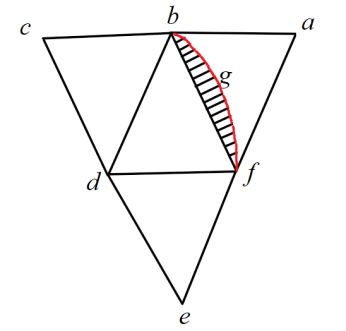}		
		\end{minipage}
	}%
	\subfigure[Approximate the curved area with a triangle.]{\label{fig:triangle-edge-bend-2}
		\begin{minipage}[t]{0.4\linewidth}
			\centering
			\includegraphics[width=3cm]{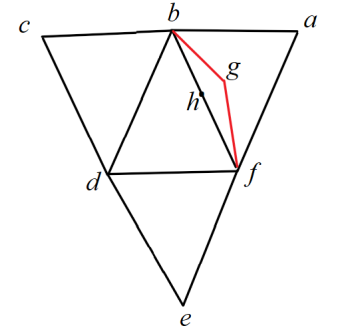}
		\end{minipage}
	}%
	\caption{Triangle mesh edge bending diagram}
	\label{fig:triangle-edge-bend}
\end{figure}

Inspired by the above analysis, we attribute the stiffness of a triangle to the fact that the edges of the triangle cannot do bending motion. Taking \reffig{fig:physics-shock-3} as an example, if the edge $ ae $ and $ de $ of the triangle $ ade $ can do bending motion, the edge $ ae $ and $ de $ will bend outward with the increase of pressure in the triangle $ ade $, which will compensate for the decrease of the area of the curved triangle $ ade $.  Thus, the oscillation of physical quantity is alleviated (This also tells us that in principle, if we adopt a Lagrangian method \cite{Barlow-ALE} which allows the grid to bend, the oscillation of physical quantities can be alleviated). From this point of view, this paper proposes a method of matter compensation flow to approximate the effect of triangular side bending motion, so as to alleviate checkerboard oscillation. The basic idea of this method is shown in \reffig{fig:triangle-edge-bend}. In the \reffig{fig:triangle-edge-bend-1}, let triangle $bdf$ pressure be greater than triangle $abf$. If the sides of a triangle can bend, under pressure differential, the edge $bf$ will become a curved $bgf$. Because in the usual Lagrangian simulation, the mesh is actually not allowed to bend, so we can consider using the material compensation flow between cells to replace the effect of the edge bending motion (\reffig{fig:triangle-edge-bend-1} the area of the shadow part determines the amount of matter flow). Because the curved edge  $ bfg $  is not easy to obtain, in order to calculate conveniently, the shadow part is approximated to a triangle, as shown in \reffig{fig:triangle-edge-bend-2}. The triangle can be considered to be formed by the midpoint $ h $ of the edge $ bf $ moving to node $ g $ under the action of pressure difference.

\begin{figure}[h]
	\centering
	\subfigure[$\Delta x \geq 0$]{\label{fig:matterflow-1}
		\begin{minipage}[t]{0.4\linewidth}
			\centering
			\includegraphics[width=4cm]{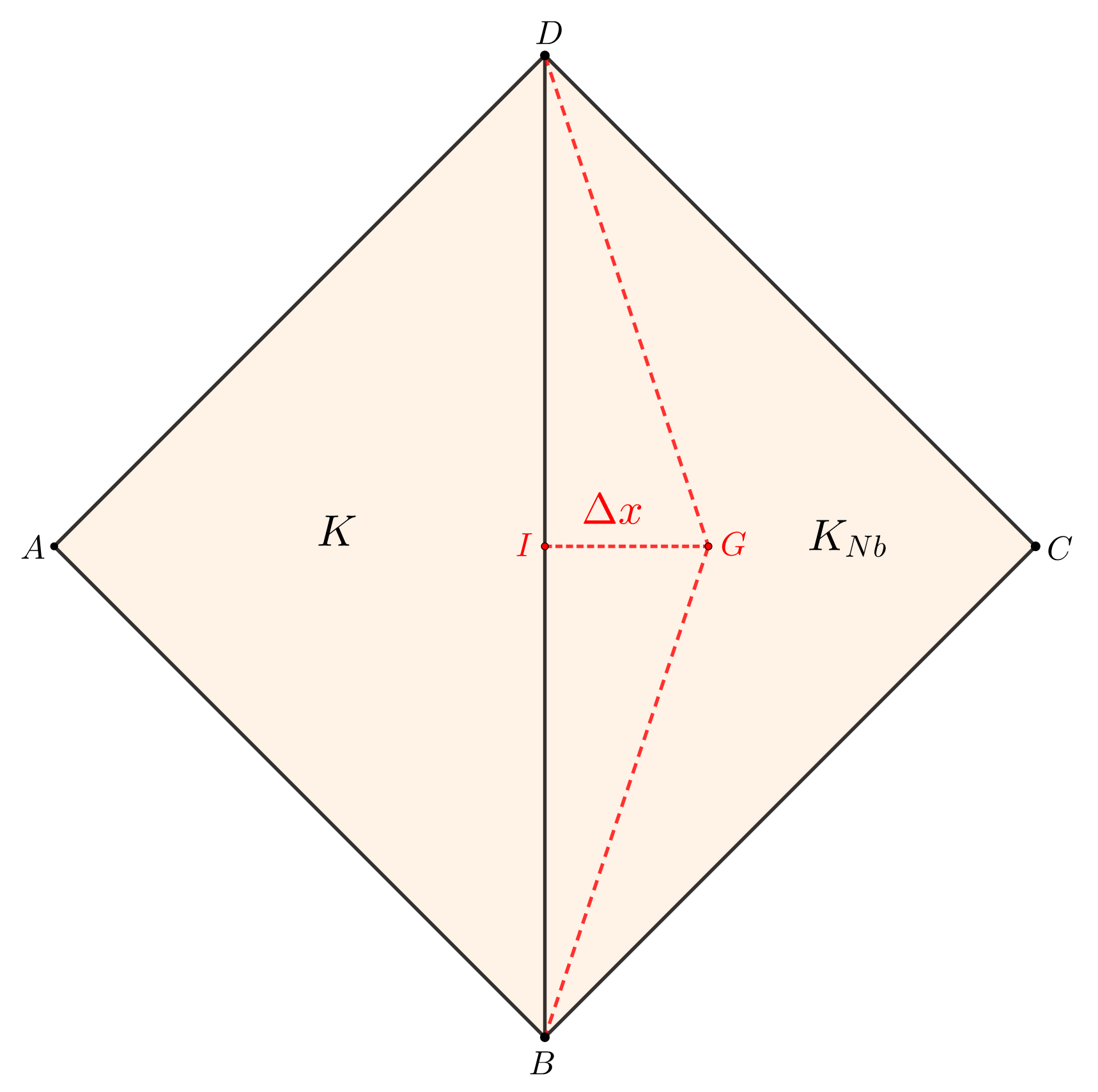}
		\end{minipage}
	}%
	\subfigure[$\Delta x < 0$]{\label{fig:matterflow-2}
		\begin{minipage}[t]{0.4\linewidth}
			\centering
			\includegraphics[width=4cm]{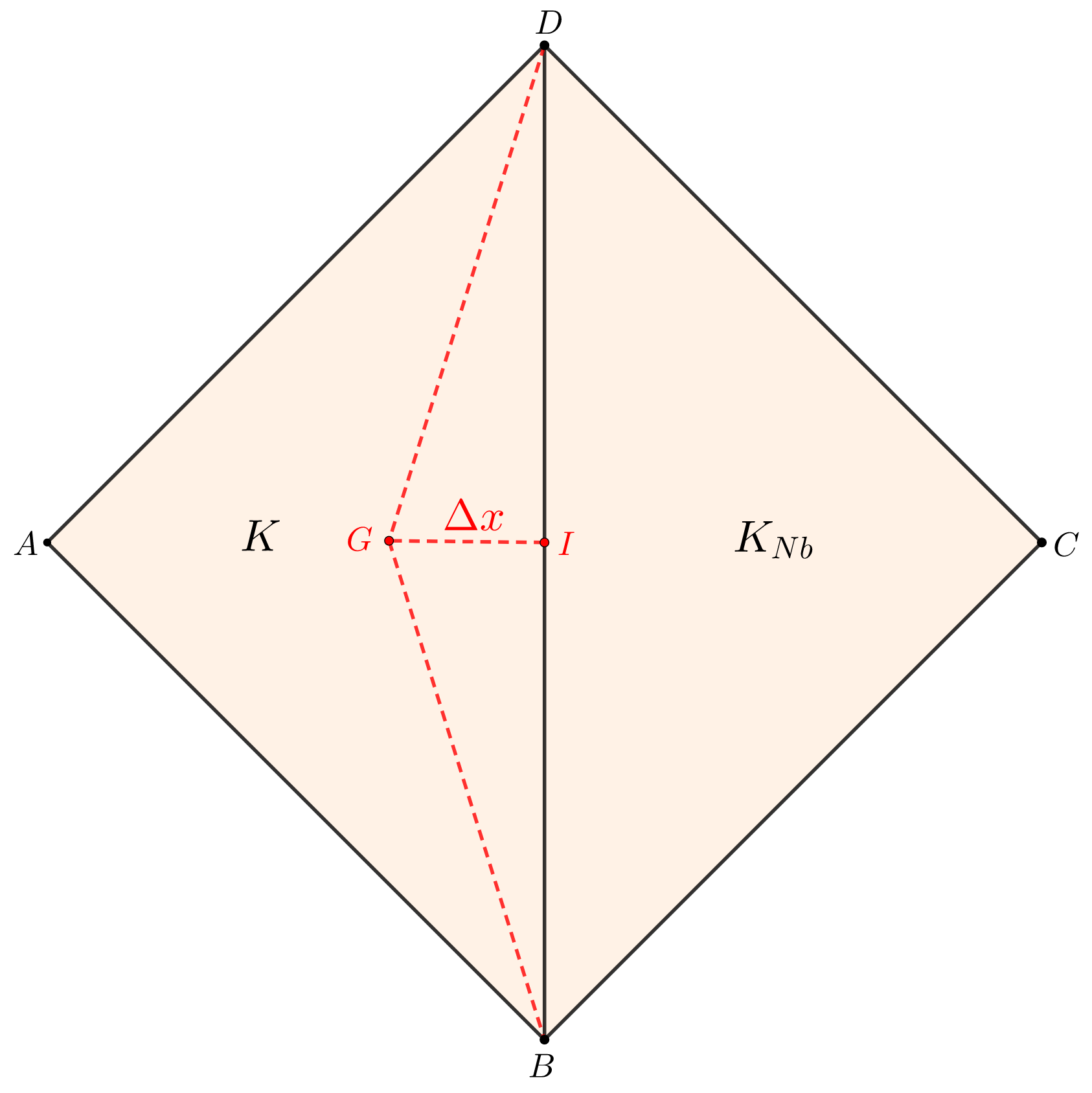}
		\end{minipage}
	}%
	\caption{Illustration of the matter compensation flow method. }\label{fig:matterflow}
\end{figure}

There are three effects related to the matter compensation flow. The first effect is that the mass, momentum and energy carried by the matter are transferred from the grid cell to the adjacent grid cell in the process of flow. The second effect is that the specific volume of the grid element is changed due to the "squeezing in" and "extrusion" of the matter from the grid cell, which will produce work effect on the original matter. The third effect is that the volume strain rate of the grid element is also affected by the matter flow, which leads to the change of the viscous stress of the grid element, which will eventually affect the evolution of the internal energy. Based on the above three effects, the steps of the matter compensation flow method are designed as follows (\reffig{fig:matterflow} gives an illustration and defines some of the symbols for the description of the steps).

\begin{enumerate}	
	\item [\textbf{Step 1}] Using the accelerations of node $B$ and node $D$ to calculate the component of the average acceleration in the normal direction of the midpoint $ I $ of the edge $ BD $:
	\begin{equation}\label{eq:bar_a_I}
	{\bar{a}}_{I} = \bm{s}_n \cdot \frac{\bm{a}_1 + \bm{a}_2}{2}.
	\end{equation}
	where $\bm{s}_n$ is the out of unit normal vector of the edge $ BD $ in the element $K$, $\bm{a}_1$ is the acceleration of node $B$, $\bm{a}_2$ is the acceleration of node $D$.

	\item [\textbf{Step 2}]	The acceleration of node $ I $ is also calculated by the pressure difference between triangular element $ K $ and $K_{Nb}$:
	\begin{equation}\label{eq:a_I}
	{a}_{I} = \frac{ p_{K}-p_{K_{Nb}} }{m_I},
	\end{equation}	
	where $p_{K}$ and $p_{K_{Nb}}$ are the pressures of elements $ K $ and $K_{Nb}$ respectively, and the mass of node $ I $ is:
	\begin{equation*}
	m_I = \frac{m_{K} + m_{K_{Nb}}}{ 4 }.     
	\end{equation*}     
	$m_{K}$ is the mass of element $ K $ and $ m_{K_{Nb}}$ is the mass of $K_{Nb}$.

	\item [\textbf{Step 3}]	The difference between $a_I$ and $\bar{a}_I$ determines the acceleration of matter flow:
	\begin{equation}\label{eq:a_flow}
	{a}_{flow} = {a}_{I} - {\bar{a}}_{I},
	\end{equation}
	\begin{equation*}
	\bm{a}_{flow} = {a}_{flow} \bm{s}_n.
	\end{equation*}
	where the direction of $\bm{a}_{flow}$ is the normal direction of the edge $ BD $.

	\item [\textbf{Step 4}]	$ The matter flow acceleration {a}_{flow} $ determines the imaginary movement of the node $ I $:
	\begin{equation}\label{eq:DeltaT-uFlow}
	\begin{array}{*{20}{l}}
	~~\Delta x     &=& u_{flow}^{n} \Delta t   + \frac{1}{2} a_{flow}^{n} (\Delta t)^{2}, \\
	u_{flow}^{n+1} &=& (u_{flow}^{n}   +  a_{flow}^{n} \Delta t)(1 - C_{diss}). 
	\end{array}
	\end{equation}	
	where $u_{flow}^{n}$ is the size of the matter flow velocity of the $t^n$ moment, ${a}_{flow}^{n}$ is the amount of matter flow acceleration size at the $ t^n $ moment, $ \Delta t $ is the time step, and $ C_{diss} $ represents the artificial dissipation factor of the flow velocity, which is proportional to the viscous coefficient:
	\[ C_{diss} = 
	\max \Big\{ \frac{3}{S_{K}} c^{K}_{visc}, \frac{3}{ S_{K_{Nb}} } c^{K_{Nb}}_{visc} \Big\} \] 
	here, $ c^{K}_{visc} $ and $ S_{K} $ are the viscosity coefficient and area of the element $ K $, and $ c^{K_{Nb}}_{visc} $ and $ S_{K_{Nb}} $ are the viscosity coefficient and area of the element $ K_{Nb} $, respectively.

	\begin{rmk}
		The direction of $\Delta x$ and $u_{flow}^{n+1}$ is the normal direction of edge BD.
	\end{rmk}

	\item [\textbf{Step 5}]	Mass compensation, according to conservation of mass:
	\begin{equation}\label{eq:mass-conservation}
	\begin{array}{*{20}{l}}
	m'_{K} = m_{K} - \Delta M	\\
	m'_{K_{Nb}} = m_{K_{Nb}} + \Delta M	
	\end{array}
	\end{equation}
	where $\Delta M $ is the mass carried by the matter flow, and its calculation formula is
	\begin{equation}\label{eq:MassFlowOut}
	\Delta M = \left\{ {\begin{array}{*{20}{l}}
		{ \frac{1}{2} \Delta x L\rho_K }&{,\Delta x \geq 0}\\
		{ \frac{1}{2} \Delta x L{\rho_{K_{Nb} }}}&{,\Delta x < 0}
		\end{array}} \right.	
	\end{equation}
	$ L $ is the length of edge $ BD $, and the $ \rho_K  $ and $ \rho_{K_{Nb} } $ are the density of $ K $ and $ K_{Nb} $, respectively.
	
	According to the definition of node mass, the mass of nodes $ A $ and $ C $ is modified as
	\begin{equation}\label{eq:node-mass-conservation}
	\begin{array}{*{20}{l}}
	m'_{A} = m_{A} - \Delta m	\\
	m'_{C} = m_{C} + \Delta m
	\end{array}
	\end{equation}
	where $  \Delta m := \frac{1}{3} \Delta M $, the mass of nodes $ B $ and $ D $ does not change.

	\item [\textbf{Step 6}]	The change of internal energy should not only consider the $ \Delta E$ carried by the matter flow, but also the extra work $ W_{extra} $ caused by the change of element volume.
	\begin{equation}\label{eq:internal-energy-conservation}
	\begin{array}{*{20}{l}}
	E'_K = E_K - \Delta E - W_{extra}	\\
	E'_{K_{Nb}} = E_{K_{Nb}} + \Delta E + W_{extra}
	\end{array}
	\end{equation}
	where $ E_K  $ and $ E_{K_{Nb}} $ are the internal energies of element $ K $ and $K_{Nb}$ respectively, and the calculation formula of $ \Delta E$ is
	\begin{equation}\label{eq:InternalEnergyOut}
	\Delta E = \left\{ {\begin{array}{*{20}{l}}
		{\frac{{\Delta M}}{m_K} E_K}				 	&	{,\Delta M \ge 0}	\\
		{\frac{{\Delta M}}{ m_{K_{Nb}} }{E_{K_{Nb}}}}	&	{,\Delta M < 0}
		\end{array}} \right.	
	\end{equation}
	$ W_{extra} $ formula (See \textbf{Remark \ref{remark:extra-work}} for details) is
	\begin{equation}\label{eq:extra-work}
	W_{extra} = \frac{1}{2} \Big[ (p_K + q_K) \frac{\Delta M }{\rho_K} +  (p_{K_{Nb}}+q_{K_{Nb}}) \frac{\Delta M }{\rho_{K_{Nb}}} \Big].
	\end{equation}
	here $ q_K $ is the viscous force of element $K $ and $ q_{K_{Nb}} $ is the viscous force of element $ {K_{Nb}} $
	
	\begin{rmk}\label{remark:extra-work}	
		The extra work done by the movement of matter to cause changes in element volume is:
		\begin{equation*}\label{eq:matter-6.1}
		W_{extra} = (p + q) \Delta v,
		\end{equation*}
		where $p$ and $q$ are the pressure and viscous force of the element. $\Delta v$ is the volume change of the element, and its calculation formula is:
		\begin{equation*}\label{eq:matter-6.2}
		\Delta v = \frac{\Delta M}{m} v = \frac{\Delta M }{\rho},
		\end{equation*}
		So the extra work can be written as:
		\begin{equation*}\label{eq:matter-6.3}
		W_{extra} = (p + q) \frac{\Delta M }{\rho},
		\end{equation*}
		To maintain conservation of energy, the extra work is averaged over the element $K $ and $ K_{Nb} $, and then written as:
		\begin{equation*}\label{eq:matter-6.4}
		W_{extra} = \frac{1}{2} [ (p_K + q_K) \frac{\Delta M }{\rho_K} +  (p_{K_{Nb}}+q_{K_{Nb}}) \frac{\Delta M }{\rho_{K_{Nb}}} ].
		\end{equation*}
	\end{rmk}

	\item [\textbf{Step 7}]	The change of volume relative rate caused by matter flow is as follows:
	\begin{equation}\label{eq:matter-5}
	\Delta \dot{v} = \left\{ {\begin{array}{*{20}{l}}
		\frac{\Delta M}{m_K} \frac{1}{\Delta t}				 	&	{,\Delta M \ge 0}	\\
		\frac{\Delta M}{m_{K_{Nb}}} \frac{1}{\Delta t}			&	{,\Delta M < 0}
		\end{array}} \right.	
	\end{equation}
	 The change of volume relative rate affects the value of viscous by \eqref{eq:qc}, and then affects the change of internal energy and other physical quantities.

	\item [\textbf{Step 8}]	Velocity compensation:
\begin{equation}\label{eq:node-velocity}
\begin{array}{*{20}{l}}
\bm{u}'_A =  \cfrac{\bm{p}'_A}{m'_A}	\\
\bm{u}'_C =  \cfrac{\bm{p}'_C}{m'_C}
\end{array}
\end{equation}
where $ \bm{p}'_A = \bm{p}_A - \Delta \bm{p}$, $ \bm{p}'_C = \bm{p}_C + \Delta \bm{p}$, $ \bm{p}_A $ and $ \bm{p}_C $ are the momentum of nodes $ A $ and $ C $, respectively. $ \Delta \bm{p} $ satisfies the following optimization problems:
\begin{equation}\label{eq:optimal-problem-s}
\left\{ \begin{array}{l}
\Delta \bm{p} = \arg\min \limits _{\bm{p} \in R^2}  \big\{ |\bm{p} - \bm{\lambda}| \big\} \\
a \langle \bm{p},\bm{p} \rangle + \langle \bm{b}, \bm{p} \rangle + c = 0.
\end{array} \right.
\end{equation}
here $ \bm{\lambda} = \Delta m \frac{\bm{u}_B + \bm{u}_D}{2} $, $ a = \frac{1}{m'_A}+\frac{1}{m'_C}> 0$, $ \bm{b} =  \frac{2}{m'_C} \bm{p}_C - \frac{2}{m'_A}\bm{p}_A$, $ c  = (\frac{1}{m'_A}-\frac{1}{m_A}) \langle \bm{p}_A, \bm{p}_A \rangle + (\frac{1}{m'_C}-\frac{1}{m_C}) \langle \bm{p}_C, \bm{p}_C \rangle $.

\textbf{The derivation of formula \eqref{eq:optimal-problem-s} is given below}. Take \reffig{fig:matterflow-1} as an example (\reffig{fig:matterflow-2} has the same result). We suppose that node $ A $ transports the momentum of $ \Delta \bm{p} $ (to be solved) to node $ C $, then the conservation of momentum is used
\begin{equation}\label{eq:momentum-sonservation}
\begin{array}{*{20}{l}}
\bm{p}'_A = \bm{p}_A - \Delta \bm{p},	\\
\bm{p}'_C = \bm{p}_C + \Delta \bm{p}.
\end{array}
\end{equation}

According to conservation of kinetic energy
\begin{equation}\label{kinetic-energy-conservation}
\frac{1}{2}m_A \langle \bm{u}_A,\bm{u}_A \rangle + \frac{1}{2}m_C \langle \bm{u}_C,\bm{u}_C \rangle = \frac{1}{2}m'_A \langle \bm{u}'_A,\bm{u}'_A \rangle+ \frac{1}{2}m'_C \langle \bm{u}'_C, \bm{u}'_C \rangle
\end{equation}

Since $ \bm{p} = m \bm{u} $, \eqref{kinetic-energy-conservation} can be written as		
\begin{equation}\label{eq:kinetic-energy}
\frac{1}{m_A} \langle \bm{p}_A,\bm{p}_A \rangle + \frac{1}{m_C} \langle \bm{p}_C,\bm{p}_C \rangle = \frac{1}{m'_A} \langle \bm{p}'_A,\bm{p}'_A \rangle + \frac{1}{m'_C} \langle \bm{p}'_C,\bm{p}'_C \rangle 
\end{equation}

We substitute \eqref{eq:momentum-sonservation} into \eqref{eq:kinetic-energy}, and we get
\begin{equation}\label{eq:kinetic-energy-1}
\frac{1}{m_A} \langle \bm{p}_A,\bm{p}_A \rangle + \frac{1}{m_C} \langle \bm{p}_C,\bm{p}_C \rangle = \frac{1}{m'_{A}} \langle \bm{p}_A - \Delta \bm{p},\bm{p}_A - \Delta \bm{p} \rangle + \frac{1}{ m'_{C}} \langle \bm{p}_C + \Delta \bm{p},\bm{p}_C + \Delta \bm{p} \rangle 
\end{equation}	

We can get from \eqref{eq:kinetic-energy-1}
\begin{equation}\label{eq:kinetic-energy-3}
a \langle \Delta \bm{p},\Delta \bm{p} \rangle + \langle \bm{b}, \Delta \bm{p}\rangle + c = 0
\end{equation}	
where $ a = \frac{1}{m'_A}+\frac{1}{m'_C} > 0$, $ \bm{b} =  \frac{2}{m'_C}\bm{p}_C-\frac{2}{m'_A}\bm{p}_A$, $ c = (\frac{1}{m'_A}-\frac{1}{m_A}) \langle \bm{p}_A, \bm{p}_A \rangle + (\frac{1}{m'_C}-\frac{1}{m_C}) \langle \bm{p}_C, \bm{p}_C \rangle $.

The difference between the momentum carried by the matter flow and the momentum $ I $ the midpoint of the edge $ BD $ reaches a minimum, i.e. $ \min \limits \big\{ |\Delta \bm{p} - \bm{\lambda}| \big\}$, in the $ \Delta \bm{p} $ satisfying \eqref{eq:kinetic-energy-3}.
 
Thus, we can get the optimization problem \eqref{eq:optimal-problem-s}.

\begin{lem}\label{lemma:lemma1}
	The binary quadratic function  $ f(\bm{p}) = a \langle \bm{p},\bm{p} \rangle + \langle \bm{b}, \bm{p} \rangle + c $~($a>0$), $\bm{p} \in R^2$, if $ \exists~ \bm{p}_0 $, satisfies $ f(\bm{p}_0) \le 0$, then $ f(\bm{p}) = 0$ must have a real solution.
\end{lem}

\begin{thm}\label{theorem:throrem1}
	The optimal solution exists in optimization problem \eqref{eq:optimal-problem-s}, and the expression of the optimal solution is:	
	\begin{equation}\label{eq:optimal-sovler-s}
	\Delta \bm{p} =
	\begin{cases} 
	\cfrac{r}{d} \; \bm{\lambda}  +  (1 - \cfrac{r}{d}) \; \bm{p}^* & ,~d \neq 0 \\
	\bm{p}^* + r \; \bm{n} & , ~d = 0 
	\end{cases}
	\end{equation}
	where $ \bm{p}^*= -\frac{1}{2a} \bm{b} $, $ r = \sqrt{\cfrac{\langle \bm{b},\bm{b} \rangle }{4a^2} - \cfrac{c}{a} }$, $ d = \sqrt{\langle \bm{\lambda} - \bm{p}^*, \bm{\lambda} - \bm{p}^* \rangle }  $, $ \bm{n} = (1, 0)^T $.	
\end{thm}

\begin{pf}
	\textbf{Firstly, the existence of solutions is proved.}  We assume that
	\begin{equation}\label{eq:function}
	f(\bm{p}) = a \langle \bm{p},\bm{p} \rangle + \langle \bm{b}, \bm{p} \rangle + c 
	\end{equation}
	where $ a = \frac{1}{m'_A}+\frac{1}{m'_C} > 0$, $ \bm{b} =  \frac{2}{m'_C}\bm{p}_C-\frac{2}{m'_A}\bm{p}_A$, $ c = (\frac{1}{m'_A}-\frac{1}{m_A}) \langle \bm{p}_A, \bm{p}_A \rangle + (\frac{1}{m'_C}-\frac{1}{m_C}) \langle \bm{p}_C, \bm{p}_C \rangle $.
	
	Let $ \bm{p}_0 = \frac{\Delta m}{m_A} \bm{p}_A$, have
	\begin{equation}
	\begin{split}
	f(\bm{p}_0) & = a \langle \bm{p}_0,\bm{p}_0 \rangle + \langle \bm{b}, \bm{p}_0 \rangle + c \\
	& = a \frac{\Delta m^2}{m^2_A} \langle  \bm{p}_A, \bm{p}_A \rangle + \frac{\Delta m}{m_A}  \langle \bm{b}, \bm{p}_A  \rangle + c  \\
	& = (\frac{1}{m'_A}+\frac{1}{m'_C}) \frac{\Delta m^2}{m^2_A} \langle  \bm{p}_A, \bm{p}_A \rangle + \frac{2 \Delta m}{m_A}  \langle \frac{1}{m'_C}\bm{p}_C-\frac{1}{m'_A}\bm{p}_A, \bm{p}_A  \rangle + \\
	&~~~~(\frac{1}{m'_A}-\frac{1}{m_A}) \langle \bm{p}_A, \bm{p}_A \rangle + (\frac{1}{m'_C}-\frac{1}{m_C}) \langle \bm{p}_C, \bm{p}_C \rangle  \\
	& =  \Big[(\frac{1}{m'_A}+\frac{1}{m'_C}) \frac{\Delta m^2}{m^2_A} - \frac{2 \Delta m}{m_A m'_A} + \frac{1}{m'_A}-\frac{1}{m_A} \Big] \langle \bm{p}_A, \bm{p}_A \rangle + \frac{2 \Delta m}{m_A m'_C}  \langle \bm{p}_A, \bm{p}_C  \rangle +  \\
	&~~~~(\frac{1}{m'_C}-\frac{1}{m_C}) \langle \bm{p}_C, \bm{p}_C \rangle  \label{eq:ge-deduce1}
	\end{split}
	\end{equation}
	
	We substitute \eqref{eq:node-mass-conservation} into \eqref{eq:ge-deduce1}, and we get
	\begin{equation*}\label{eq:ge-deduce2}
	\begin{split}
	f(\bm{p}_0)
	& = -\frac{m_C \Delta m}{m^2_A(m_C+\Delta m)} \langle \bm{p}_A, \bm{p}_A \rangle + \frac{2 \Delta m}{m_A (m_C+\Delta m)}  \langle \bm{p}_A, \bm{p}_C  \rangle -\frac{\Delta m}{m_C(m_C + \Delta m)} \langle \bm{p}_C, \bm{p}_C  \rangle \\
	& = - \frac{\Delta m}{m^2_A m_C (m_C+\Delta m)}\Big( m^2_C \langle \bm{p}_A, \bm{p}_A \rangle -2m_A m_C \langle \bm{p}_A, \bm{p}_C  \rangle + m^2_A \langle \bm{p}_C, \bm{p}_C  \rangle \Big) \\
	& = - \frac{\Delta m}{m^2_A m_C (m_C+\Delta m)}\Big( m_C \bm{p}_A - m_A \bm{p}_C \Big)^2 
	\end{split}
	\end{equation*}
	
	Since $ \Delta m > 0 $, $ m_A > 0 $, $ m_C > 0 $, then $ f(\bm{p}_0) \le 0 $. According to Lemma \ref{lemma:lemma1}, we know that there must be a real solution in formula \eqref{eq:constraint}.
	\begin{equation}\label{eq:constraint}
	a \langle \bm{p},\bm{p} \rangle + \langle \bm{b}, \bm{p} \rangle + c = 0 
	\end{equation}

	\textbf{Then the expression of the optimal solution $\Delta \bm{p}$ is derived.}
	For the convenience of derivation, we set $ \bm{p} = (x,y)^T $, $ \bm{b} = (b_1, b_2)^T $ in \eqref{eq:constraint}, have
	\begin{equation}\label{eq:circle}
	a(x^2 + y^2) + b_1 x + b_2 y +  c = 0
	\end{equation}
	
	We can get \eqref{eq:circle-1} from formula \eqref{eq:circle}.
	\begin{equation}\label{eq:circle-1}
	(x + \frac{b_1}{2a})^2 + (y + \frac{b_2}{2a})^2 = \frac{b_1^2 + b_2^2}{4a^2} - \frac{c}{a}
	\end{equation}
	
	Because there is a real solution, then $ \frac{b_1^2 + b_2^2}{4a^2} - \frac{c}{a} \geq 0$. We introduce the notation $ r := \sqrt{\frac{b_1^2 + b_2^2}{4a^2} - \frac{c}{a} }$, $ \bm{p}^* := -\frac{1}{2a} \bm{b} = (-\frac{b_1}{2a},-\frac{b_2}{2a})^T  $, $ d := |\bm{\lambda} - \bm{p}^*| = \sqrt{\langle \bm{\lambda} - \bm{p}^*, \bm{\lambda} - \bm{p}^* \rangle }$.
	
	\begin{enumerate}[(1)]
		\item If $ r = 0 $, it is easy to know that $ \Delta \bm{p} =  \bm{p}^*$.
		
		\item If $ r > 0 $, then the trajectory of the \eqref{eq:circle-1} formula represents a circle with the center of the circle as the $\bm{p}^* $, radius as the $ r $. The geometric meaning of the solution of the optimization problem \eqref{eq:optimal-problem-s} indicates that the distance from the point on the circle to the $ \bm{\lambda} $ is the minimum.
		\begin{figure}[!h]
			\subfigure[$ d = 0$]{\label{fig:Circle-ex1}
				\begin{minipage}[t]{0.25\linewidth}
					\centering
					\includegraphics[width=3cm]{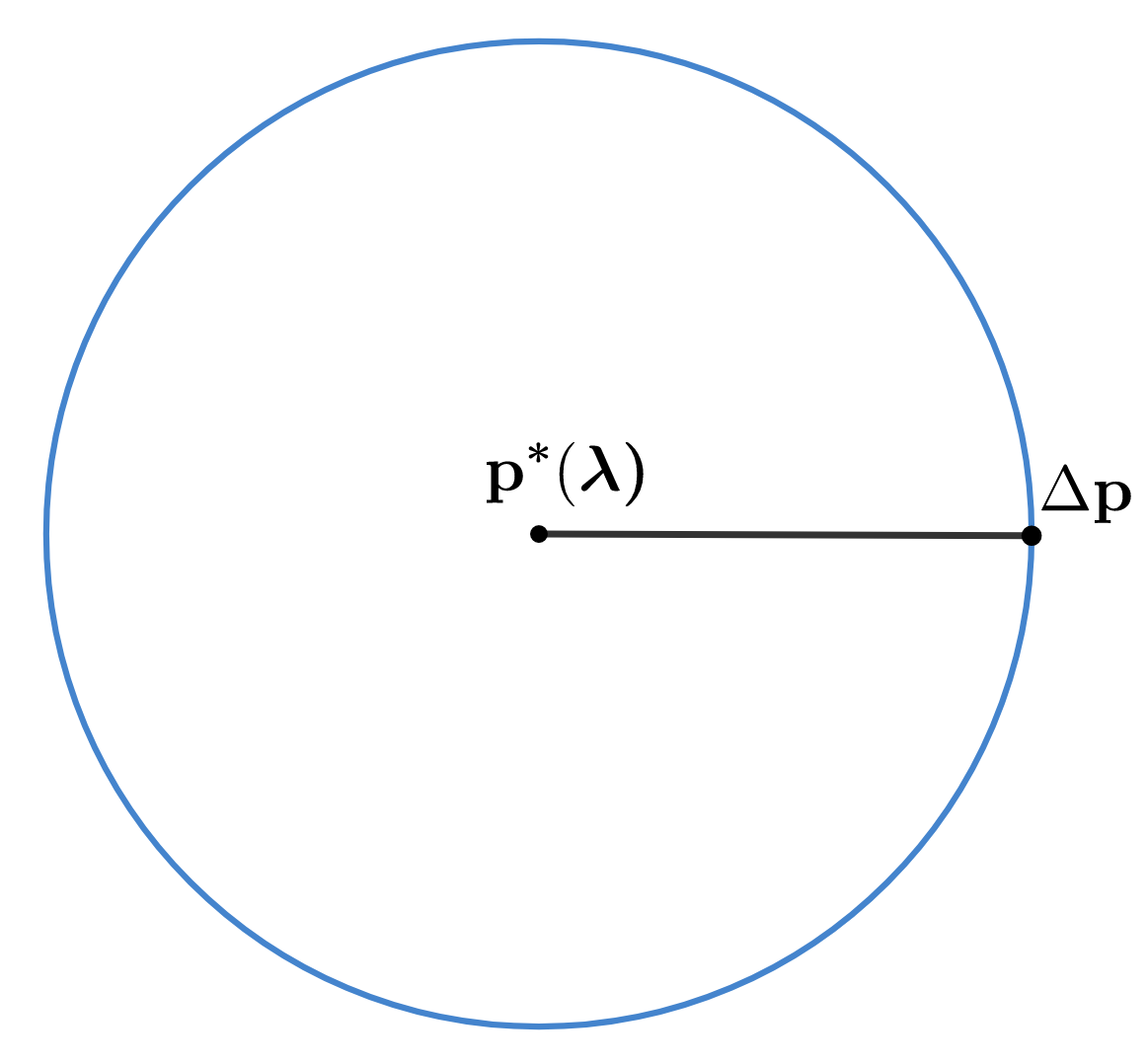}
				\end{minipage}
			}%
			\subfigure[$ 0 < d< r $]{\label{fig:Circle-ex2}
				\begin{minipage}[t]{0.25\linewidth}
					\centering
					\includegraphics[width=3cm]{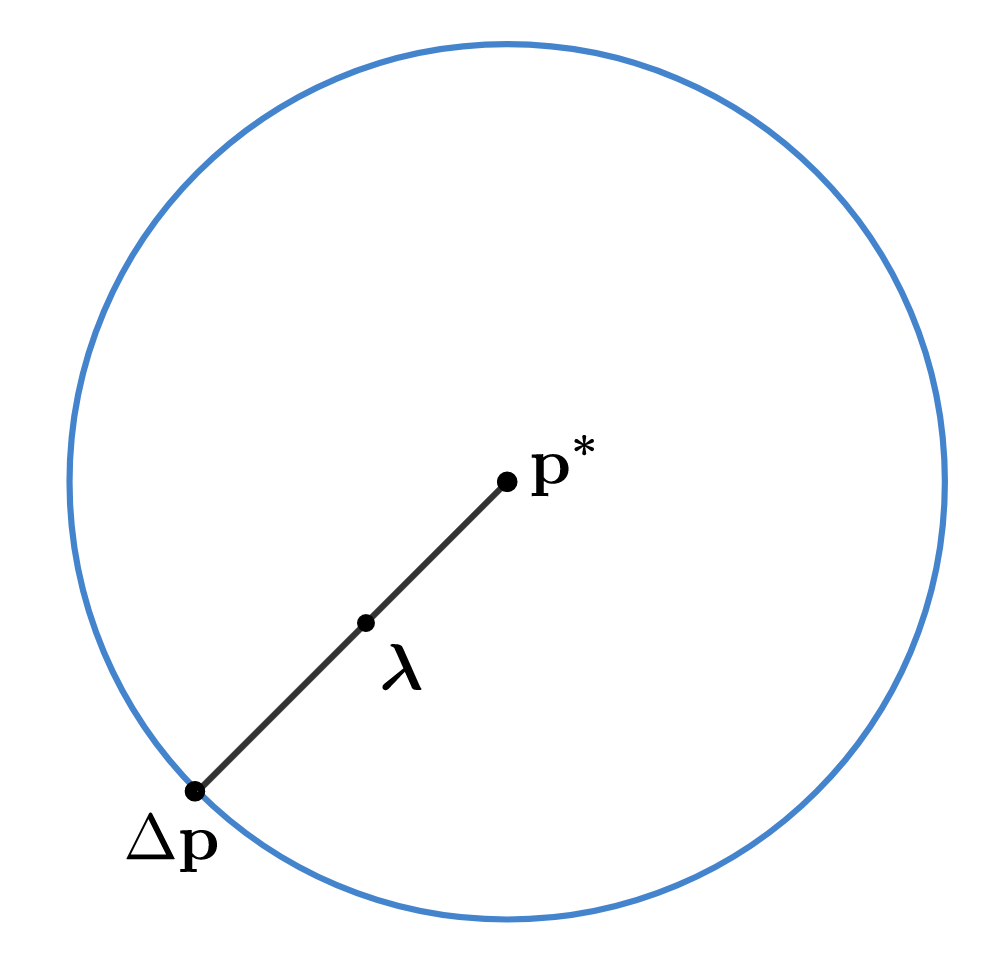}
				\end{minipage}
			}%
			\subfigure[$ d = r$]{\label{fig:Circle-ex3}
				\begin{minipage}[t]{0.25\linewidth}
					\centering
					\includegraphics[width=3cm]{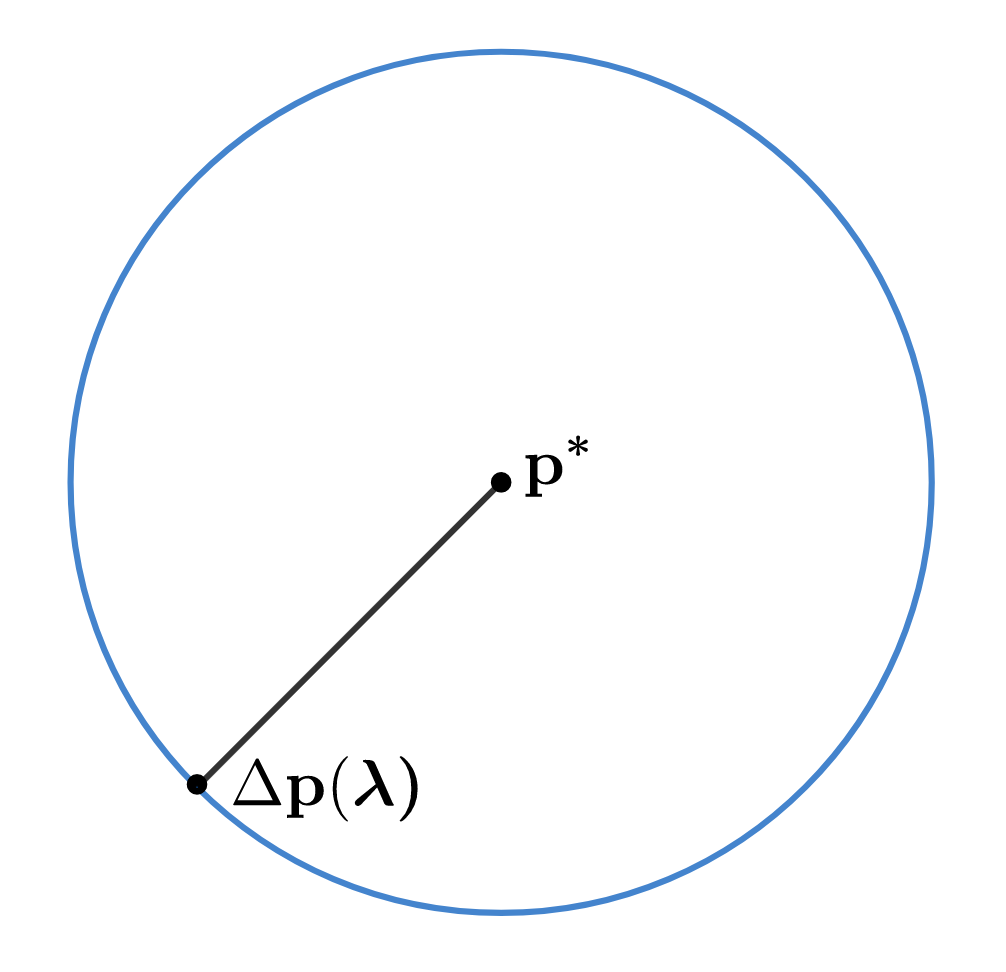}
				\end{minipage}
			}%
			\subfigure[$ d > r $]{\label{fig:Circle-ex4}
				\begin{minipage}[t]{0.25\linewidth}
					\centering
					\includegraphics[width=3cm]{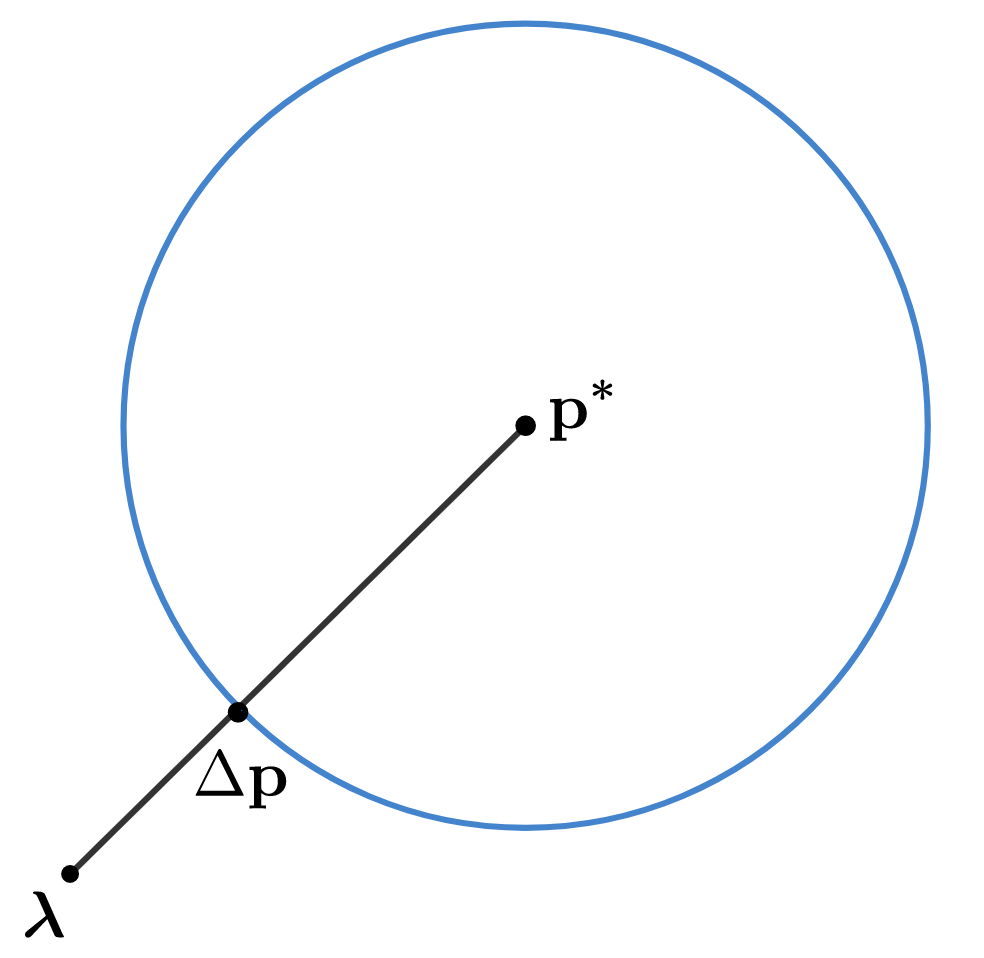}
				\end{minipage}
			}%
			\caption{Diagram of a circle}\label{fig:Circle}
		\end{figure}
		
		\begin{enumerate}
			\item [1)] If $ d = 0 $ (i.e. the center of the circle $\bm{p}^* = \bm{\lambda} $, see \reffig{fig:Circle-ex1}), then any point on the circle is the solution of optimization problem \eqref{eq:optimal-problem-s}. For simplicity, let's take $ \Delta \bm{p} = \bm{p}^* + (r, 0)^T $.
			
			\item [2)] If $ 0 < d < r $ (i.e. $ \bm{\lambda} $ is in the circle, see \reffig{fig:Circle-ex2}), then $ \Delta \bm{p} = \cfrac{r}{d} \; \bm{\lambda} + (1 - \cfrac{r}{d}) \bm{p}^* $.
			
			\item [3)] If $d = r $ (i.e. $ \bm{\lambda} $ is on a circle, see \reffig{fig:Circle-ex3}), then $ \Delta \bm{p} =  \bm{\lambda}$.
			
			\item [4)] If $d > r $ (i.e. $ \bm{\lambda} $ is outside the circle, see \reffig{fig:Circle-ex4}), then $ \Delta \bm{p} = \cfrac{r}{d} \; \bm{\lambda} + (1 - \cfrac{r}{d}) \bm{p}^* $.	
		\end{enumerate}
		
	\end{enumerate}
	
	In summary, the optimal solution of optimization problem \eqref{eq:optimal-problem-s} is as follows:	
	\begin{equation}\label{eq:optimal-sovler-a}
	\Delta \bm{p} =
	\begin{cases} 
	\cfrac{r}{d} \; \bm{\lambda} + (1 - \cfrac{r}{d}) \; \bm{p}^* & ,~d \neq 0 \\
	\bm{p}^* + r \bm{n} & , ~d = 0 
	\end{cases}
	\end{equation}
	where $ r = \sqrt{\cfrac{\langle \bm{b},\bm{b} \rangle }{4a^2} - \cfrac{c}{a} }$, $ \bm{p}^*= -\frac{1}{2a} \bm{b} $, $ d = \sqrt{\langle \bm{\lambda} - \bm{p}^*, \bm{\lambda} - \bm{p}^* \rangle } $, $ \bm{n} = (1, 0)^T $.		\hfill $ \square $
\end{pf}
\end{enumerate}
The above eight steps are the complete operation process of the matter compensation flow method.

\section{Parallel implementation of matter compensation flow based on OpenMP}\label{sec:four}
\begin{figure}[h]
	\centering
	\includegraphics[width=8cm]{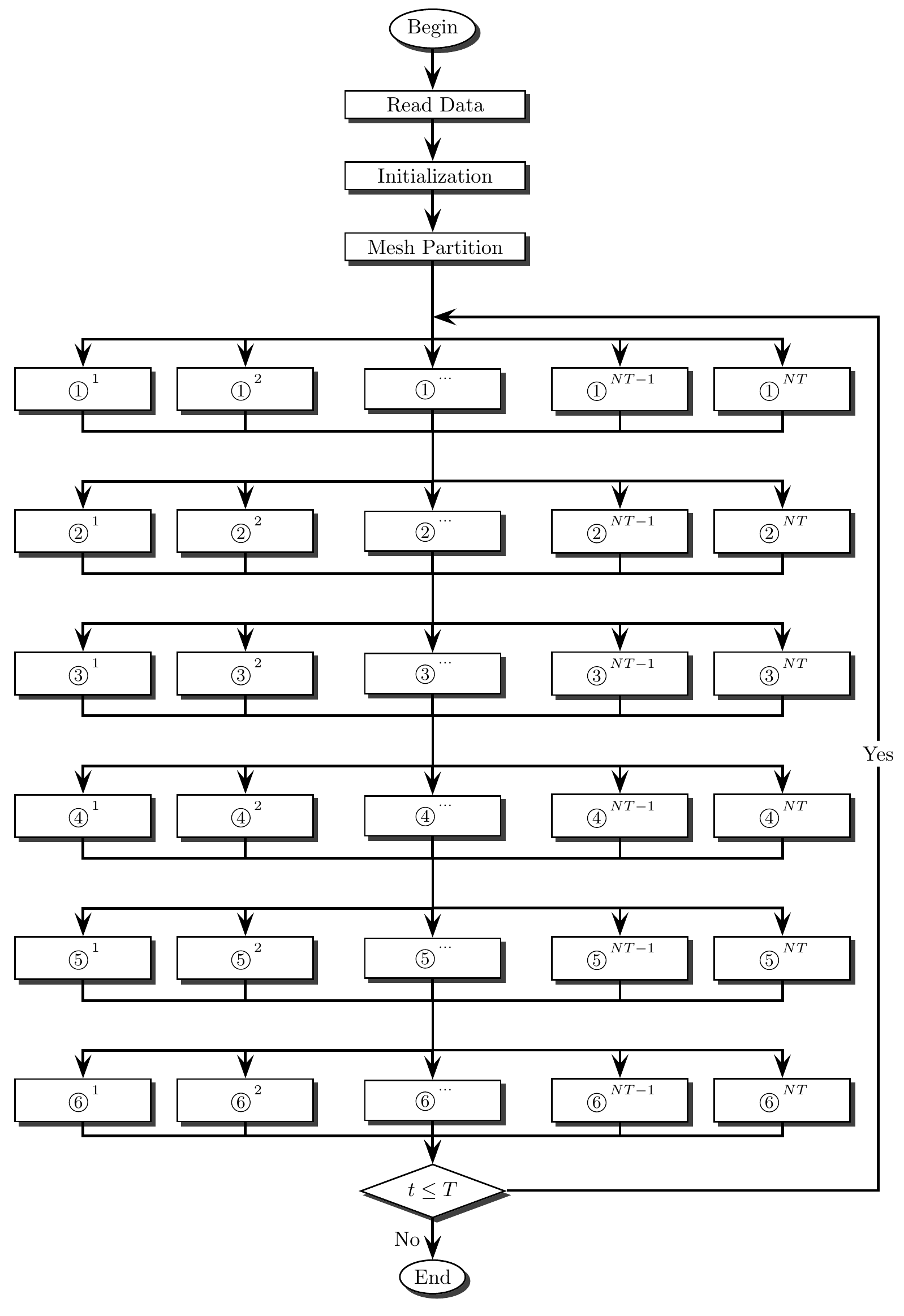}
	\caption{P-MFL algorithm flow chart}
	\label{fig:p-flow}
\end{figure}

OpenMP is a thread level parallel Application Programming Interface (API) based on shared memory. It is composed of a set of compilation guidance, run-time routines and environment variables. It has the advantages of simple programming, portability and expansibility, and is widely used in the field of scientific computing. In this paper, a Parallel Matter Flow Lagrangian (P-MFL) algorithm is designed for SGH Lagrangian simulation with matter flow based on OpenMP. The flow chart of the algorithm is shown in \reffig{fig:p-flow}.

\reftab{tab:parallel-evolution} introduces the functions of the six modules in \reffig{fig:p-flow}.
\begin{table}[h]									
	\centering									
	\caption{Function description of the six core modules in the P-MFL}\label{tab:parallel-evolution}									
	\begin{tabular}{|l|l|}									
		\hline									
		Core modules      						&	  Function		\\
		\hline	
		\circled{1} P-DetermineDeltT 			&	 Parallel computing time step $ \Delta t $\\  
		\circled{2} P-DynamicEvolve 			&	 Parallel evolution of a time step \\  
		\circled{3} P-SetAllDependentVariables  &	 Parallel setting of all dependent variables on the element\\   
		\circled{4} P-CalculateMatterFlowAcc    &	 Parallel calculation of matter flow acceleration on three edges of triangular element\\      
		\circled{5} P-MatterFlowEvolve 		    &	 Parallel evolution of matter compensation flow\\    
		\circled{6} P-CalculateVertexForce 	    &	 Parallel calculation of node forces\\     		  		       		
		\hline									
	\end{tabular}									
\end{table}

\begin{figure}[h]
	\centering
	\includegraphics[width=3.5cm]{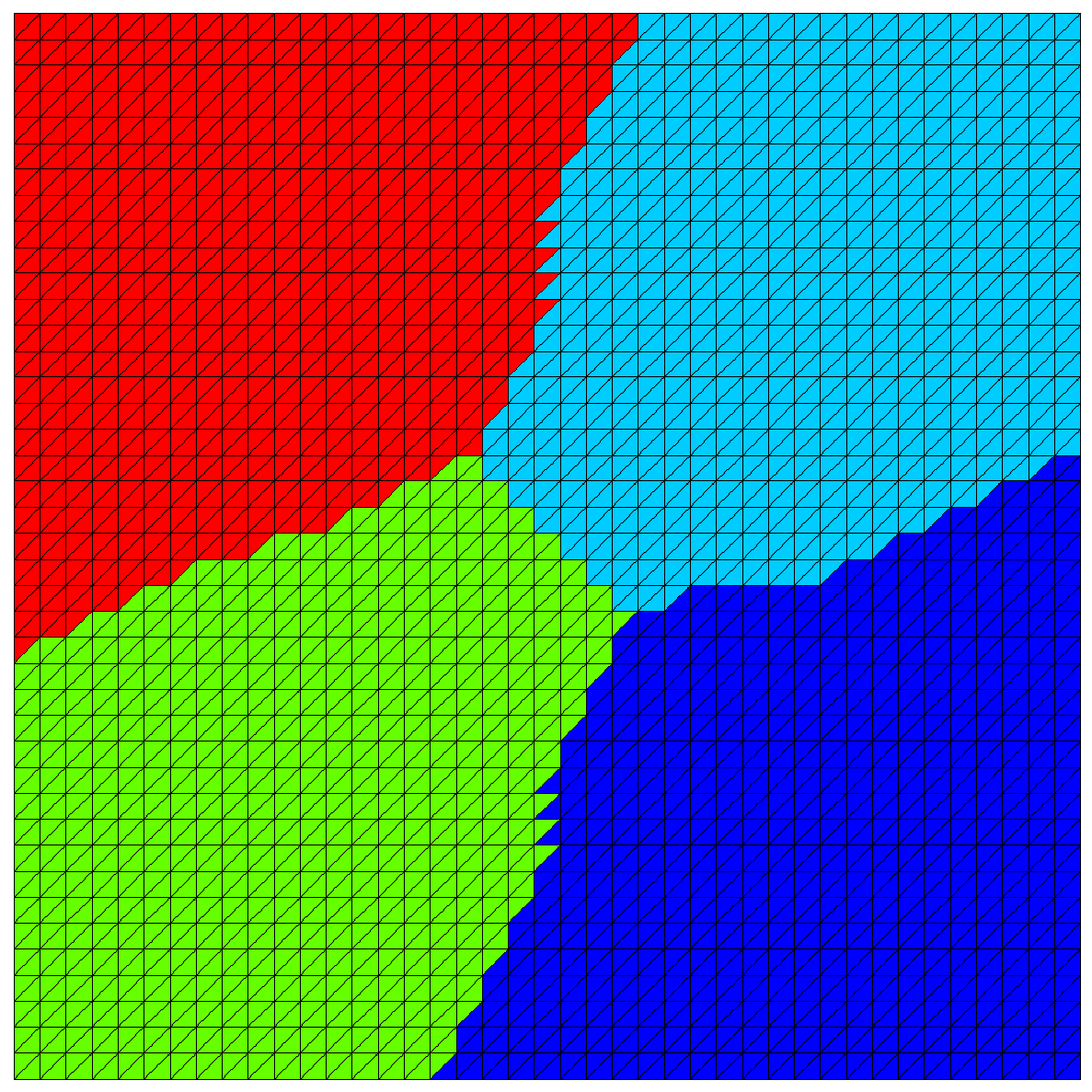}
	\caption{Mesh partition}
	\label{fig:mesh-partition}
\end{figure}

In the Mesh Partition link in the P-MFL algorithm flow chart, this paper uses the graph partition toolbox METIS\cite{metis-url} to decompose the grid $ \mathcal{T} $ into $ NT $ sub-grids $ \mathcal{T}_i $ ($i = 1, \cdots, NT$). \reffig{fig:mesh-partition} shows the mesh partition diagram for $ NT = 4 $. The number of cells in the subgrid generated by METIS is almost the same, and the sum of all the subgrid boundaries satisfies the minimum principle, so the parallel partition often has good parallel performance. Let's suppose that the cell index set in $ \mathcal{T} $ is $ \mathscr{C} $ and the node index set is $ \mathscr{N} $, and the index set in sub grid $ \mathcal{T}_i $ is $ \mathscr{C}_i $ and the node index set is $ \mathscr{N}_i $, then (1)~$ \mathscr{C} = \bigcup\limits_{i=1}^{NT} \mathscr{C}_i$, $ \mathscr{N} = \bigcup\limits_{i=1}^{NT} \mathscr{N}_i$; (2)~$  \mathscr{C}_i \cap \mathscr{C}_j = \phi$,  $  \mathscr{N}_i \cap \mathscr{N}_j = \phi$, $ \forall~ i \neq j,~i,j = 1, 2, \cdots , NT $.

\begin{figure}[h]
	\centering
	\subfigure[ "Fork-Join" mode]{\label{fig:fork-join}
		\begin{minipage}[t]{0.4\linewidth}
			\centering
			\includegraphics[width=4cm]{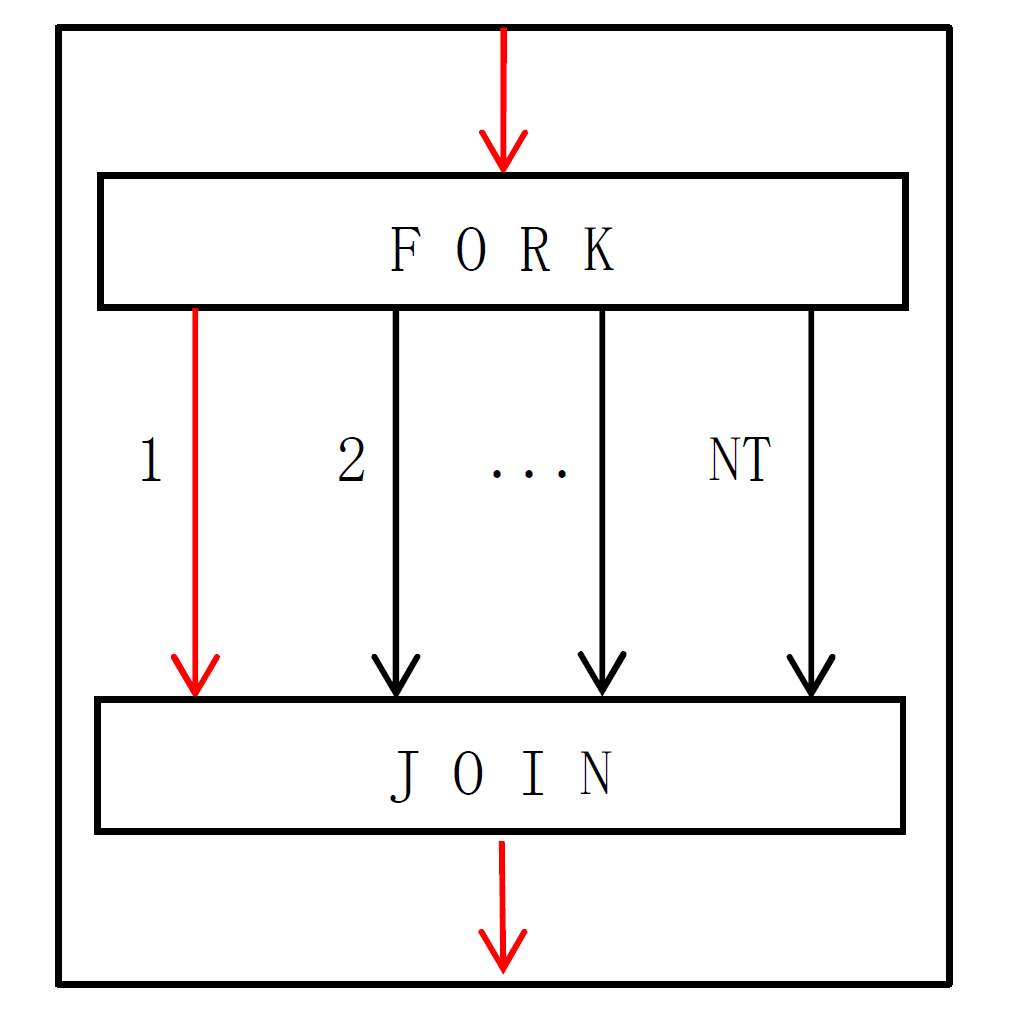}	
		\end{minipage}
	}%
	\subfigure[Diagram of interface point $ p $]{\label{fig:parallel-point-partition}
		\begin{minipage}[t]{0.4\linewidth}
			\centering
			\includegraphics[width=4.5cm]{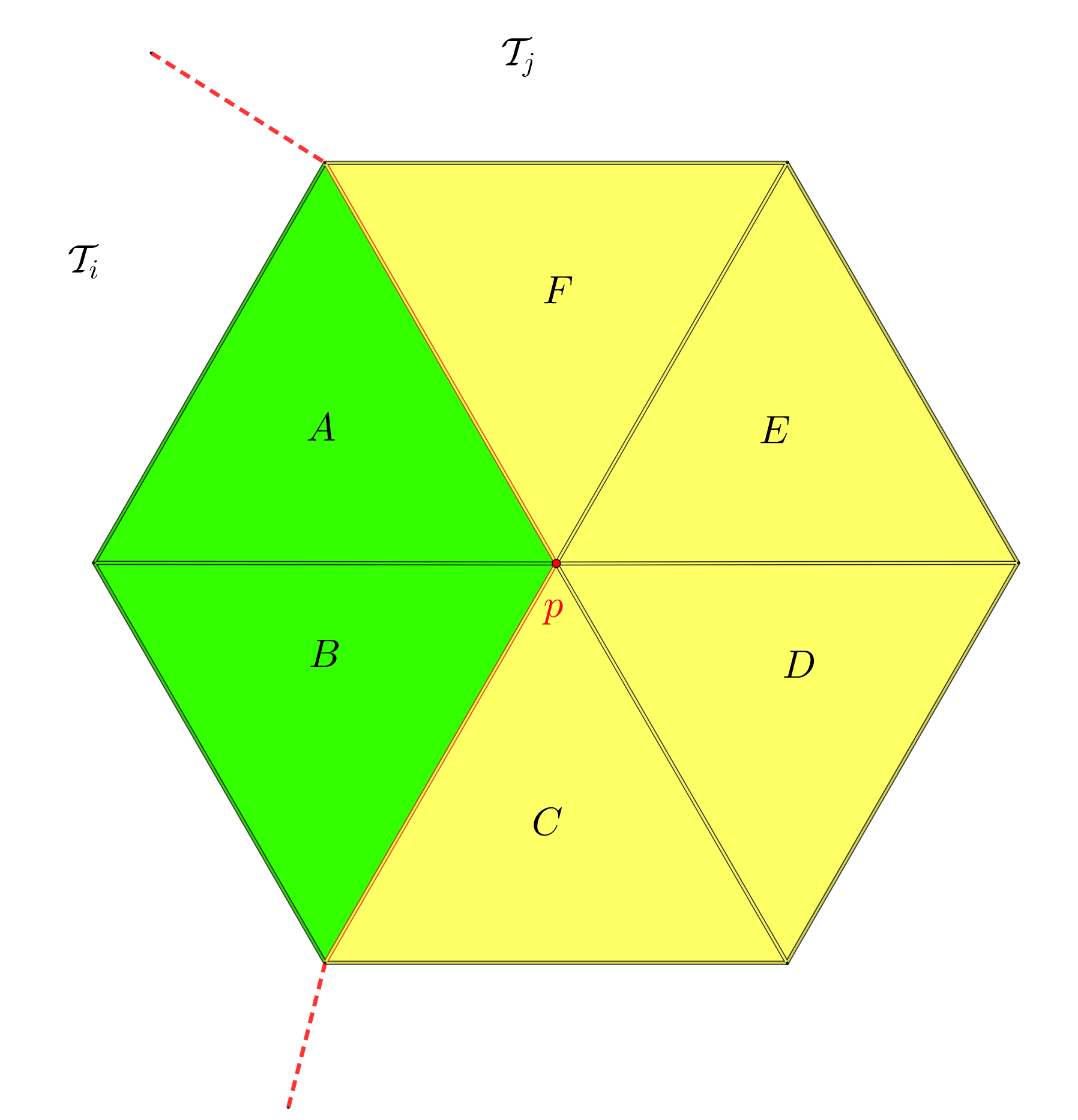}
		\end{minipage}
	}%
	\caption{"Fork-Join" mode and Diagram of interface point $ p $}\label{fig:abc}
\end{figure}

"Fork-Join" is the standard parallel mode of OpenMP, as shown in \reffig{fig:fork-join}. The code is divided into serial region and parallel region. The serial region is executed by the main thread. When executing to the parallel region, the slave thread is forked by the system. In the parallel region, the parallel task is completed by the main thread and the slave thread. After the calculation of the parallel region is finished, all threads will join together again. The derived slave thread will exit or block, no longer work, and control the flow return to the main thread and proceed to the next task.

In parallel region, thread $i$ ($ i=1,\cdots, NT $) is responsible for computing tasks in subgrid $ \mathcal{T}_i $, that is, only cell index set $ \mathscr{C}_i $ and node index set $ \mathscr{N}_i $ are considered, and the calculation task of each thread is about $ 1/NT $ of serial task. There will be data competition for shared memory cells at the subgrid boundary that may lead to inaccurate calculation results, such as computing node force at interface $p$ point. Let's consider two subgrids $\mathcal{T}_i $ (green area) and $\mathcal{T}_j$ (yellow area), as shown in \reffig{fig:parallel-point-partition}. Assuming that thread $ i $ is calculating the force $ \bm{f}_p^A $ of element $ A $ on node $ p $, and thread $ j $ is also calculating the force $ \bm{f}_p^F $ of element $ F $ on node $ p $, and they read the data $ \bm{f}_p $ of the shared memory unit at the same time, then the updated node force will either take the value $ \bm{f}'_p = \bm{f}_p + \bm{f}_p^A$ in thread $ i $ or $ \bm{f}''_p = \bm{f}_p + \bm{f}_p^F$ in thread $ j $, while the correct result should be $ \bm{f}''_p = \bm{f}'_p + \bm{f}_p^F$. To ensure the correctness of the calculation results and the security of the data, techniques such as "atomic operation" and "critical region" are used. Because the subgrid interface is short and the cell aggregation is strong in the Mesh Partition step, the probability of the above situation is very small (the larger the scale, the smaller the possibility), which almost does not affect the parallel efficiency of the program.

\section{Example and analysis}\label{sec:five}
This section examines the previous matter flow method based on Saltzman Piston Problem\cite{Saltzman1, Saltzman2}, Noh Implosion Problem\cite{Noh} and Sedov Explosion Problem\cite{Sedov}. All three examples contain a highly transient shock, and the orientation of the wavefront is inconsistent with that of the grid. The conventional Lagrangian method is easy to appear the physical quantity oscillation problem caused by the stiffness of the triangular mesh.

\subsection{Saltzman Piston Problem}
\begin{example}
Consider model question \eqref{eq:mass-equation}-\eqref{eq:state-equation}, domain $ \Omega = [0,1] \times [0, 0.1] $, simulation time $ t \in [0,0.5] $, gas adiabatic $ \gamma = \frac{5}{3} $. Initial condition: initial density is 1, and the pressure is 0. Boundary condition: the left boundary adopts piston boundary condition (i.e. the boundary moves to the right at constant unit velocity), and the right and upper and lower boundaries adopt solid wall boundary condition (i.e. normal velocity or displacement is 0). 	
\end{example}

\begin{figure}[!htpb]
	\subfigure[Mesh $\mathcal{T}_{1}^{Sa,I}$]{\label{fig:SaltzmanI-100x10-mesh-t0}
		\begin{minipage}[t]{1\linewidth}
			\centering
			\includegraphics[width=14cm]{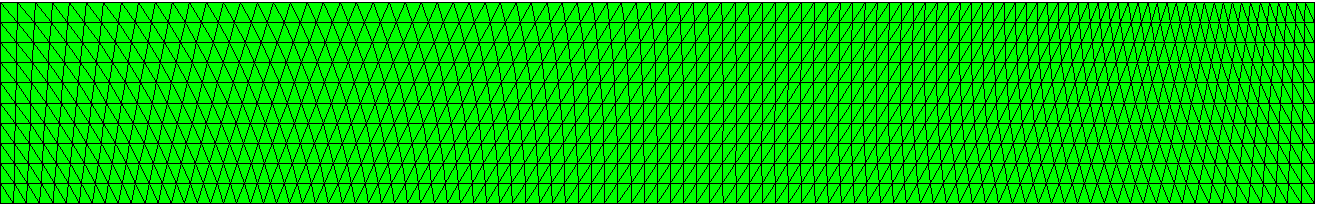}		
		\end{minipage}
	}%
	
	\subfigure[Mesh $\mathcal{T}_{1}^{Sa,II}$]{\label{fig:SaltzmanII-100x10-mesh-t0}
		\begin{minipage}[t]{1\linewidth}
			\centering
			\includegraphics[width=14cm]{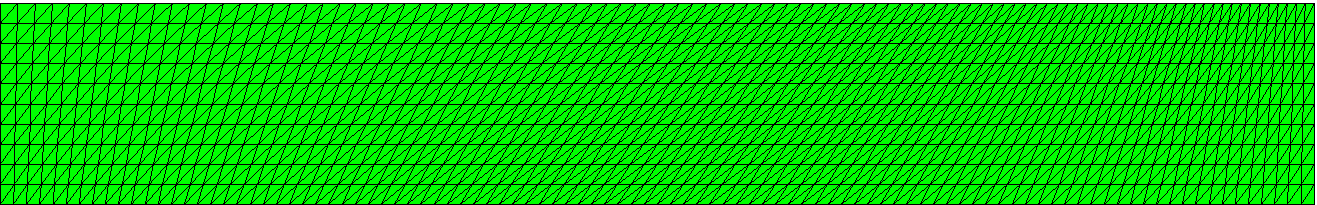}		
		\end{minipage}
	}%
	\caption{The type I grid with $\mathcal{T}_{1}^{Sa,I}$ of 100$\times$10 and the type II grid with $\mathcal{T}_{1}^{Sa,II}$ of 100$\times$10}\label{fig:Saltzman-typeI-II-initial-mesh}
\end{figure}

Two grid types of type I and type II, as shown in \reffig{fig:SaltzmanI-100x10-mesh-t0} and \reffig{fig:SaltzmanII-100x10-mesh-t0} respectively, are used to simulate the Saltzman piston problem.

Saltzman piston problem with the initial mesh $\mathcal{T}_{1}^{Sa,I}$ and $\mathcal{T}_{1}^{Sa,II}$ is simulated with both the regular SGH Lagrangian method (labeled as "no-matterflow") and the method implementing the matter flow (labeled as "matterflow") in order to test the effectiveness of the proposed matter flow method.

\begin{figure}[!htpb]
	\setlength{\abovecaptionskip}{-0cm}
	\setlength{\belowcaptionskip}{-0cm} 
	\subfigure[Mesh, no-matterflow, $\mathcal{T}_{1}^{Sa,I}$]{\label{fig:SaltzmanI-c-no-mf}
		\begin{minipage}[t]{0.5\linewidth}
			\centering
			\includegraphics[width=6cm]{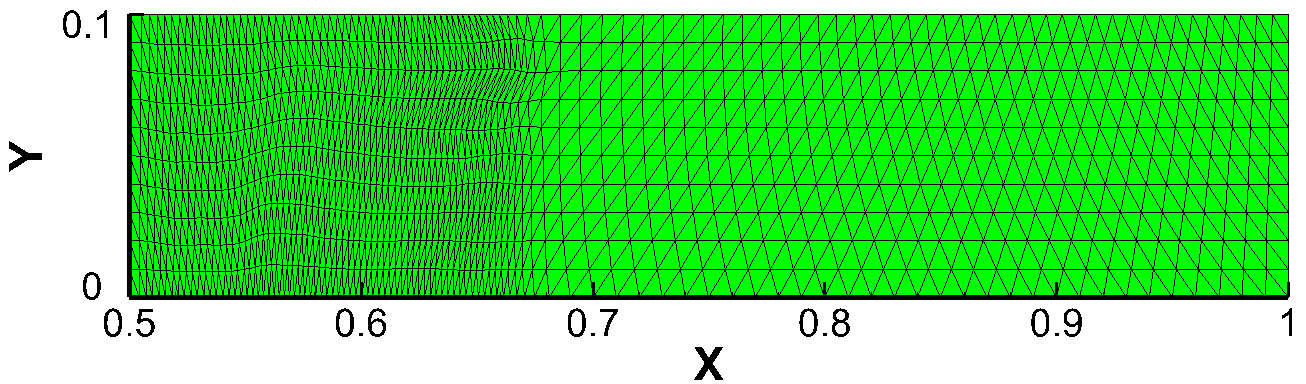}		
		\end{minipage}
	}%
	\subfigure[Mesh, matterflow, $\mathcal{T}_{1}^{Sa,I}$]{\label{fig:SaltzmanI-c-mf}
		\begin{minipage}[t]{0.5\linewidth}
			\centering
			\includegraphics[width=6cm]{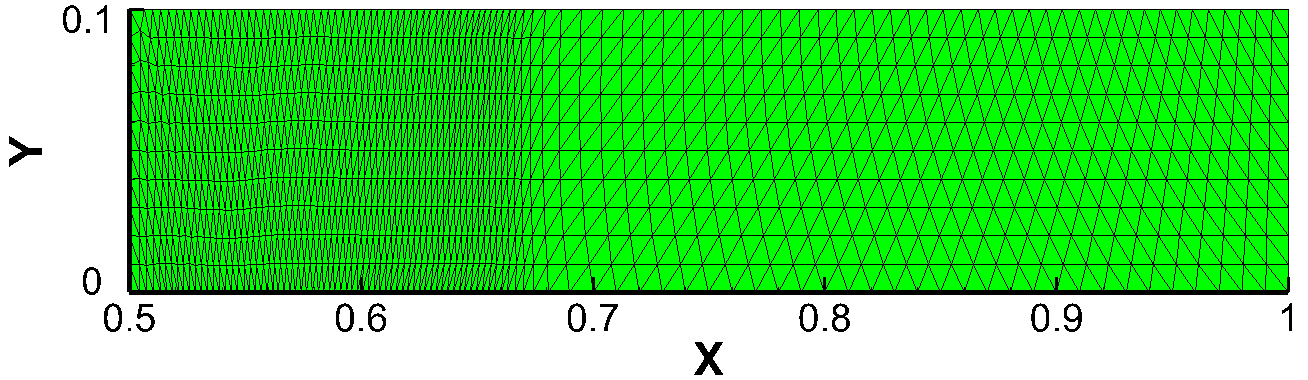}	
		\end{minipage}
	}\vspace{-0.4cm}
	\subfigure[Density, no-matterflow, $\mathcal{T}_{1}^{Sa,I}$]{\label{fig:SaltzmanI-density-no-mf}
		\begin{minipage}[t]{0.5\linewidth}
			\centering
			\includegraphics[width=7cm]{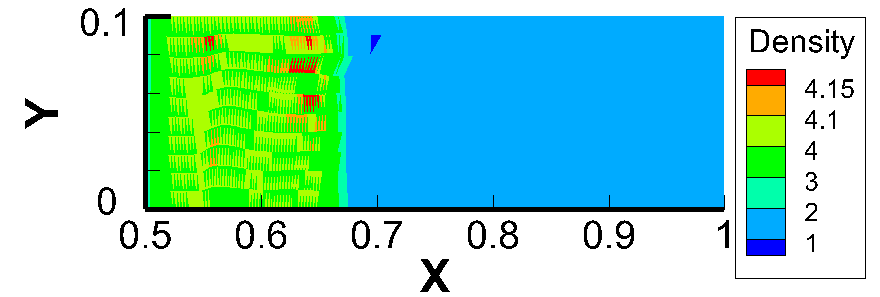}		
		\end{minipage}
	}%
	\subfigure[Density, matterflow, $\mathcal{T}_{1}^{Sa,I}$]{\label{fig:SaltzmanI-density-mf}
		\begin{minipage}[t]{0.5\linewidth}
			\centering
			\includegraphics[width=7cm]{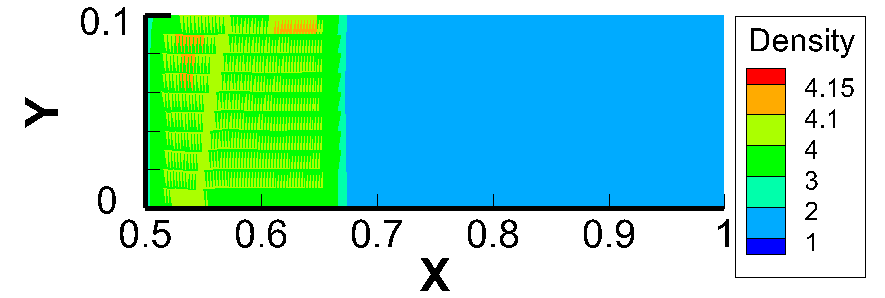}		
		\end{minipage}
	}\vspace{-0.4cm}
	\subfigure[Pressure, no-matterflow, $\mathcal{T}_{1}^{Sa,I}$]{\label{fig:SaltzmanI-pressure-no-mf}
		\begin{minipage}[t]{0.5\linewidth}
			\centering
			\includegraphics[width=7cm]{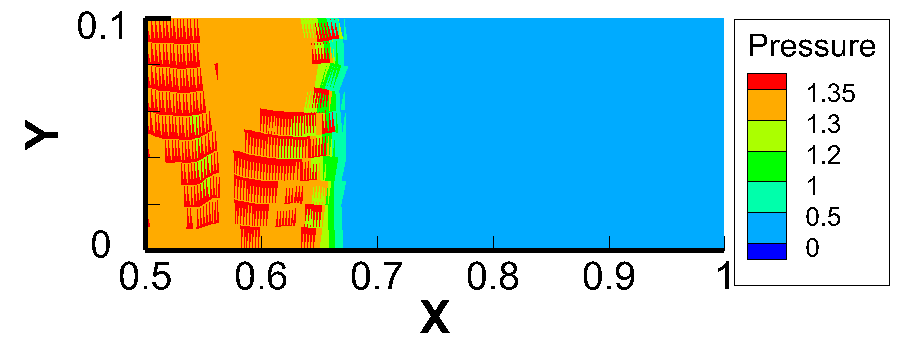}		
		\end{minipage}
	}%
	\subfigure[Pressure, matterflow, $\mathcal{T}_{1}^{Sa,I}$]{\label{fig:SaltzmanI-pressure-mf}
		\begin{minipage}[t]{0.5\linewidth}
			\centering
			\includegraphics[width=7cm]{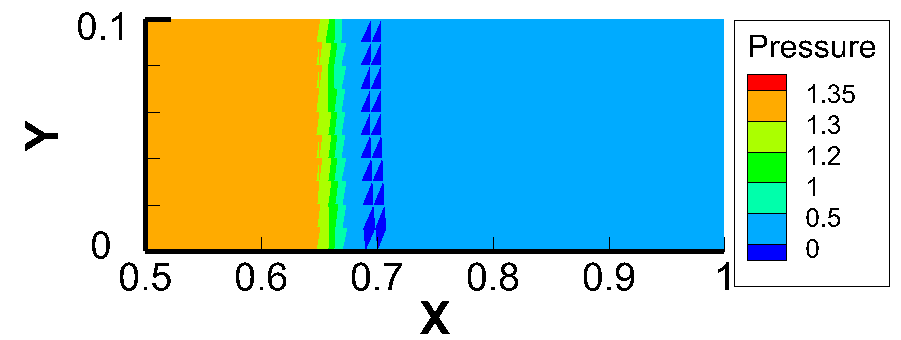}		
		\end{minipage}
	}
	\caption{$\mathcal{T}_{1}^{Sa,I}$ mesh diagram, density and pressure contour diagram at $ t =0.5 $.}\label{fig:SaltzmanI-compare-dist-mf}
\end{figure}

\begin{figure}[!htpb]
	\setlength{\abovecaptionskip}{-0cm} 
	\setlength{\belowcaptionskip}{-0cm} 
	\subfigure[Mesh, no-matterflow, $\mathcal{T}_{1}^{Sa,II}$]{\label{fig:SaltzmanII-c-no-mf}
		\begin{minipage}[t]{0.5\linewidth}
			\centering
			\includegraphics[width=6cm]{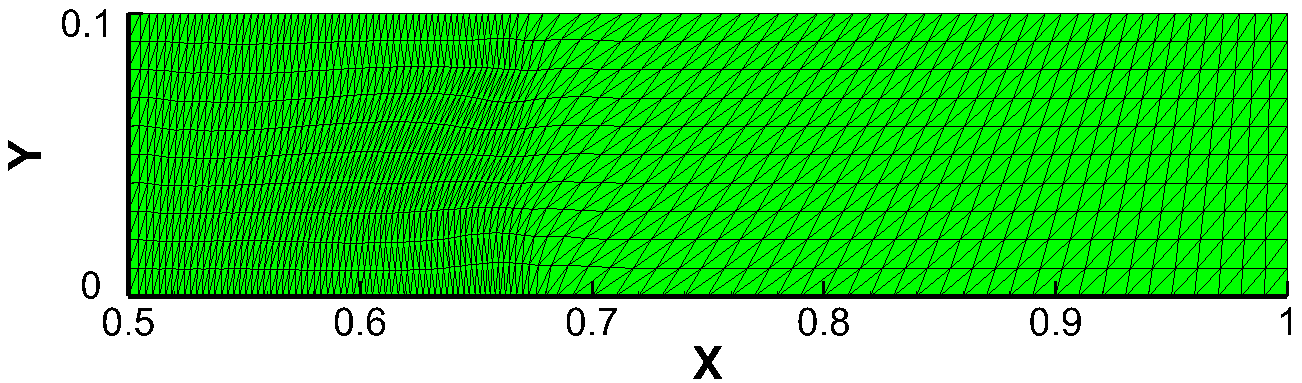}		
		\end{minipage}
	}%
	\subfigure[Mesh, matterflow, $\mathcal{T}_{1}^{Sa,II}$]{\label{fig:SaltzmanII-c-mf}
		\begin{minipage}[t]{0.5\linewidth}
			\centering
			\includegraphics[width=6cm]{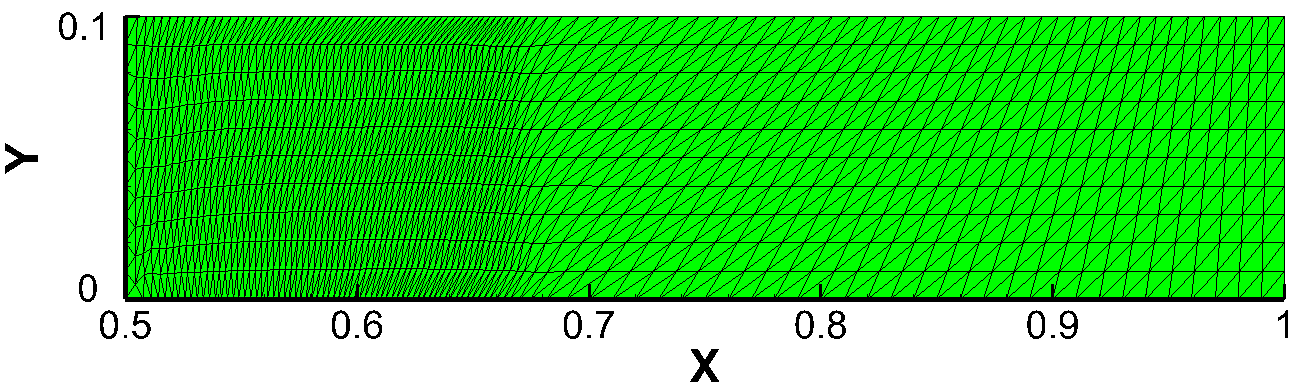}	
		\end{minipage}
	}\vspace{-0.4cm}
	\subfigure[Density, no-matterflow, $\mathcal{T}_{1}^{Sa,II}$]{\label{fig:SaltzmanII-density-no-mf}
		\begin{minipage}[t]{0.5\linewidth}
			\centering
			\includegraphics[width=7cm]{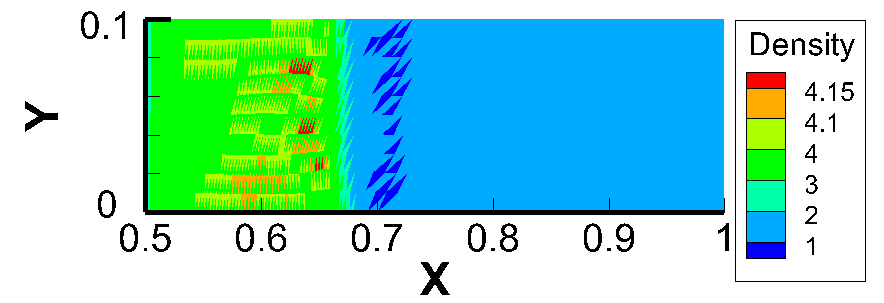}		
		\end{minipage}
	}%
	\subfigure[Density, matterflow, $\mathcal{T}_{1}^{Sa,II}$]{\label{fig:SaltzmanII-density-mf}
		\begin{minipage}[t]{0.5\linewidth}
			\centering
			\includegraphics[width=7cm]{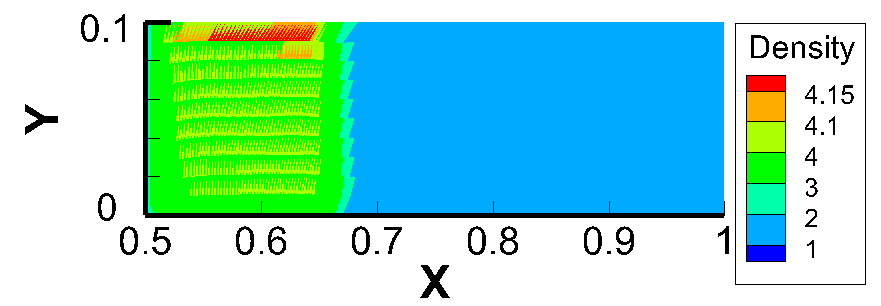}		
		\end{minipage}
	}\vspace{-0.4cm}
	\subfigure[Pressure, no-matterflow, $\mathcal{T}_{1}^{Sa,II}$]{\label{fig:SaltzmanII-pressure-no-mf}
		\begin{minipage}[t]{0.5\linewidth}
			\centering
			\includegraphics[width=7cm]{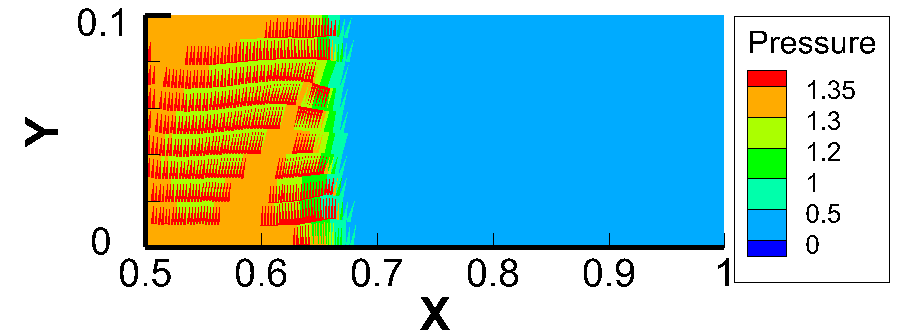}		
		\end{minipage}
	}%
	\subfigure[Pressure, matterflow, $\mathcal{T}_{1}^{Sa,II}$]{\label{fig:SaltzmanII-pressure-mf}
		\begin{minipage}[t]{0.5\linewidth}
			\centering
			\includegraphics[width=7cm]{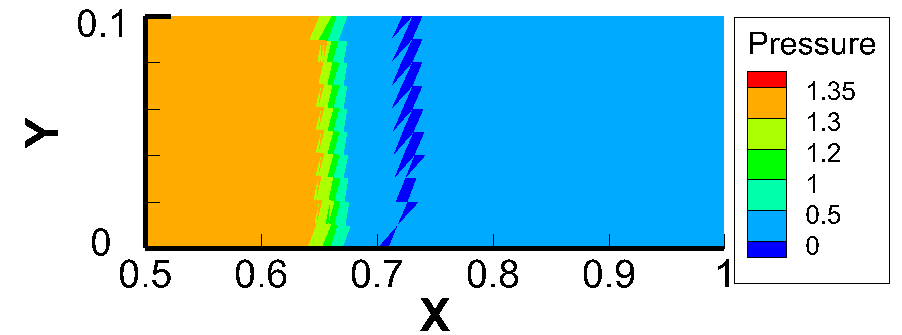}		
		\end{minipage}
	}%
	\caption{$\mathcal{T}_{1}^{Sa,II}$ mesh diagram, density and pressure contour diagram at $ t =0.5 $. }\label{fig:SaltzmanII-compare-dist-mf}
\end{figure}

\begin{figure}[!htpb]
	\subfigure[Density, $\mathcal{T}_{1}^{Sa,I}$]{\label{fig:SaltzmanI-Density-100x10-mf}
		\begin{minipage}[t]{0.33\linewidth}
			\centering
			\includegraphics[width=5.5cm]{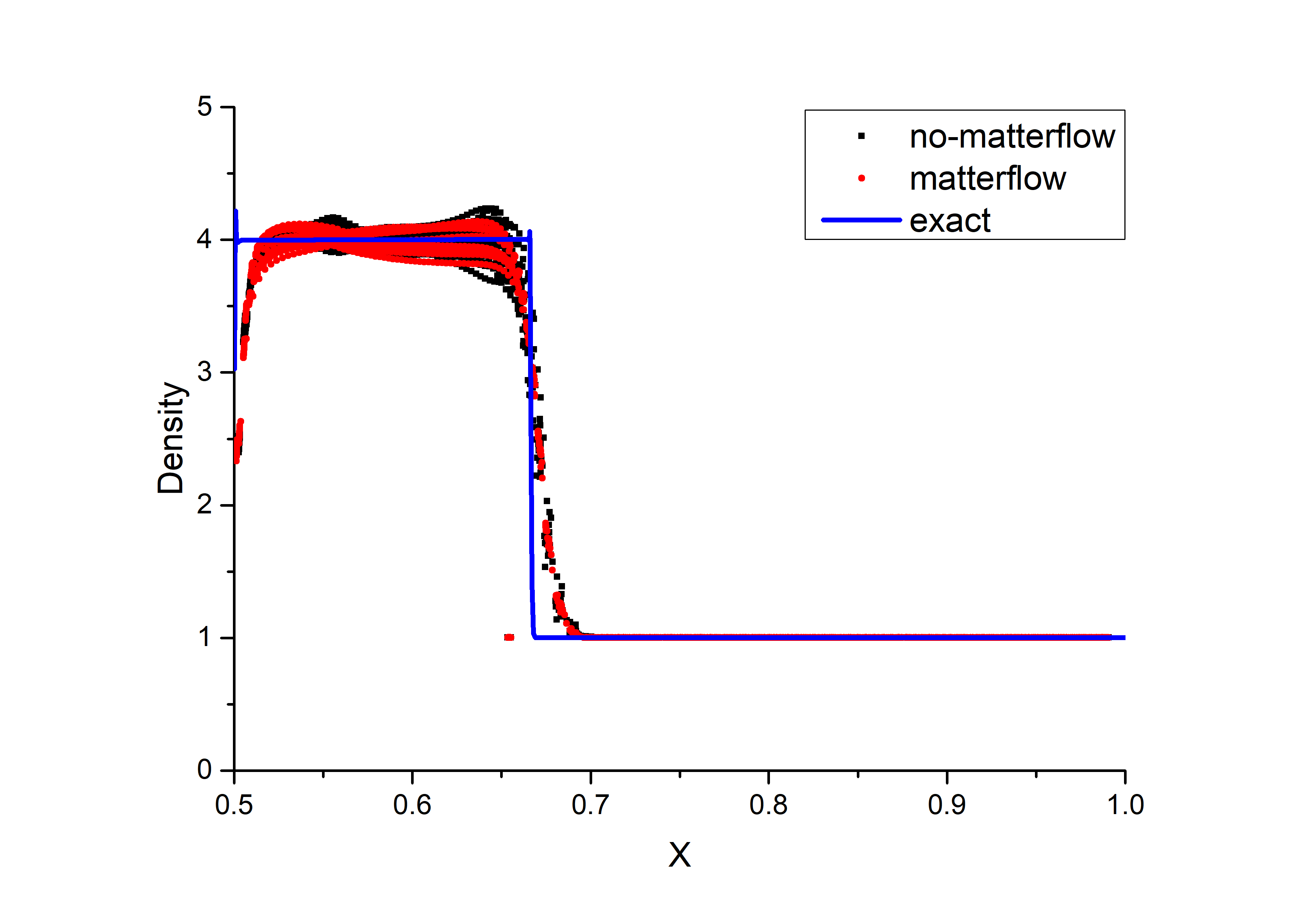}		
		\end{minipage}
	}%
	\subfigure[Pressure, $\mathcal{T}_{1}^{Sa,I}$]{\label{fig:SaltzmanI-Pressure-100x10-mf}
		\begin{minipage}[t]{0.33\linewidth}
			\centering
			\includegraphics[width=5.5cm]{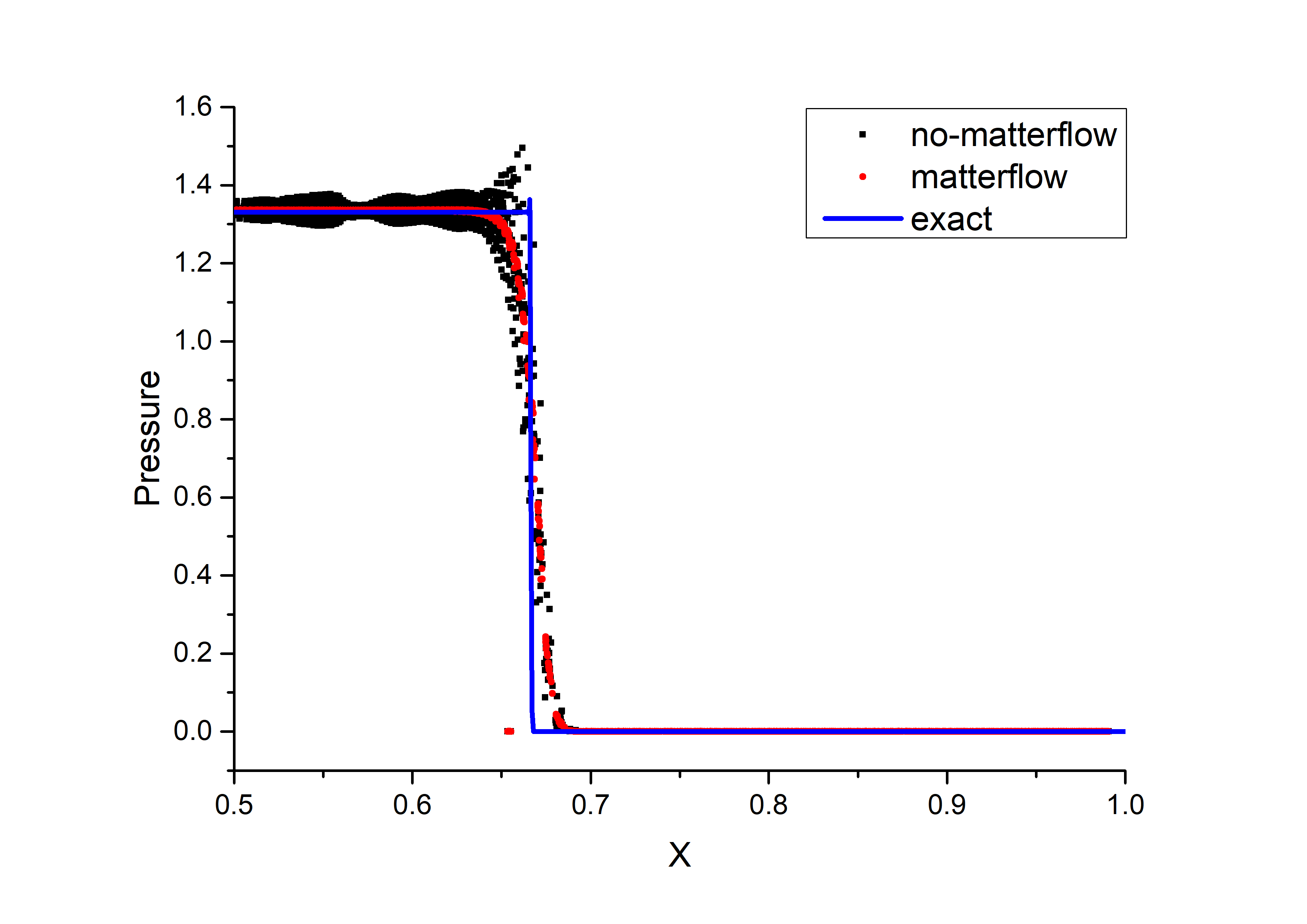}		
		\end{minipage}
	}%
	\subfigure[Velocity (x), $\mathcal{T}_{1}^{Sa,I}$]{\label{fig:SaltzmanI-Velocity-100x10-mf}
		\begin{minipage}[t]{0.33\linewidth}
			\centering
			\includegraphics[width=5.5cm]{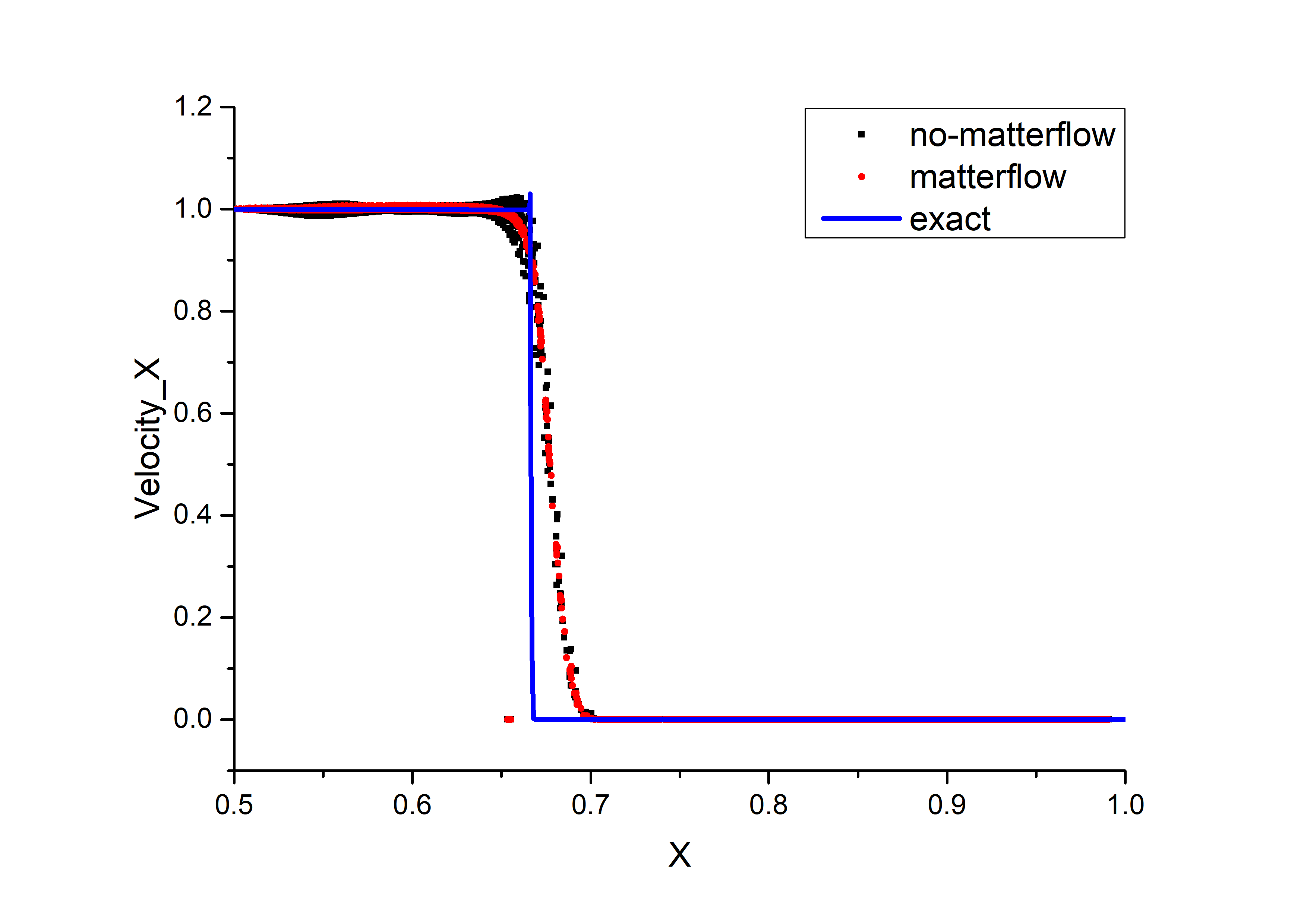}		
		\end{minipage}
	}%
	\vspace{-0.4cm}
	
	\subfigure[Density, $\mathcal{T}_{1}^{Sa,II}$]{\label{fig:SaltzmanII-Density-100x10-mf}
		\begin{minipage}[t]{0.33\linewidth}
			\centering
			\includegraphics[width=5.5cm]{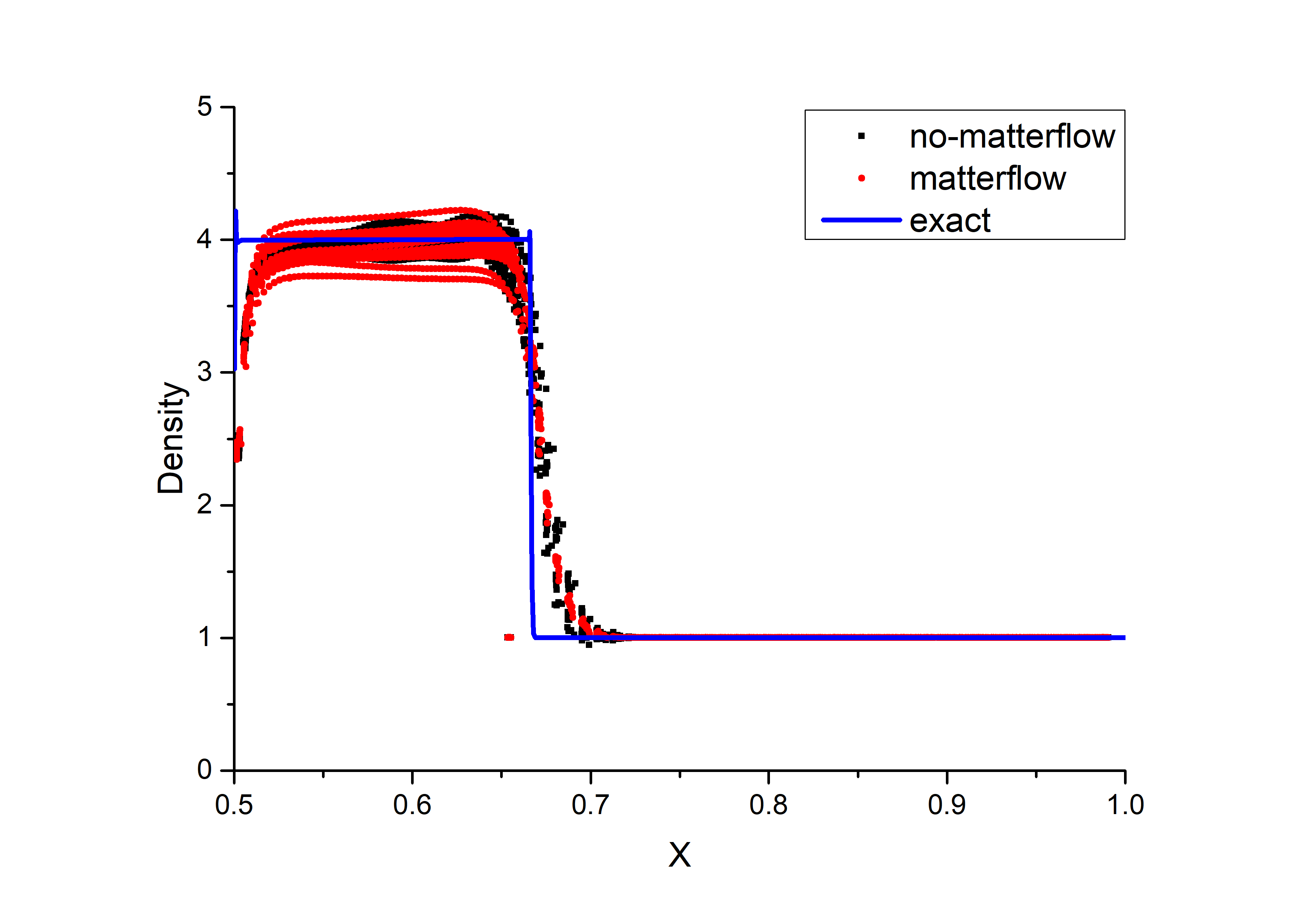}		
		\end{minipage}
	}%
	\subfigure[Pressure, $\mathcal{T}_{1}^{Sa,II}$]{\label{fig:SaltzmanII-Pressure-100x10-mf}
		\begin{minipage}[t]{0.33\linewidth}
			\centering
			\includegraphics[width=5.5cm]{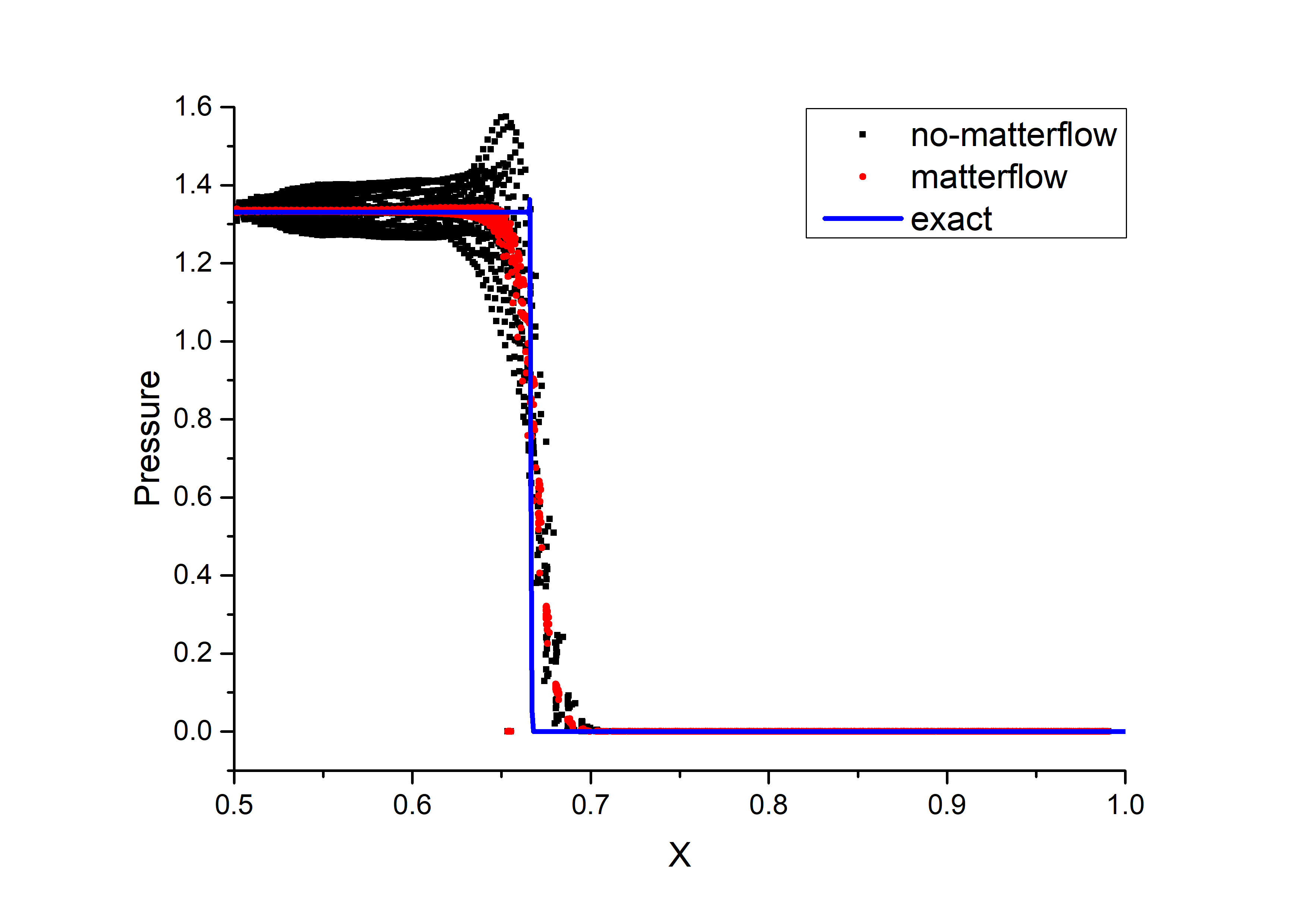}		
		\end{minipage}
	}%
	\subfigure[Velocity (x), $\mathcal{T}_{1}^{Sa,II}$]{\label{fig:SaltzmanII-Velocity-100x10-mf}
		\begin{minipage}[t]{0.33\linewidth}
			\centering
			\includegraphics[width=5.5cm]{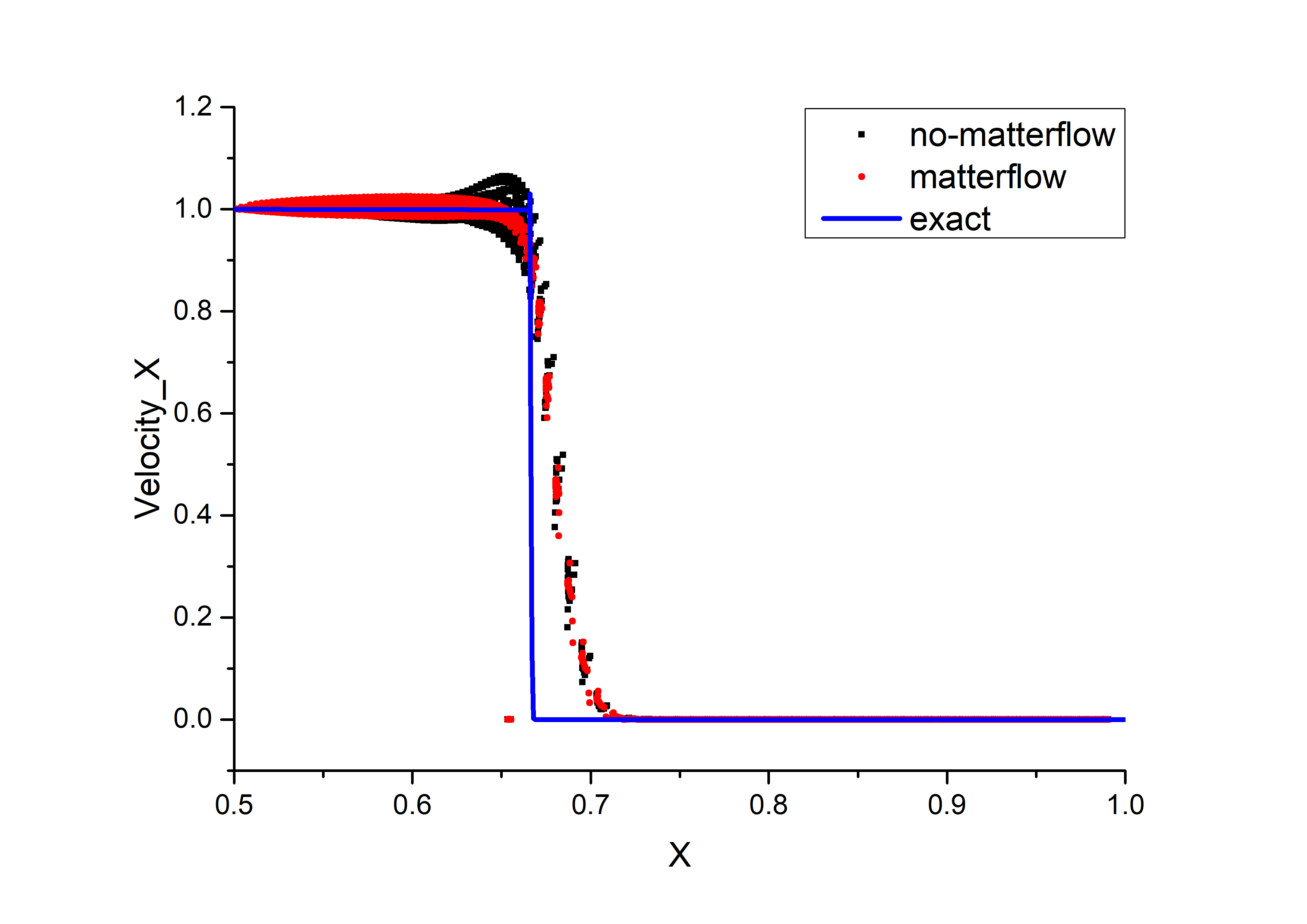}		
		\end{minipage}
	}%
	\caption{
	The scatter plot of density, pressure and velocity (x) of $\mathcal{T}_{1}^{Sa,I}$ and $\mathcal{T}_{1}^{Sa,II}$ with or without matter flow at $ t = 0.5 $.} \label{fig:Saltzman-typeI-II-compare-exact-mf}
\end{figure}

\reffig{fig:SaltzmanI-compare-dist-mf} and \reffig{fig:SaltzmanII-compare-dist-mf} show the grid diagram, density and pressure contour diagram, and \reffig{fig:Saltzman-typeI-II-compare-exact-mf} shows the scatter plots of density, pressure and velocity (x), at $\mathcal{T}_{1}^{Sa,I}$ and $\mathcal{T}_{1}^{Sa,II}$ with or without matter flow, at $ t = 0.5 $. The cell-to-cell oscillation is clear in the regular Lagrangian simulation. When the matter flow method is implemented, the pressure oscillation is flattened to a nearly perfect state, which is an expected result for the method, while the oscillation in density or velocity distributions with and without the matter flow method are with about the same size. Even though the density and velocity distributions do not gain remarkable improvement as the pressure, we can say the result is improved as a whole. 

\begin{figure}[!htpb]
	\subfigure[Density, $\mathcal{T}_{i}^{Sa,I}(i=1,2,3)$]{\label{fig:SaltzmanI-Density-multiscale-appro}
		\begin{minipage}[t]{0.33\linewidth}
			\centering
			\includegraphics[width=5.5cm]{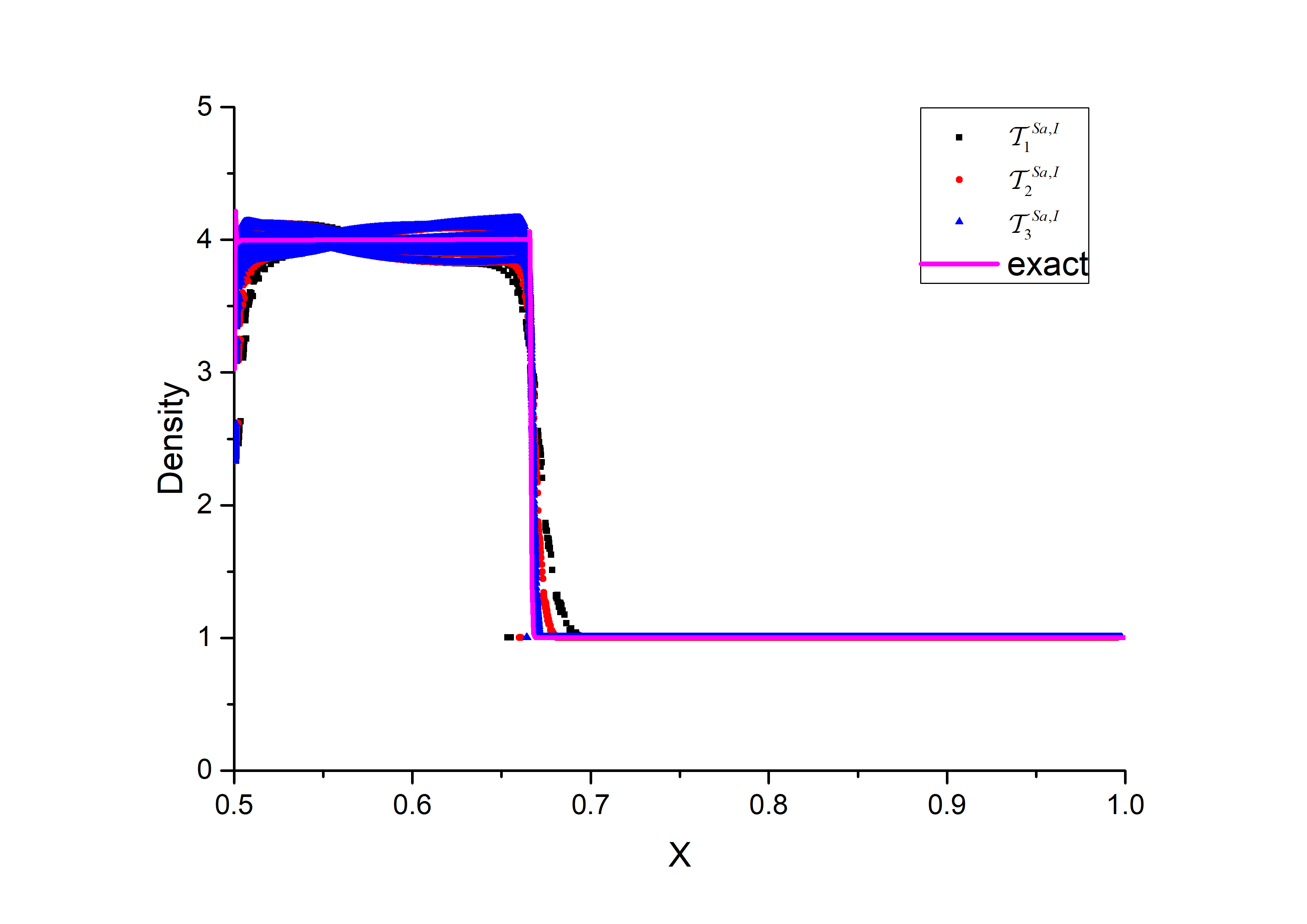}		
		\end{minipage}
	}%
	\subfigure[Pressure, $\mathcal{T}_{i}^{Sa,I}(i=1,2,3)$]{\label{fig:SaltzmanI-Pressure-multiscale-appro}
		\begin{minipage}[t]{0.33\linewidth}
			\centering
			\includegraphics[width=5.5cm]{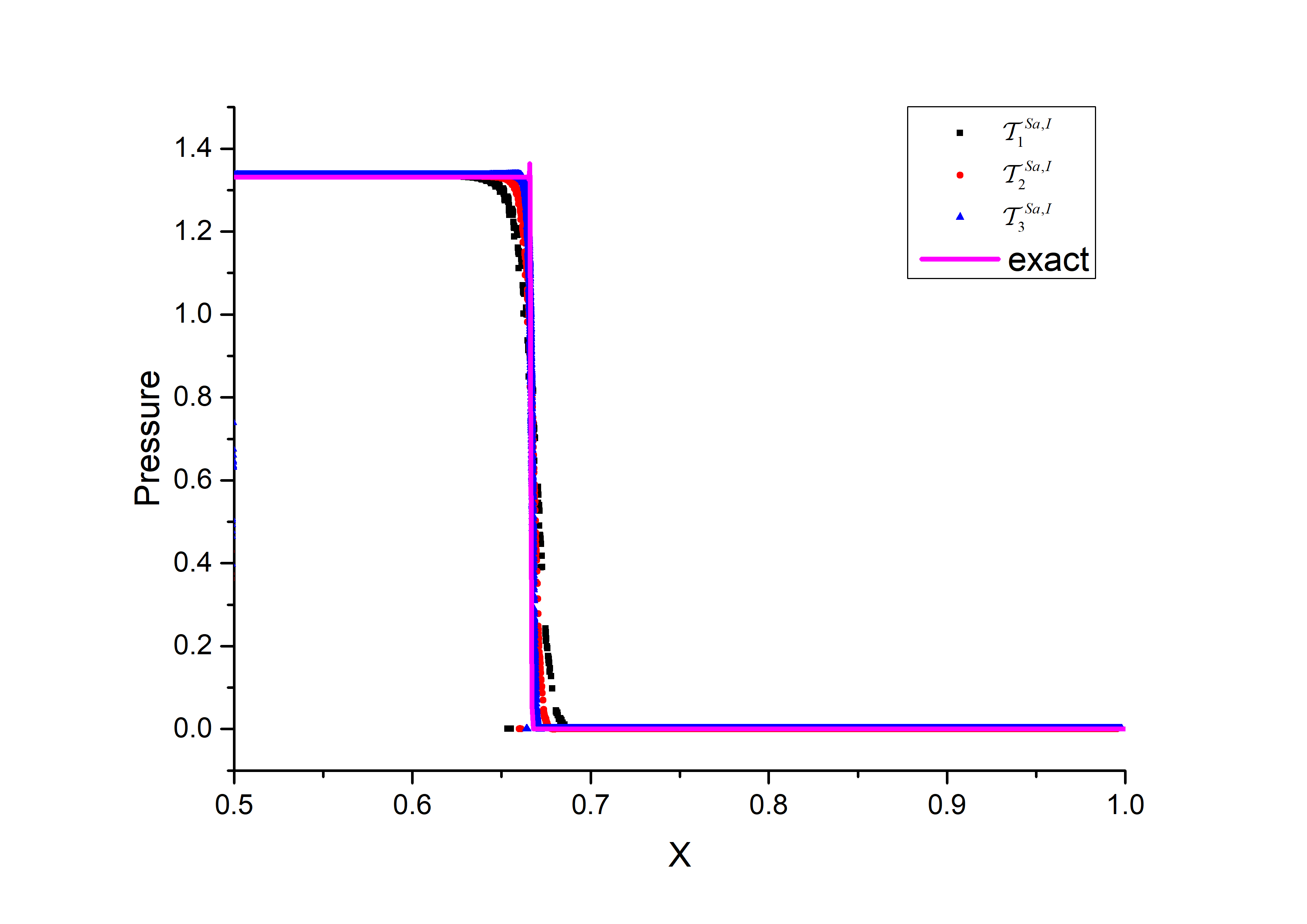}		
		\end{minipage}
	}%
	\subfigure[Velocity (x), $\mathcal{T}_{i}^{Sa,I}(i=1,2,3)$]{\label{fig:SaltzmanI-Velocity-multiscale-appro}
		\begin{minipage}[t]{0.33\linewidth}
			\centering
			\includegraphics[width=5.5cm]{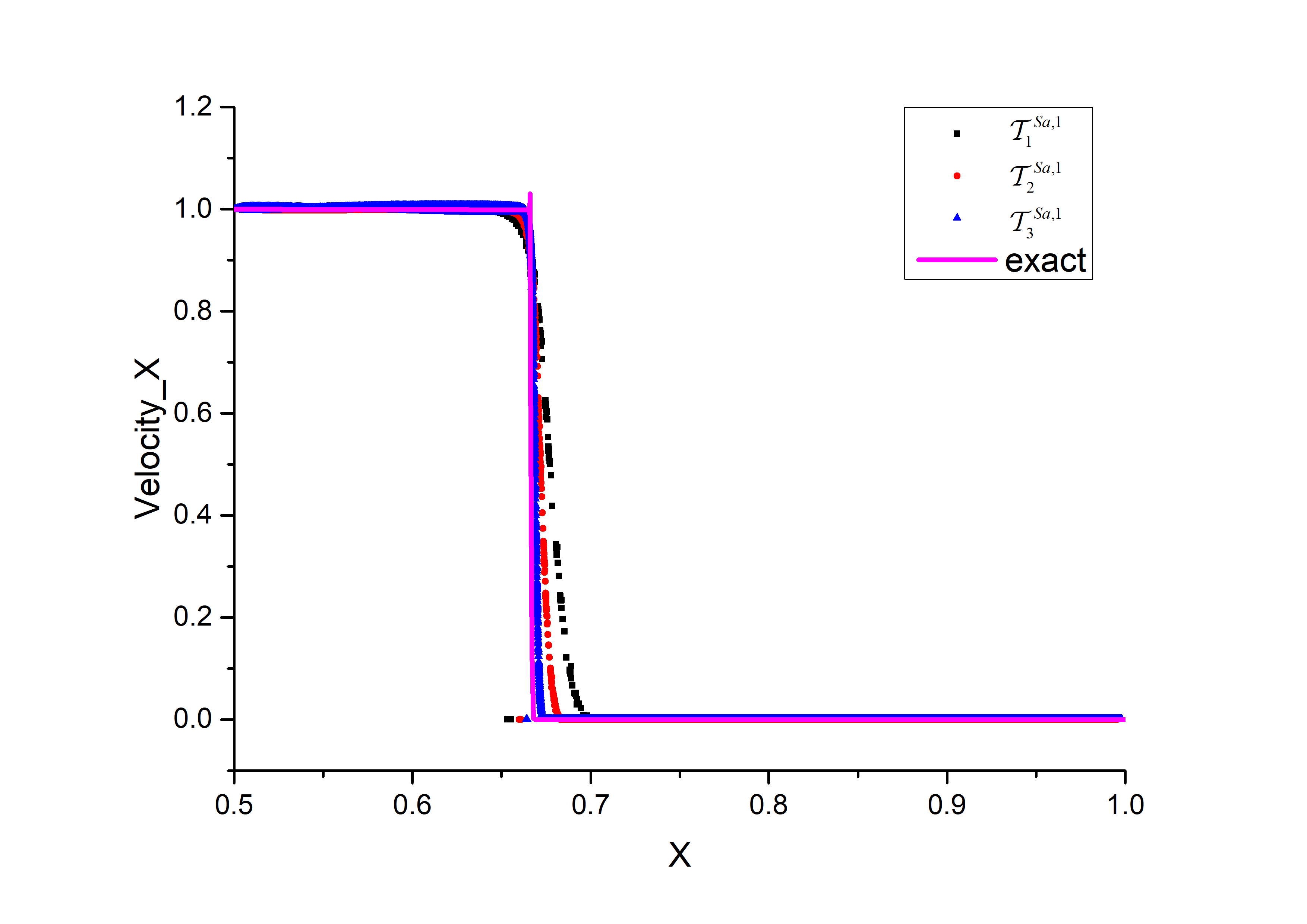}		
		\end{minipage}
	}%
	\vspace{-0.4cm}
	
	\subfigure[Density, $\mathcal{T}_{i}^{Sa,II}(i=1,2,3)$]{\label{fig:SaltzmanII-Density-multiscale-appro}
		\begin{minipage}[t]{0.33\linewidth}
			\centering
			\includegraphics[width=5.5cm]{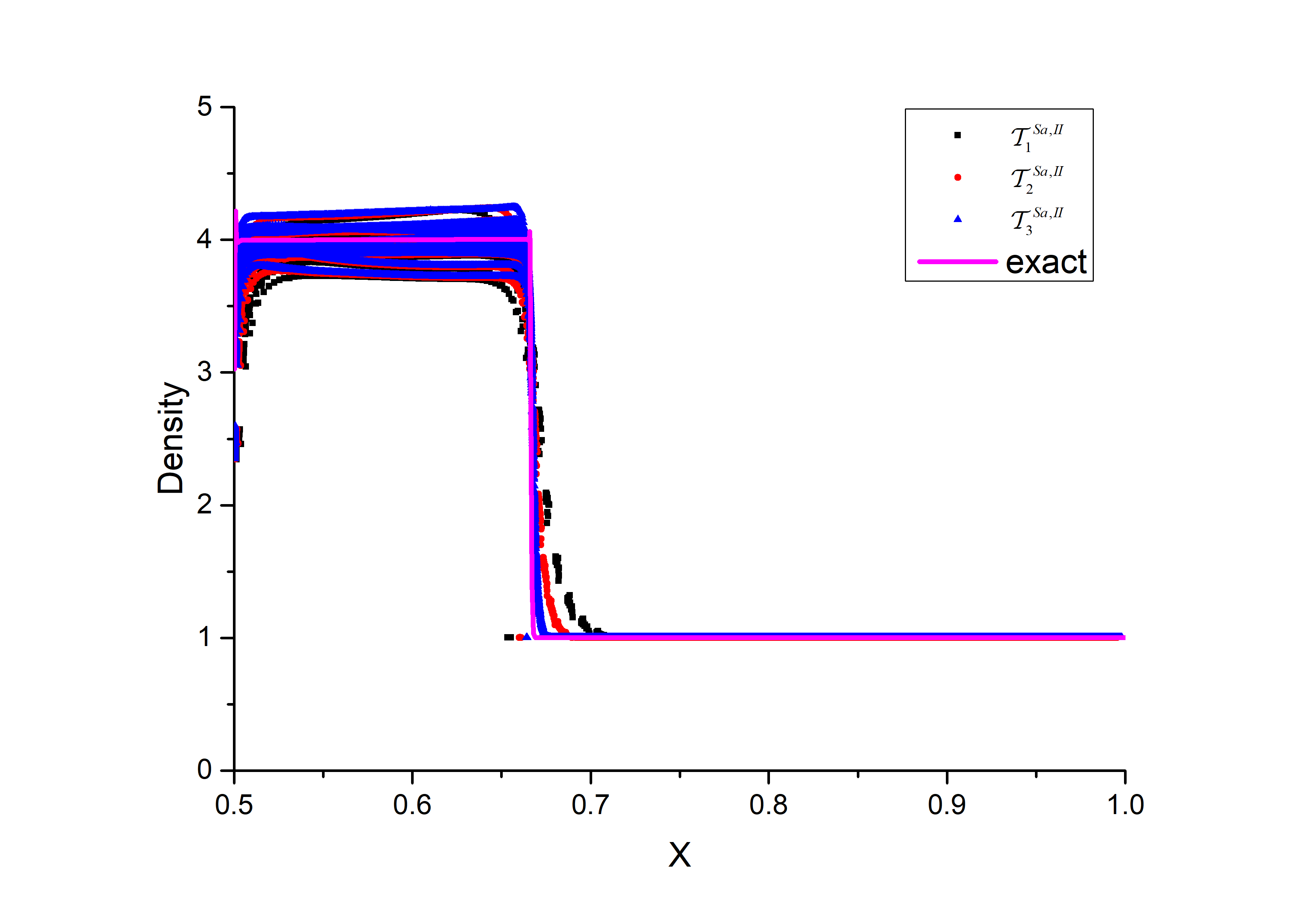}		
		\end{minipage}
	}%
	\subfigure[Pressure, $\mathcal{T}_{i}^{Sa,II}(i=1,2,3)$]{\label{fig:SaltzmanII-Pressure-multiscale-appro}
		\begin{minipage}[t]{0.33\linewidth}
			\centering
			\includegraphics[width=5.5cm]{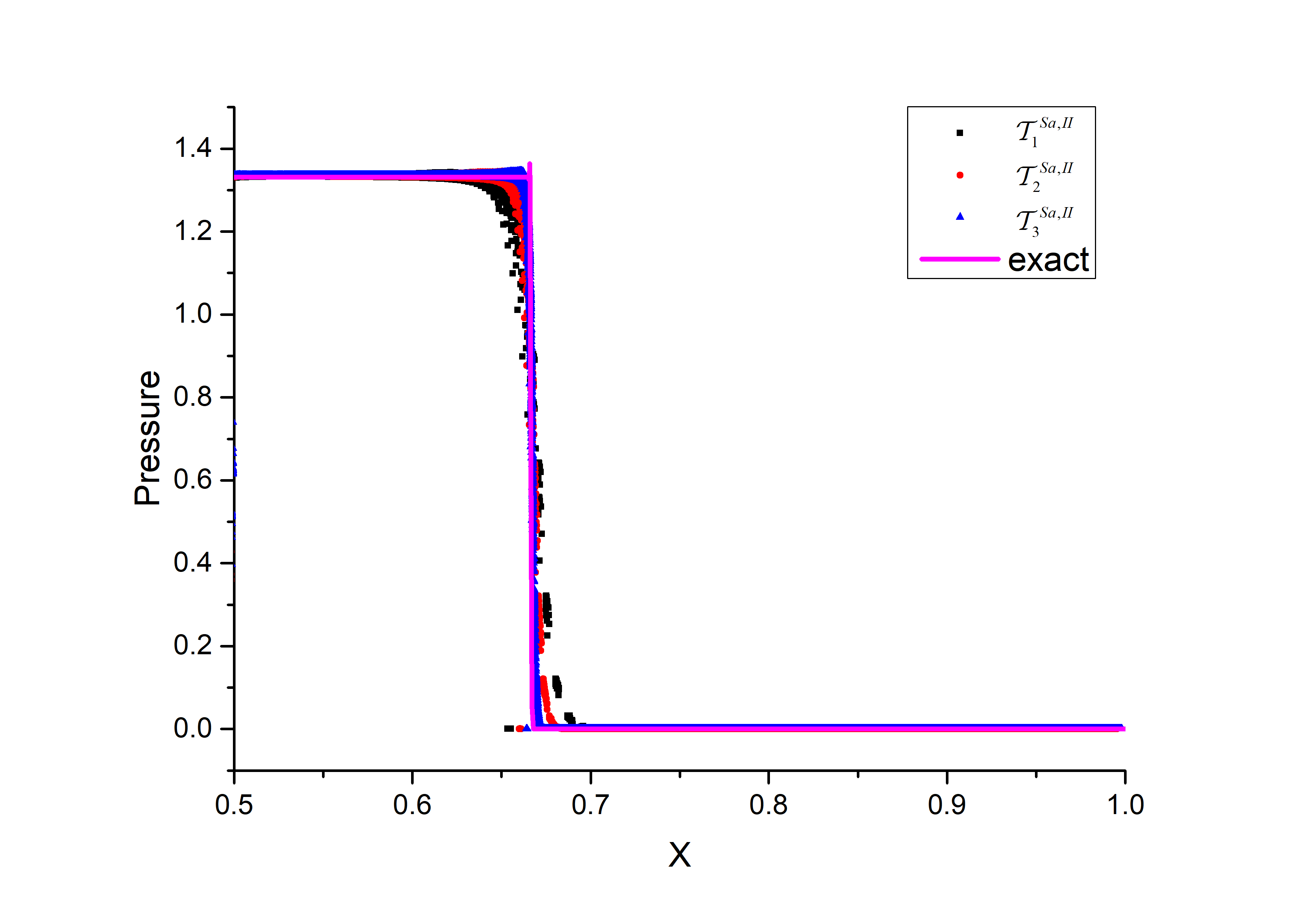}		
		\end{minipage}
	}%
	\subfigure[Velocity (x), $\mathcal{T}_{i}^{Sa,II}(i=1,2,3)$]{\label{fig:SaltzmanII-Velocity-multiscale-appro}
		\begin{minipage}[t]{0.33\linewidth}
			\centering
			\includegraphics[width=5.5cm]{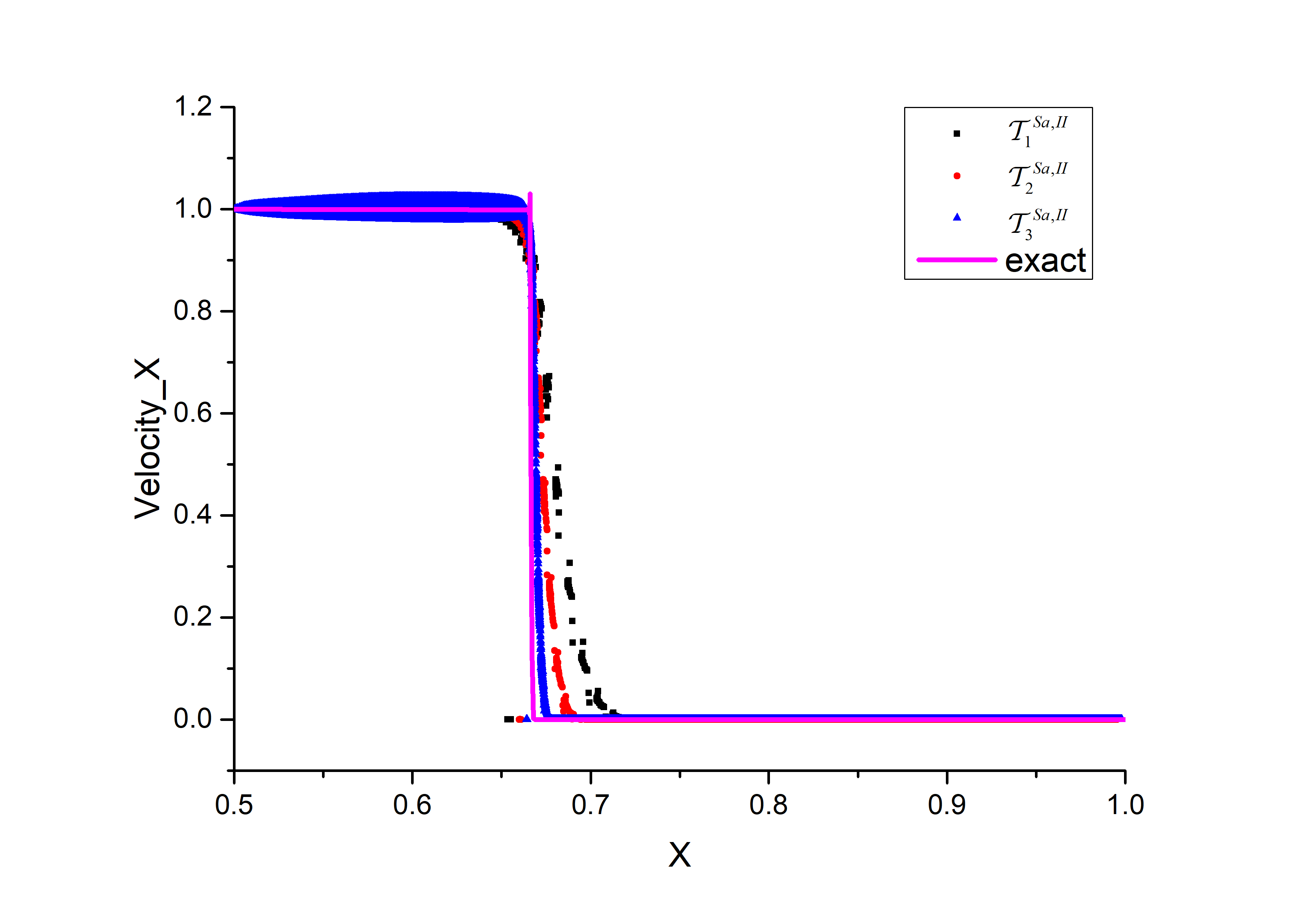}		
		\end{minipage}
	}%
	\caption{The density, pressure and velocity x scatter diagrams of $\mathcal{T}_{i}^{Sa,I}$ and $\mathcal{T}_{i}^{Sa,II}$, $i = 1,2,3 $ at the time of $t = 0.5 $.}\label{fig:Saltzman-typeI-II-multiscale-appro}
\end{figure}

To explore the dependence of the effect of the matter flow method on grid size, we have also carried out simulations of Saltzman problem using the matter flow method based on refined meshes $\mathcal{T}_{i}^{Sa,I}$ and $\mathcal{T}_{i}^{Sa,II}$, $i=2,3$ that are refined 2 and 5 times respectively on the basis of $\mathcal{T}_{1}^{Sa,I}$ and $\mathcal{T}_{1}^{Sa,II}$, as listed in \reftab{tab:saltzman-mesh-resolution}. The simulation results are shown in \reffig{fig:Saltzman-typeI-II-multiscale-appro}. As the refined mesh corresponds to a smaller viscos, the simulation results are closer to the ideal solution for refined mesh, as expected. The oscillation, on the other hand however, does not shrink with the mesh refinement. We interpret this phenomenon as follows: the effect of the matter flow method depends on the smoothness of the physical quantities distributions along cells, the smoother the better; when the mesh is refined, the viscous also becomes smaller, this leads to sharper distributions of physical quantities along space for shock wave problem, and if transferring from space to cells, it leads to physical quantity distribution smoothness approximately independent of the mesh size, and so the oscillation is also approximately independent of the mesh size. So, this phenomenon can be viewed as a nature of the matter flow method combined with the SGH Lagrangian scheme. This analysis also point a way of how to obtain result with the oscillation better alleviated than that in \reffig{fig:Saltzman-typeI-II-multiscale-appro}: to construct more refined mesh and, at the same time, increase the viscous coefficient.

\newpage

\begin{table}[!htpb]         
	\centering         
	\caption{Type I and type II grid size information}\label{tab:saltzman-mesh-resolution}         
	\begin{tabular}{|c|c|c|}         
		\hline         
		Type    & Notation                    & Mesh resolution\\
		\hline 
		&  $\mathcal{T}_{1}^{Sa,I}$   &  100x10 \\   
		Type I  &  $\mathcal{T}_{2}^{Sa,I}$   &  200x20 \\  
		&  $\mathcal{T}_{3}^{Sa,I}$   &  500x50 \\  
		\hline   
		&  $\mathcal{T}_{1}^{Sa,II}$  &  100x10  \\
		Type II &  $\mathcal{T}_{2}^{Sa,II}$  &  200x20  \\
		&  $\mathcal{T}_{3}^{Sa,II}$  &  500x50  \\
		\hline         
	\end{tabular}         
\end{table}

\newpage
\subsection{Noh Implosion Problem}
\begin{figure}[!htpb]
	\centering
	\subfigure[$\mathcal{T}_{1}^{No}$]{\label{fig:NohII-40x40-mesh-t0}
		\begin{minipage}[t]{0.33\linewidth}
			\centering
			\includegraphics[width=4cm]{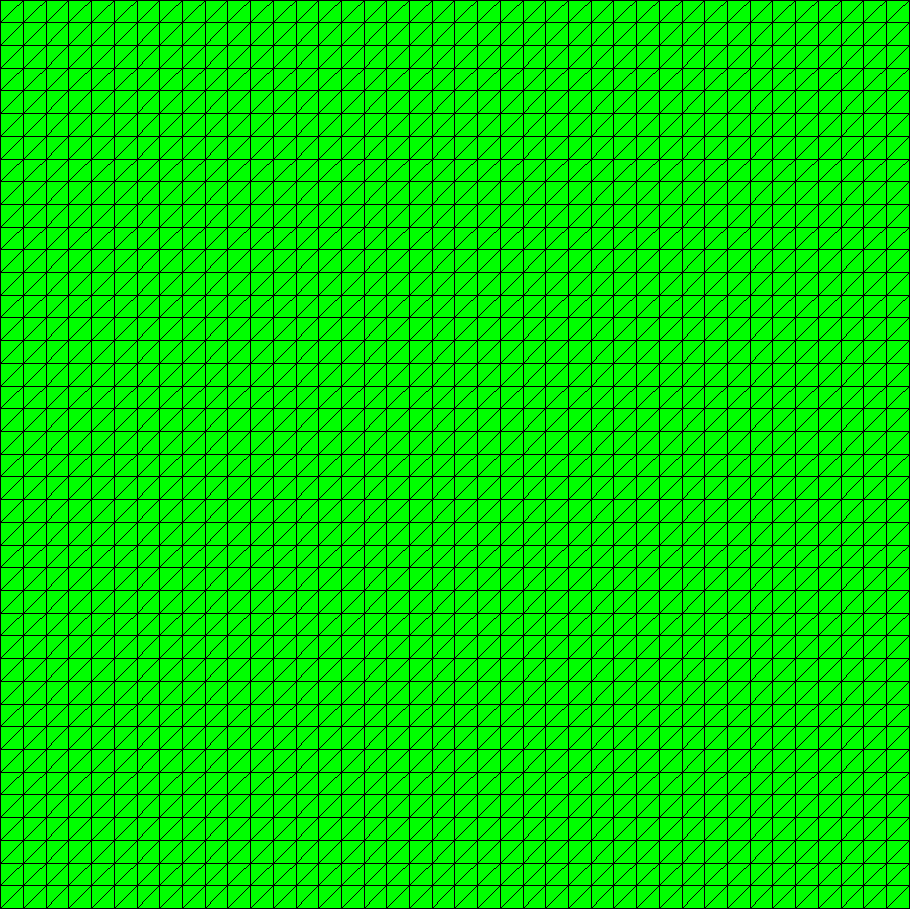}		
		\end{minipage}
	}%
	\subfigure[$\mathcal{T}_{2}^{No}$]{\label{fig:NohII-80x80-mesh-t0}
		\begin{minipage}[t]{0.33\linewidth}
			\centering
			\includegraphics[width=4cm]{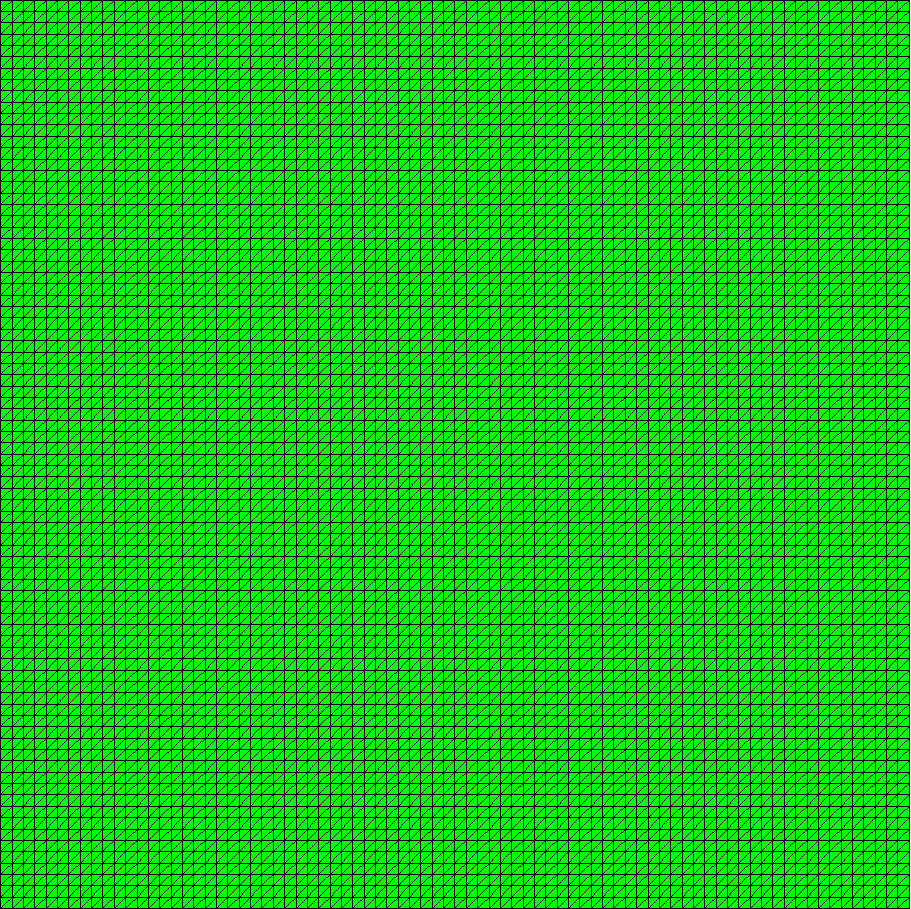}		
		\end{minipage}
	}%
	\subfigure[$\mathcal{T}_{3}^{No}$]{\label{fig:NohII-160x160-mesh-t0}
		\begin{minipage}[t]{0.33\linewidth}
			\centering
			\includegraphics[width=4cm]{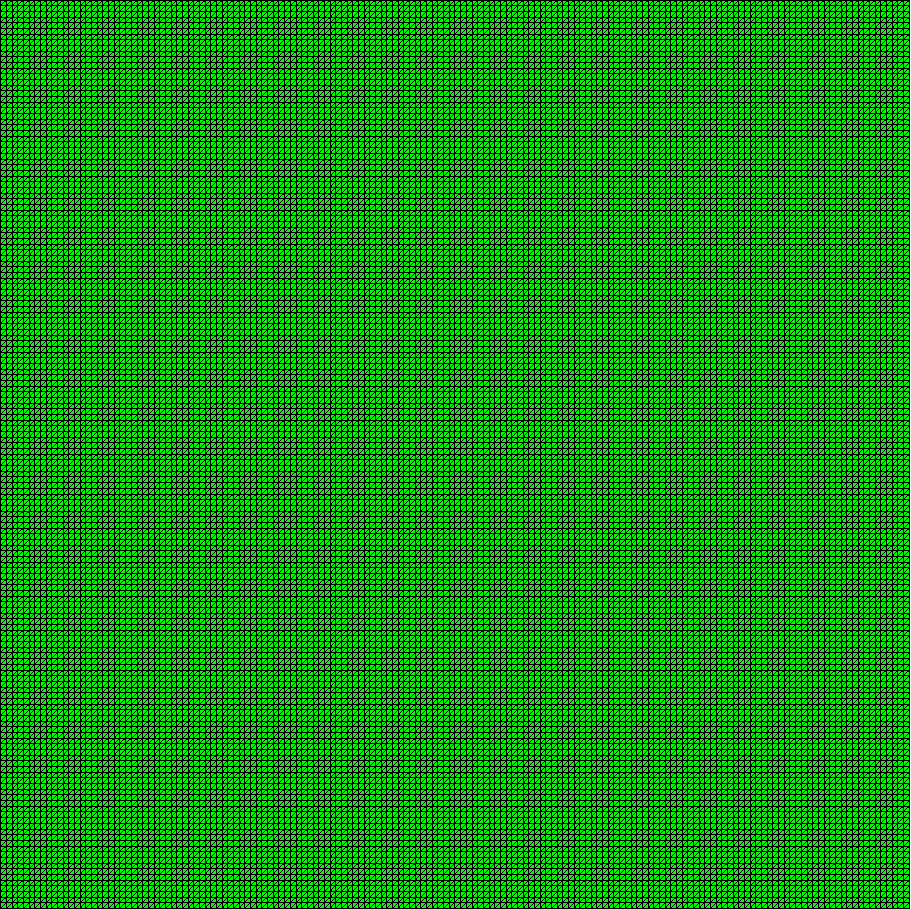}		
		\end{minipage}
	}%
	\caption{Three kinds of initial meshes}\label{fig:NohII-initial-mesh}
\end{figure}

\begin{example}
Consider the model problem \eqref{eq:mass-equation}-\eqref{eq:state-equation}, where domain $ \Omega = [0,0.8] \times [0, 0.8] $, simulation time $ t \in [0, 0.4] $ (General reference $ t=0.6 $, but the grid distortion in our calculation is too severe, we can not calculate this time, so we take $ t = 0.4 $), gas adiabatic $ \gamma = \frac{5}{3} $. Initial condition: initial density is 1, specific internal energy is 0, and velocity is 1 (the direction points to the origin, i.e. lower left corner (0, 0) position). Boundary condition: the left and lower boundaries of the domain adopts the solid wall boundary condition, and the right and upper boundaries adopt free surfaces conditions. The domain $ \Omega $ is divided by a uniform grid of 40$\times$40, 80$\times$80 and 160$\times$160 (see \reffig{fig:NohII-initial-mesh}), which are recorded as $\mathcal{T}_{1}^{No}$, $\mathcal{T}_{2}^{No}$ and $\mathcal{T}_{3}^{No}$, respectively.
\end{example}

\begin{rmk}
	$Radius$ represents the radius, i.e. $ Radius  = \sqrt{x^2 + y^2}$, and the radial velocity represents the velocity along the radius direction, which is $\sqrt{u_x^2 + u_y^2}$ and points to the origin.
\end{rmk}

\begin{figure}[!htpb]
	\centering
	\subfigure[Local mesh, no-matterflow]{\label{fig:NohII-c-local-no-mf}
		\begin{minipage}[t]{0.33\linewidth}
			\centering
			\includegraphics[width=4.4cm]{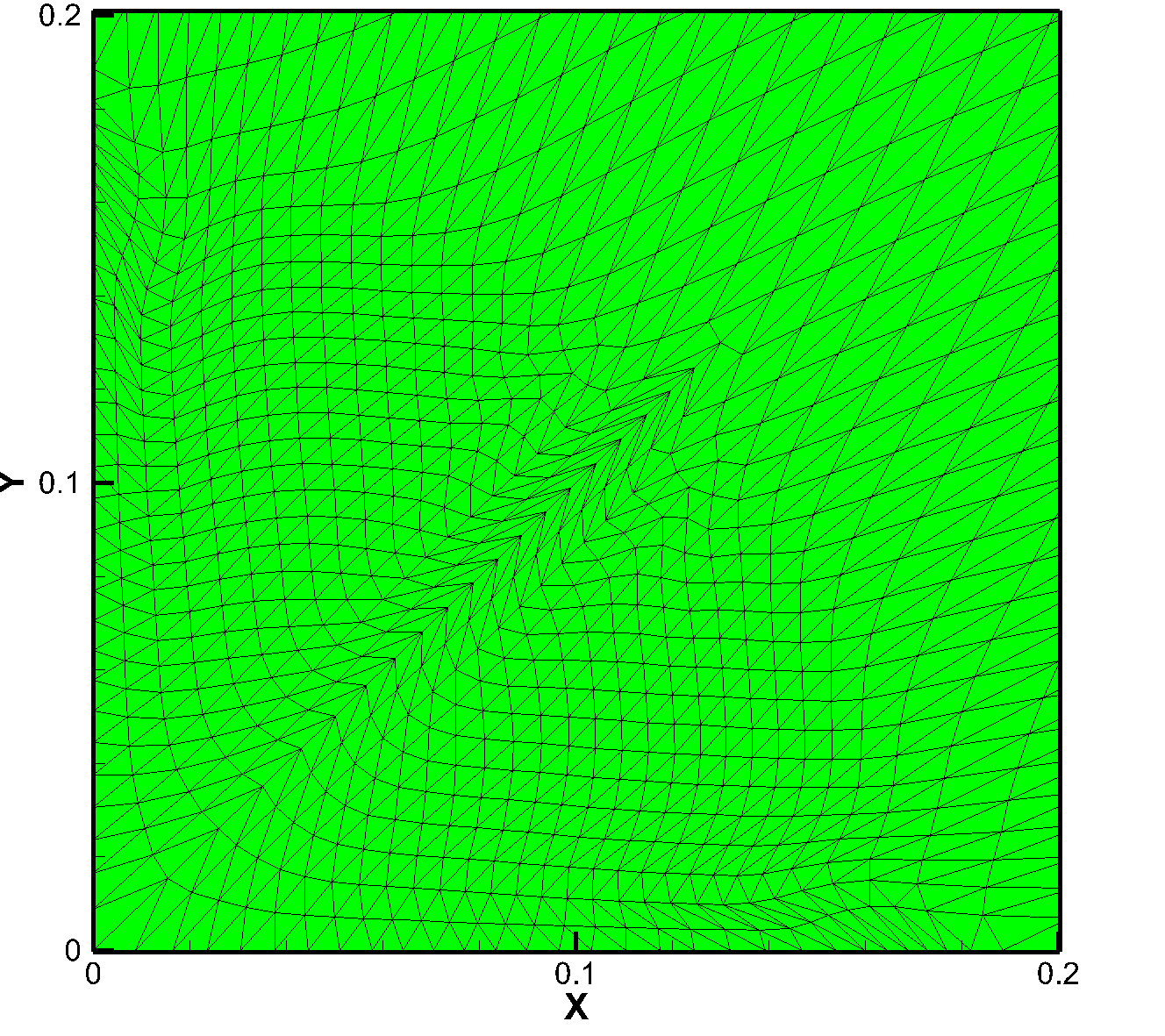}	
		\end{minipage}
	}%
	\subfigure[Density, no-matterflow]{\label{fig:NohII-density-local-no-mf}
		\begin{minipage}[t]{0.33\linewidth}
			\centering
			\includegraphics[width=4.8cm]{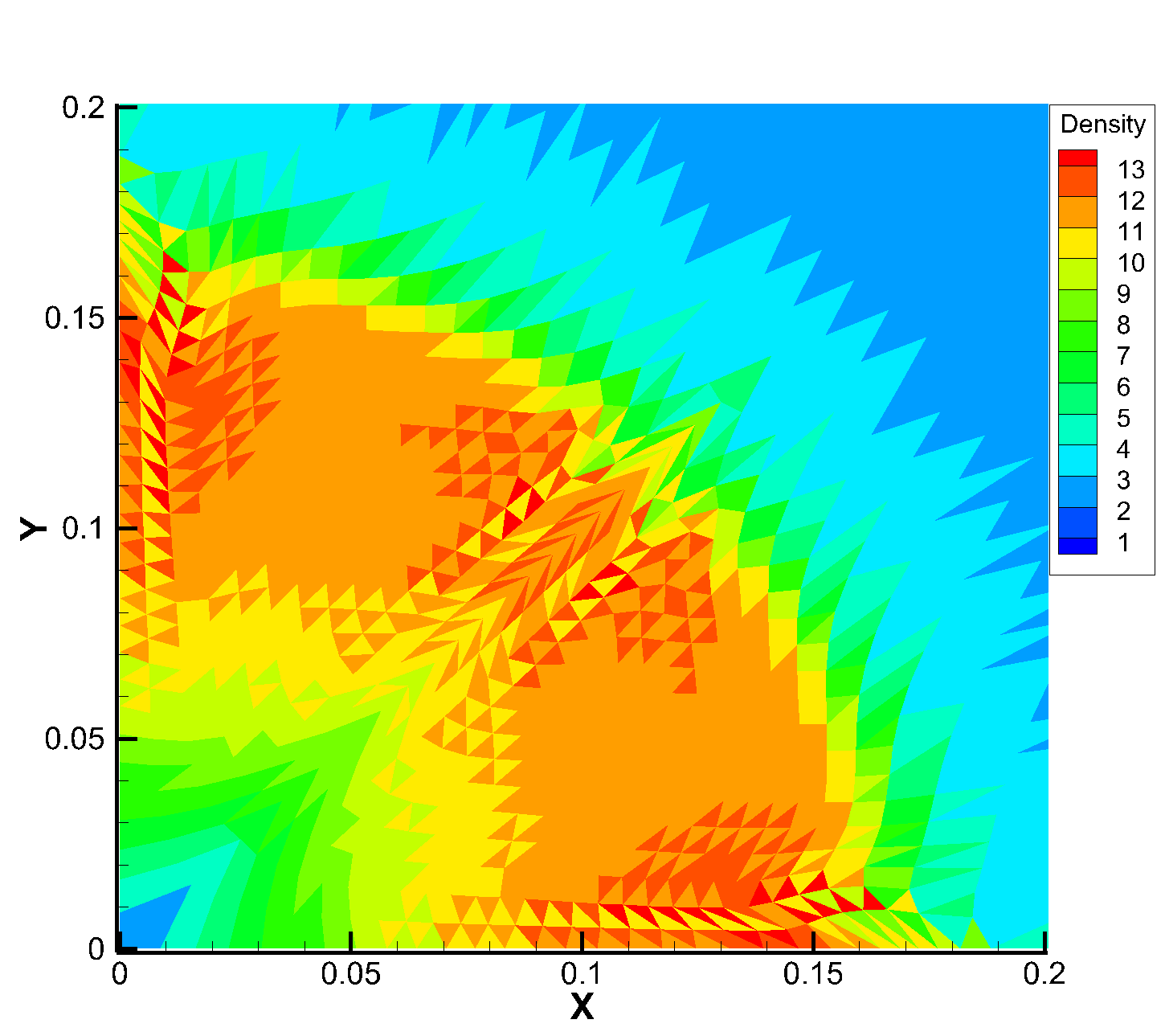}		
		\end{minipage}
	}%
	\subfigure[Pressure, no-matterflow]{\label{fig:NohII-pressure-local-no-mf}
		\begin{minipage}[t]{0.33\linewidth}
			\centering
			\includegraphics[width=4.8cm]{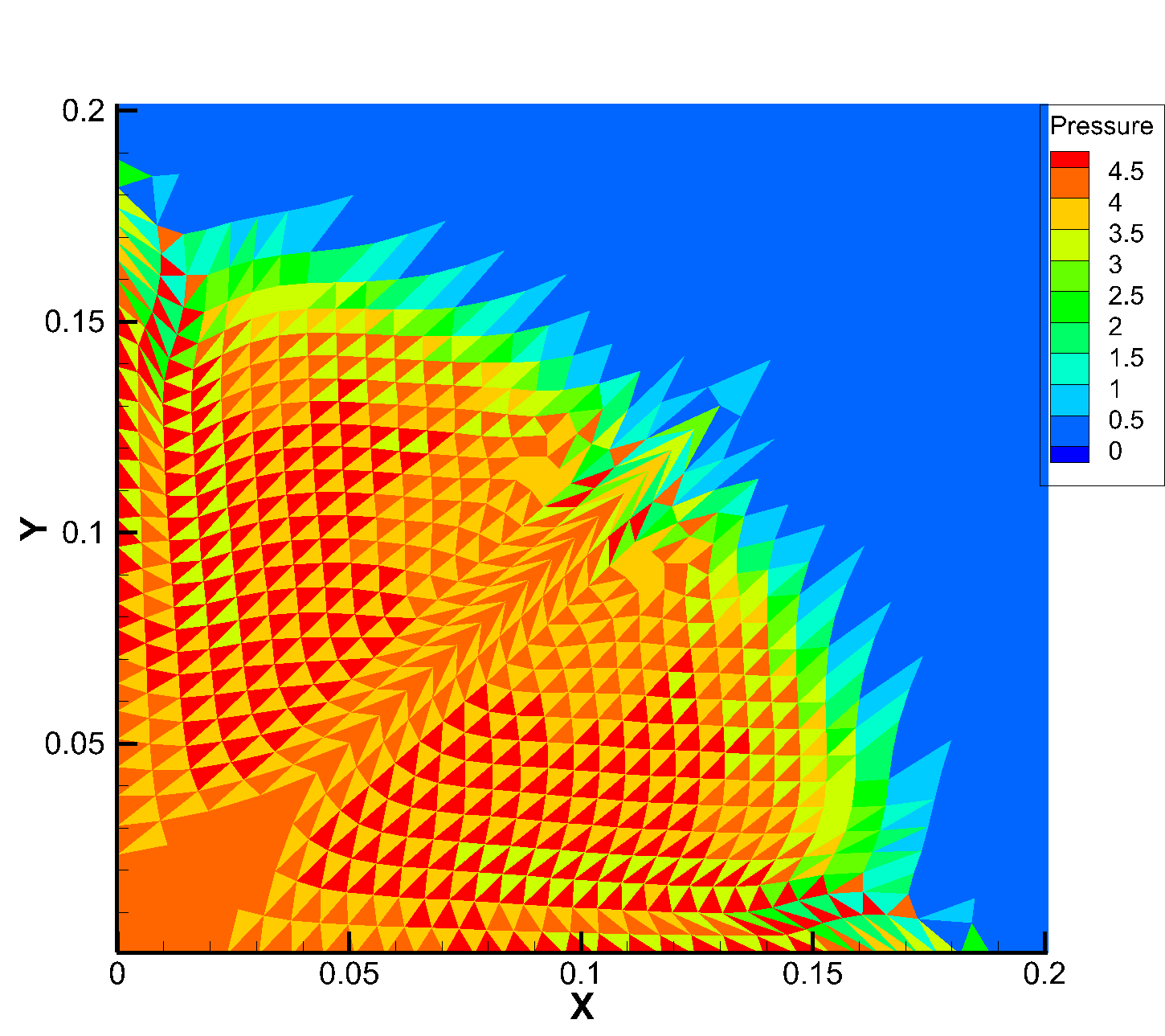}		
		\end{minipage}
	}%
	
	\subfigure[Local mesh, matterflow]{\label{fig:NohII-c-local-mf}
		\begin{minipage}[t]{0.33\linewidth}
			\centering
			\includegraphics[width=4.4cm]{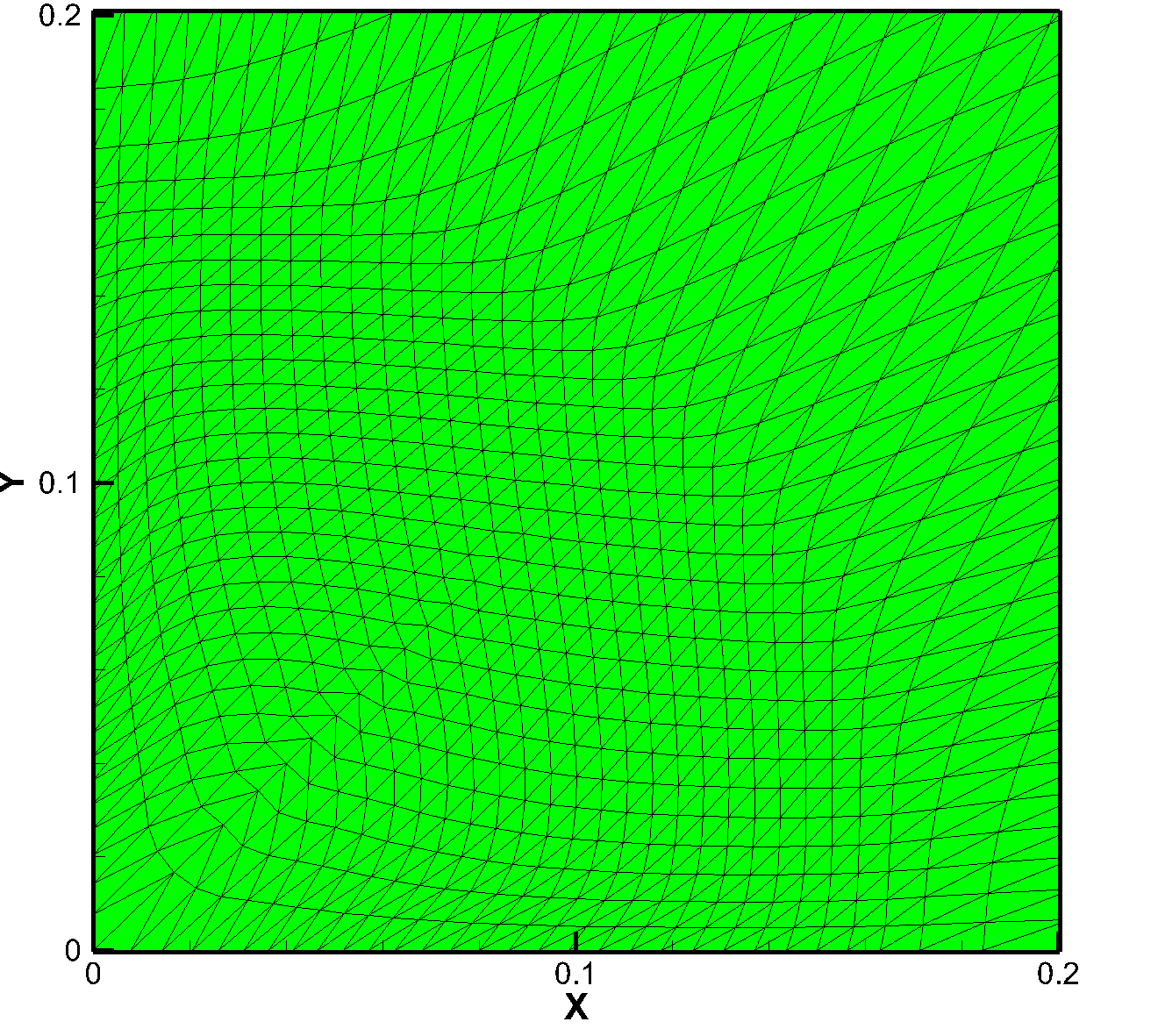}
		\end{minipage}
	}%
	\subfigure[Density, matterflow]{\label{fig:NohII-density-local-mf}
		\begin{minipage}[t]{0.33\linewidth}
			\centering
			\includegraphics[width=4.8cm]{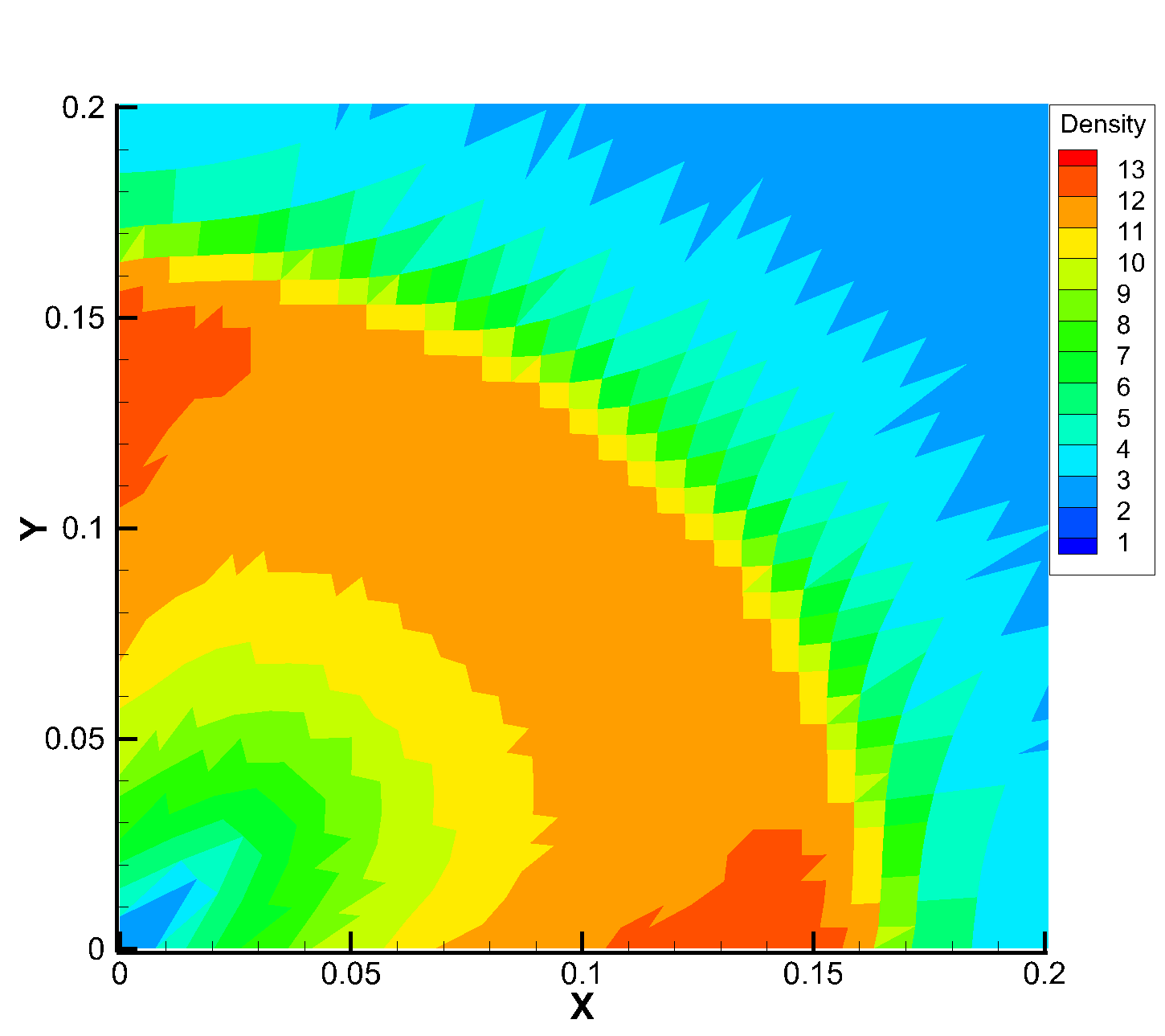}		
		\end{minipage}
	}%
	\subfigure[Pressure, matterflow]{\label{fig:NohII-pressure-local-mf}
		\begin{minipage}[t]{0.33\linewidth}
			\centering
			\includegraphics[width=4.8cm]{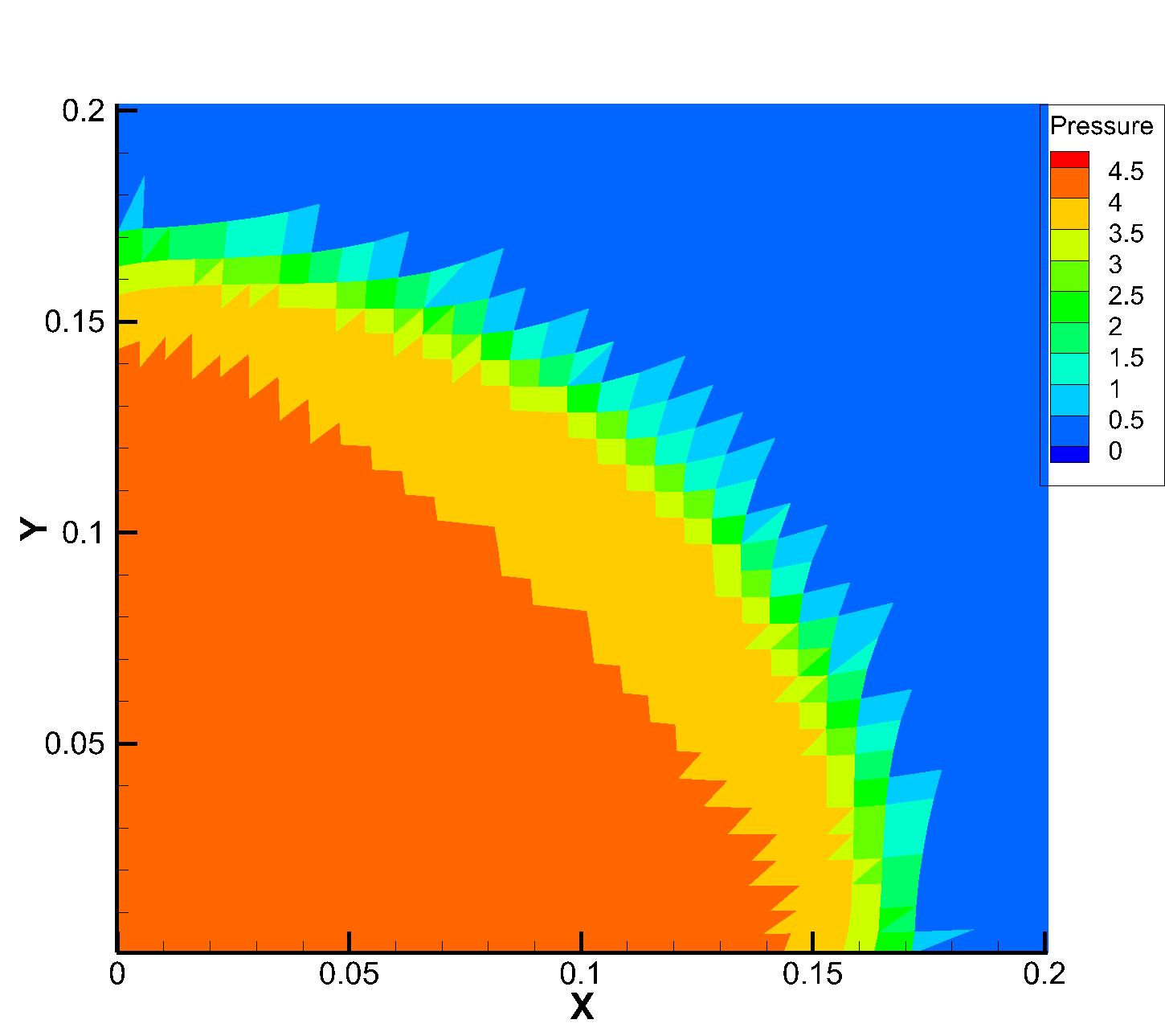}		
		\end{minipage}
	}%
	\caption{Mesh, density, pressure contour diagram of $\mathcal{T}_{1}^{No}$ at $ t = 0.4 $.
	}\label{fig:NohII-compare-dist-mf}
\end{figure}

\begin{figure}[!htpb]
	\centering
	\subfigure[Density, matterflow]{\label{fig:NohII-Density-40x40-mf}
		\begin{minipage}[t]{0.33\linewidth}
			\centering
			\includegraphics[width=5.5cm]{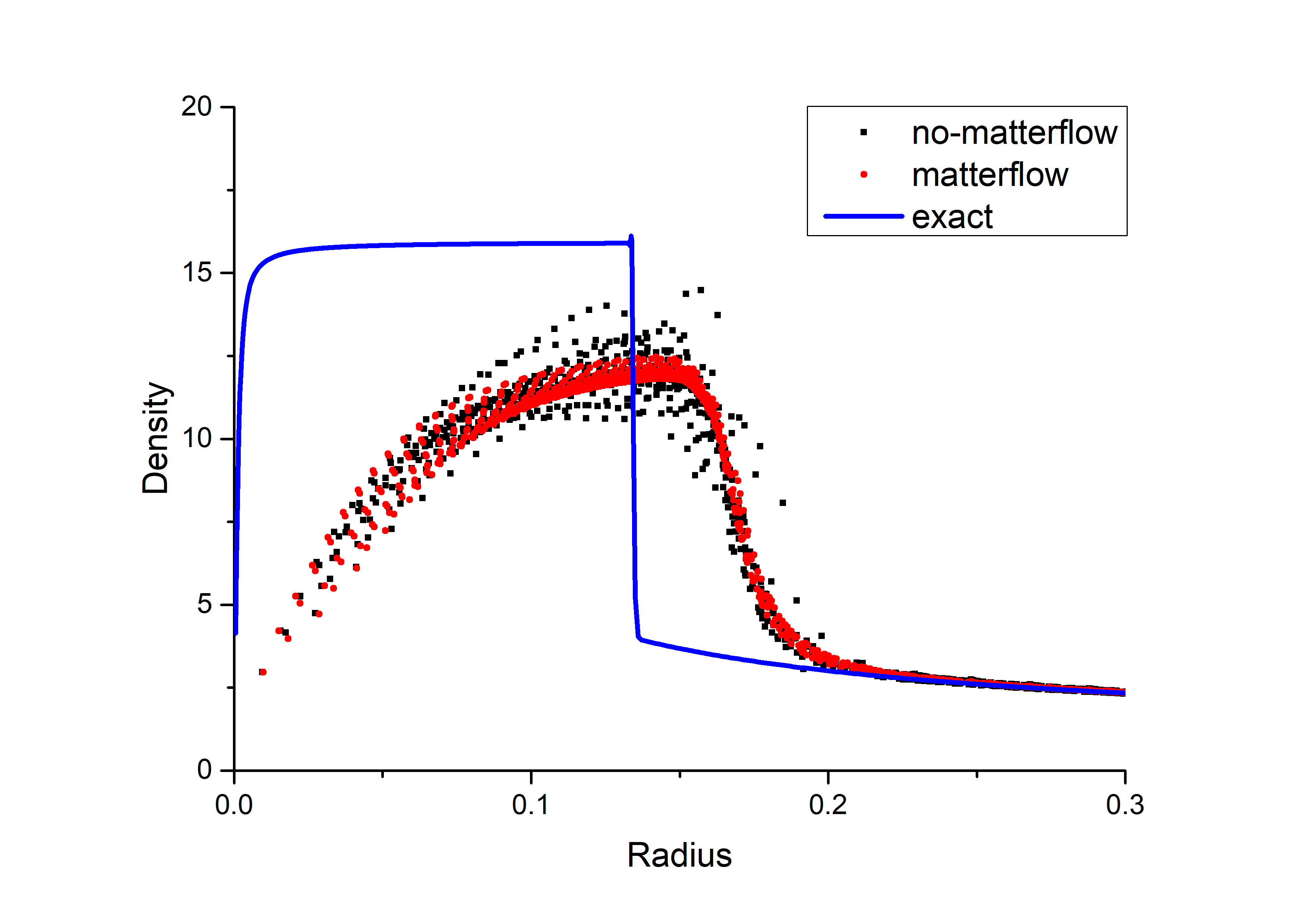}		
		\end{minipage}
	}%
	\subfigure[Pressure, matterflow]{\label{fig:NohII-Pressure-40x40-mf}
		\begin{minipage}[t]{0.33\linewidth}
			\centering
			\includegraphics[width=5.5cm]{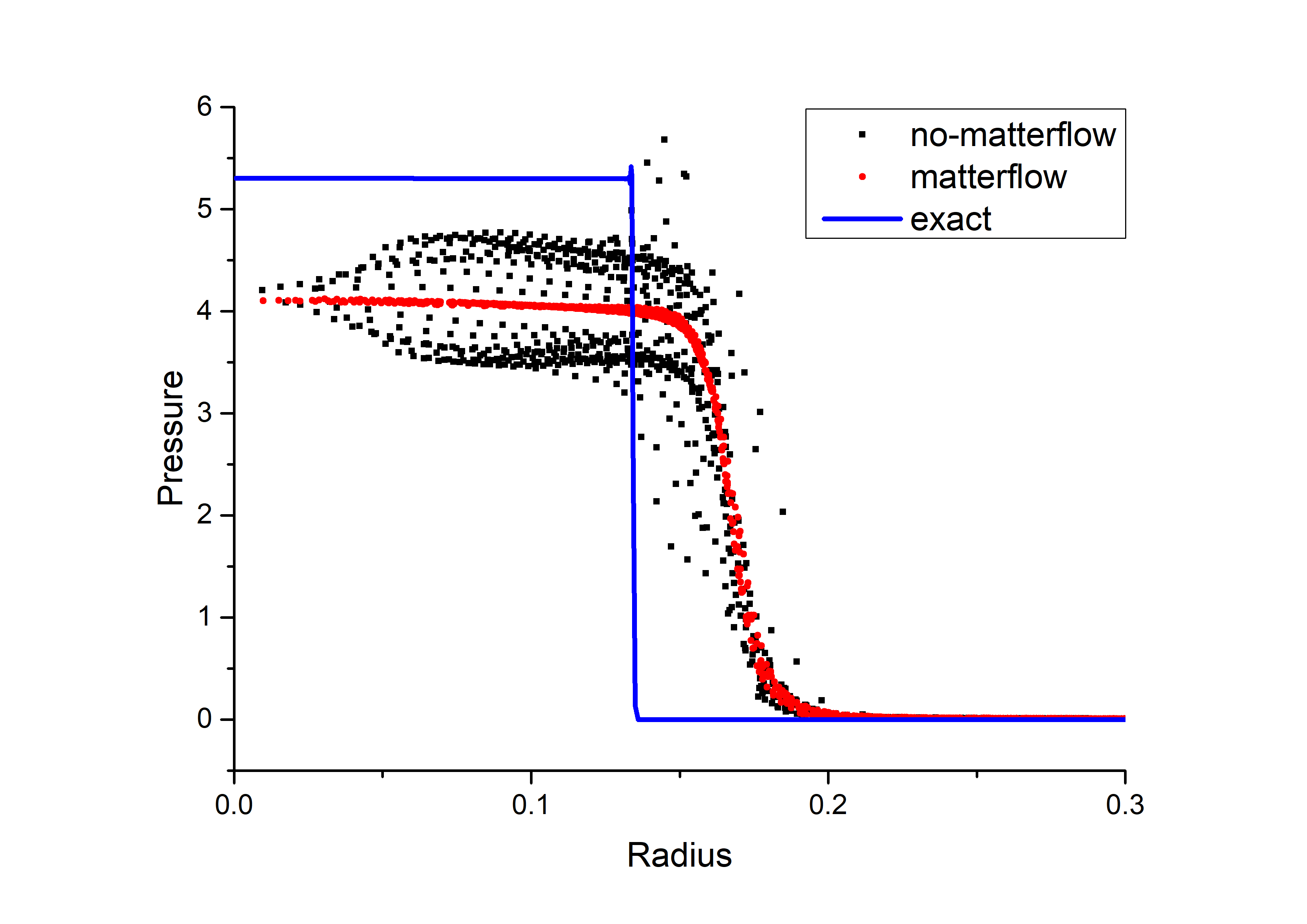}		
		\end{minipage}
	}%
	\subfigure[Velocity(r), matterflow]{\label{fig:NohII-Velocity-40x40-mf}
		\begin{minipage}[t]{0.33\linewidth}
			\centering
			\includegraphics[width=5.5cm]{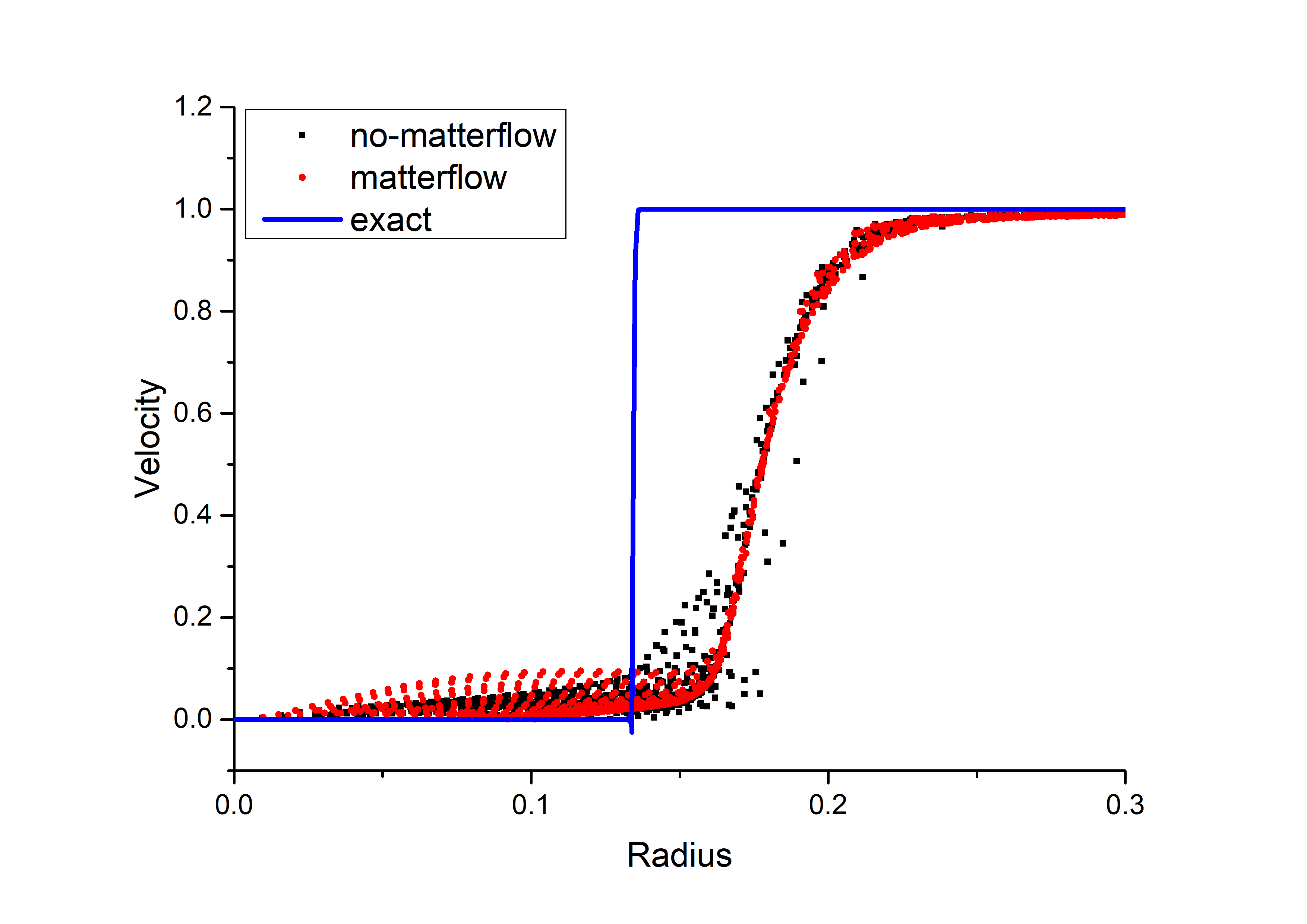}		
		\end{minipage}
	}%
	\caption{The density, pressure and radial velocity scatter diagram of $\mathcal{T}_{1}^{No}$ at $ t = 0.4 $.
	}\label{fig:NohII-compare-exact-mf}
\end{figure}

\textbf{Firstly, the influence of matter flow on the experimental results is explored for the initial mesh $\mathcal{T}_{1}^{No}$.} \reffig{fig:NohII-compare-dist-mf} shows the grid and density, pressure contour diagram at $t =0.4$. \reffig{fig:NohII-compare-exact-mf} shows the density, pressure, radial velocity scatter diagram at $t =0.4$. The big difference between the numerical solution and the ideal solution in \reffig{fig:NohII-compare-exact-mf} is due to the large artificial viscosity corresponding to the coarse mesh. As can be seen from the diagram in \reffig{fig:NohII-compare-dist-mf} and \reffig{fig:NohII-compare-exact-mf}, the introduction of the matter flow method can greatly alleviate the physical quantity oscillation in SGH Lagrange simulation. As a side effect, matter flow method also reduces the distortion of the mesh in the simulation of Noh problems.

\begin{figure}[!htpb]
	\subfigure[Density, matterflow]{\label{fig:NohII-Density-multiscale-appro}
		\begin{minipage}[t]{0.33\linewidth}
			\centering
			\includegraphics[width=5.5cm]{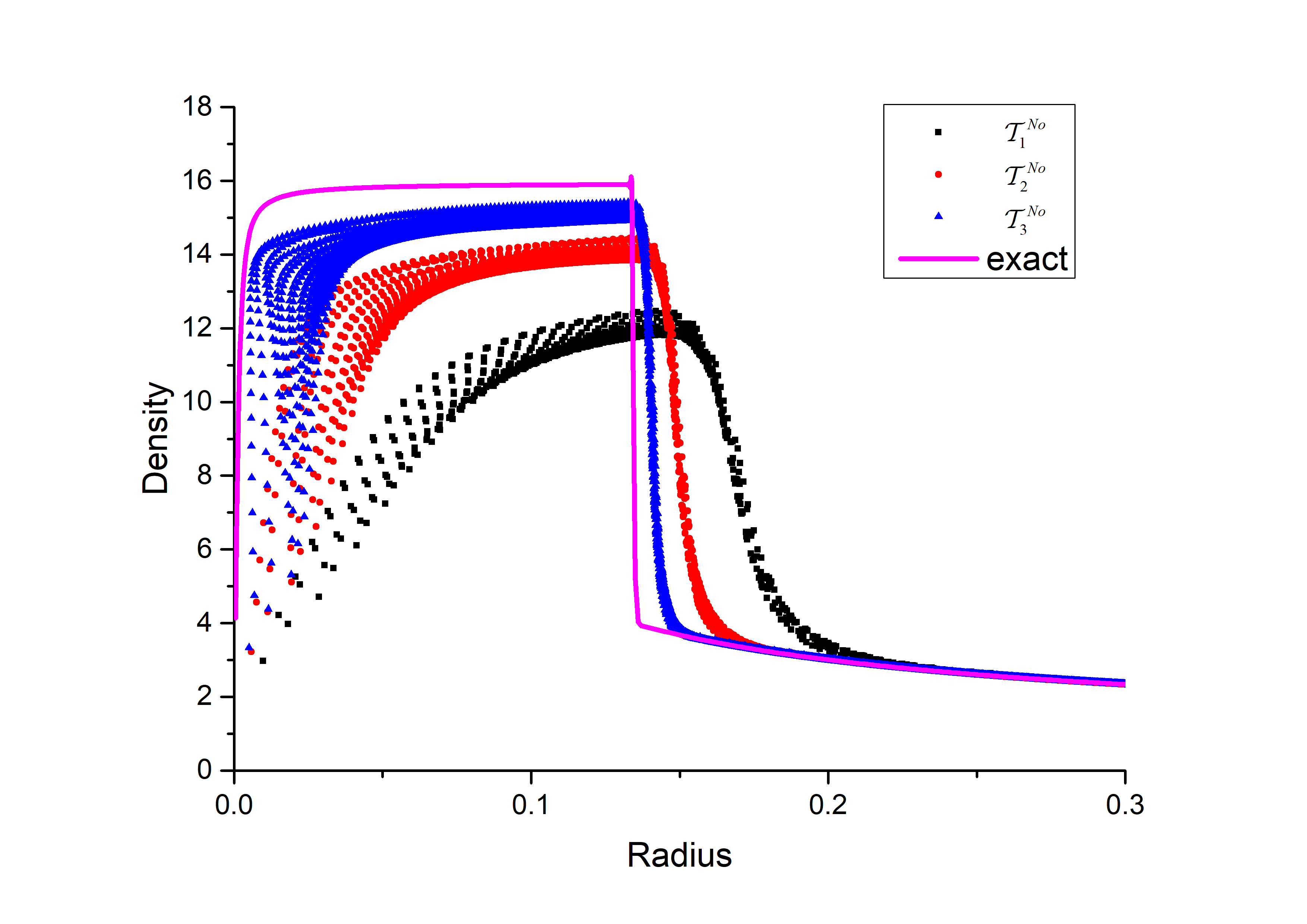}		
		\end{minipage}
	}%
	\subfigure[Pressure, matterflow]{\label{fig:NohII-Pressure-multiscale-appro}
		\begin{minipage}[t]{0.33\linewidth}
			\centering
			\includegraphics[width=5.5cm]{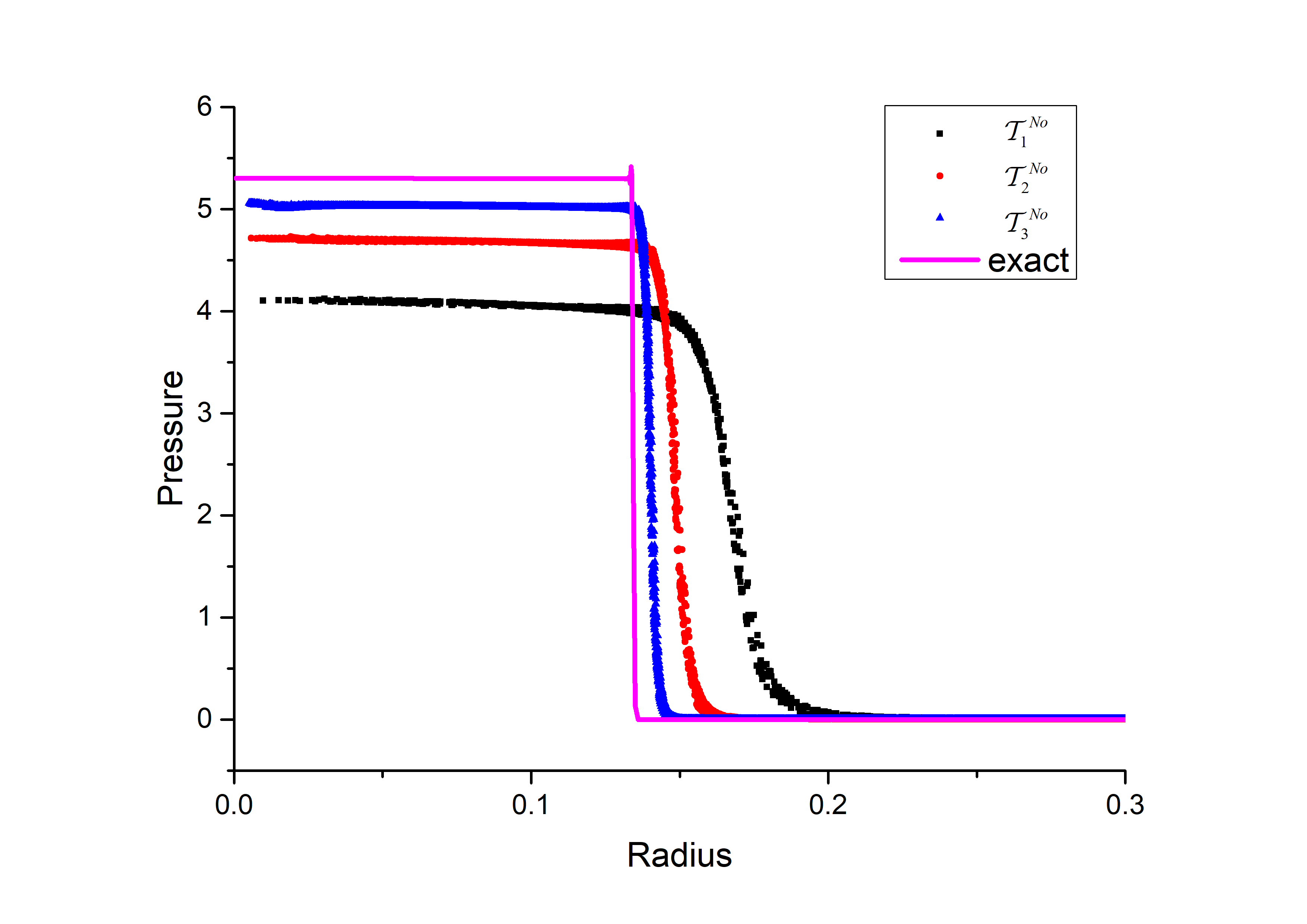}		
		\end{minipage}
	}%
	\subfigure[Velocity(r), matterflow]{\label{fig:NohII-Velocity-multiscale-appro}
		\begin{minipage}[t]{0.33\linewidth}
			\centering
			\includegraphics[width=5.5cm]{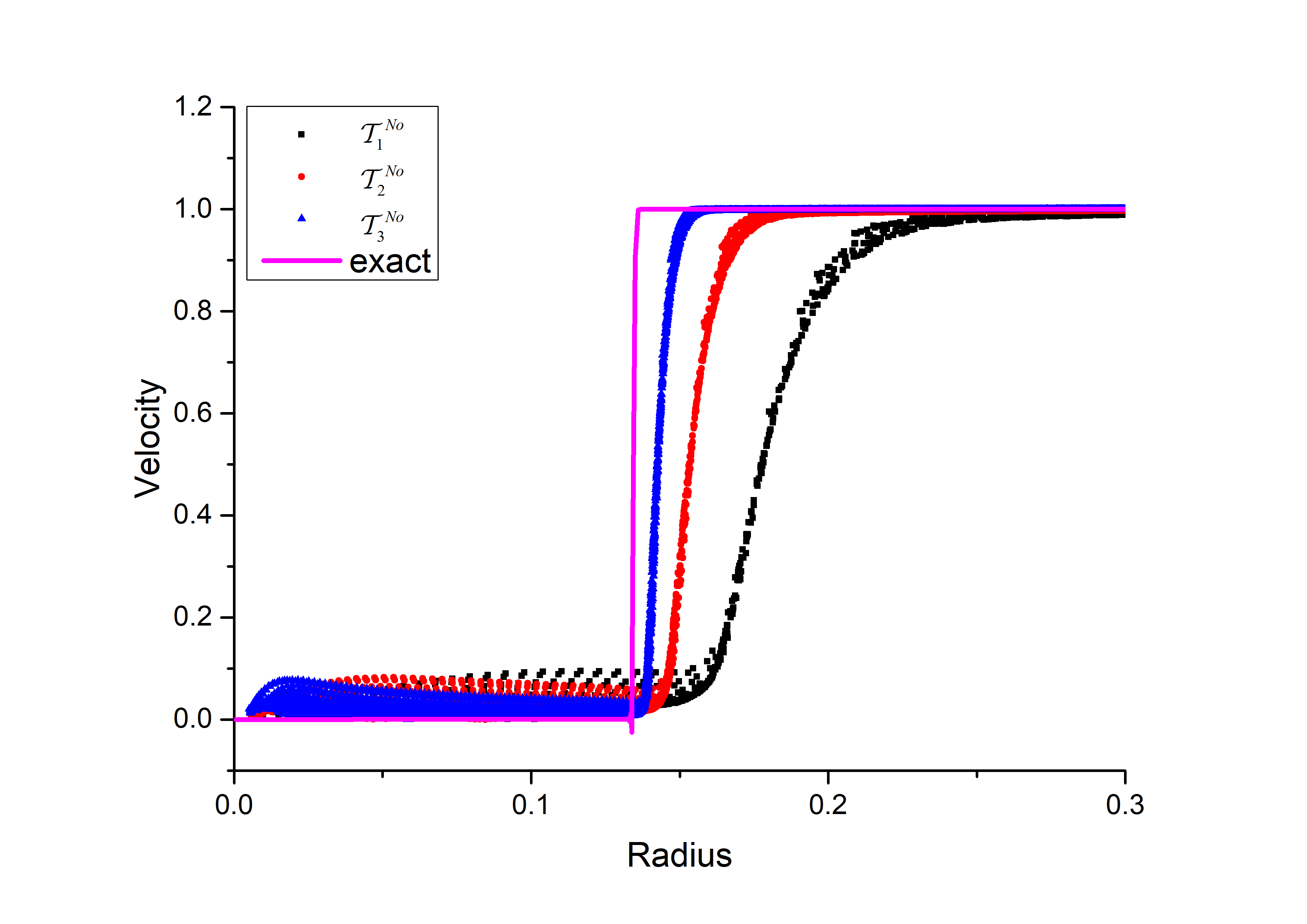}		
		\end{minipage}
	}%
	\caption{The density, pressure and velocity scatter diagram of different grid sizes at $ t = 0.4 $.
	}\label{fig:NohII-multiscale-appro}
\end{figure}

\textbf{Then, the influences of different grid sizes on the results are explored for three groups of initial meshes $\mathcal{T}_{1}^{No}$, $\mathcal{T}_{2}^{No}$ and $\mathcal{T}_{3}^{No}$.} \reffig{fig:NohII-multiscale-appro} shows the corresponding scatter diagram of density, pressure and velocity at $ t = 0.4 $ simulated using the matter flow method. Similar to that in Saltzman problem simulation, the numerical solution is closer to the exact solution when the mesh is refined, but the oscillation is almost independent of the mesh refinement.

\newpage
\subsection{Sedov Explosion Problem}
\begin{figure}[!htpb]
	\subfigure[$\mathcal{T}_{1}^{Se}$]{\label{fig:SedovII-40x40-mesh-t0}
		\begin{minipage}[t]{0.33\linewidth}
			\centering
			\includegraphics[width=4.2cm]{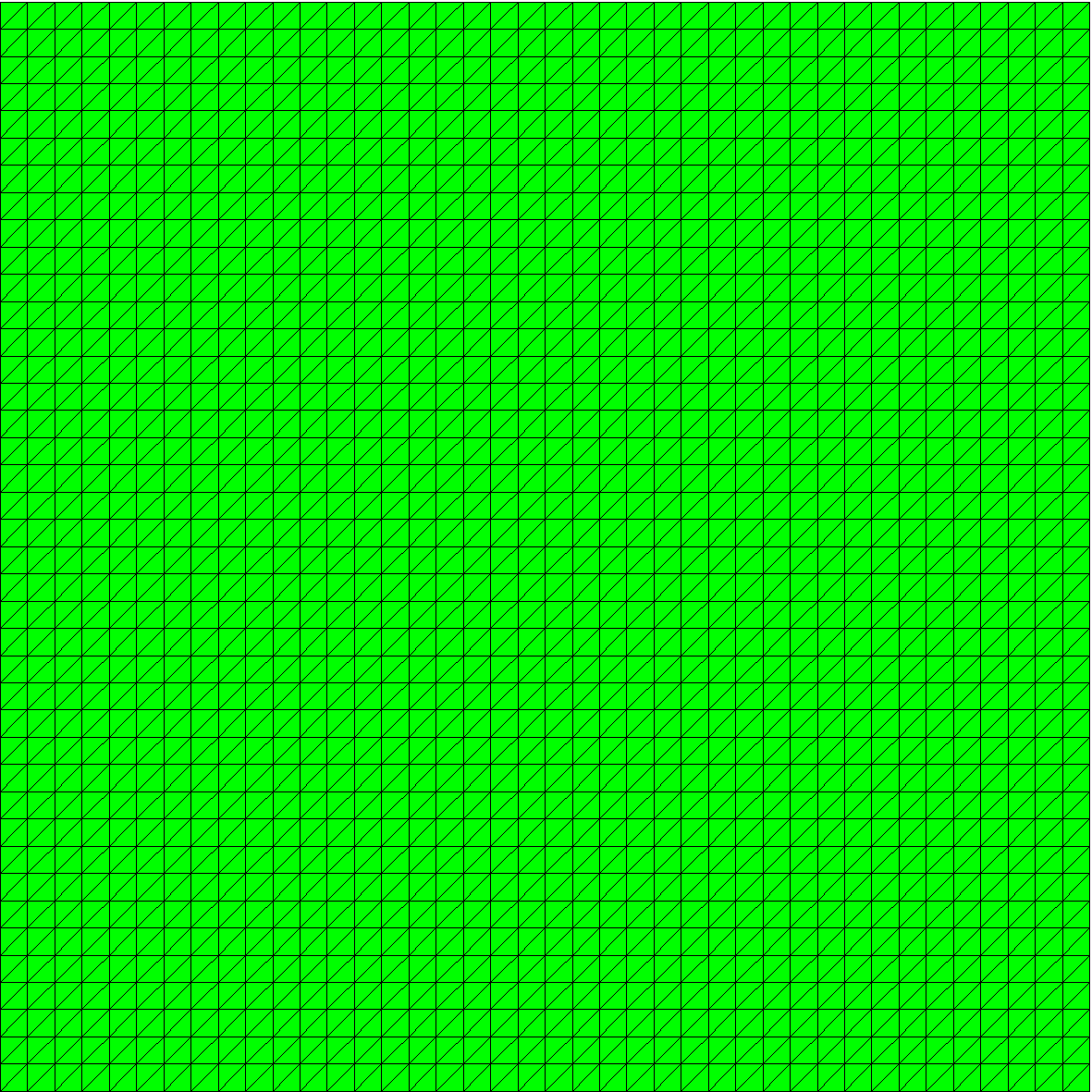}		
		\end{minipage}
	}%
	\subfigure[$\mathcal{T}_{2}^{Se}$]{\label{fig:SedovII-80x80-mesh-t0}
		\begin{minipage}[t]{0.33\linewidth}
			\centering
			\includegraphics[width=4.2cm]{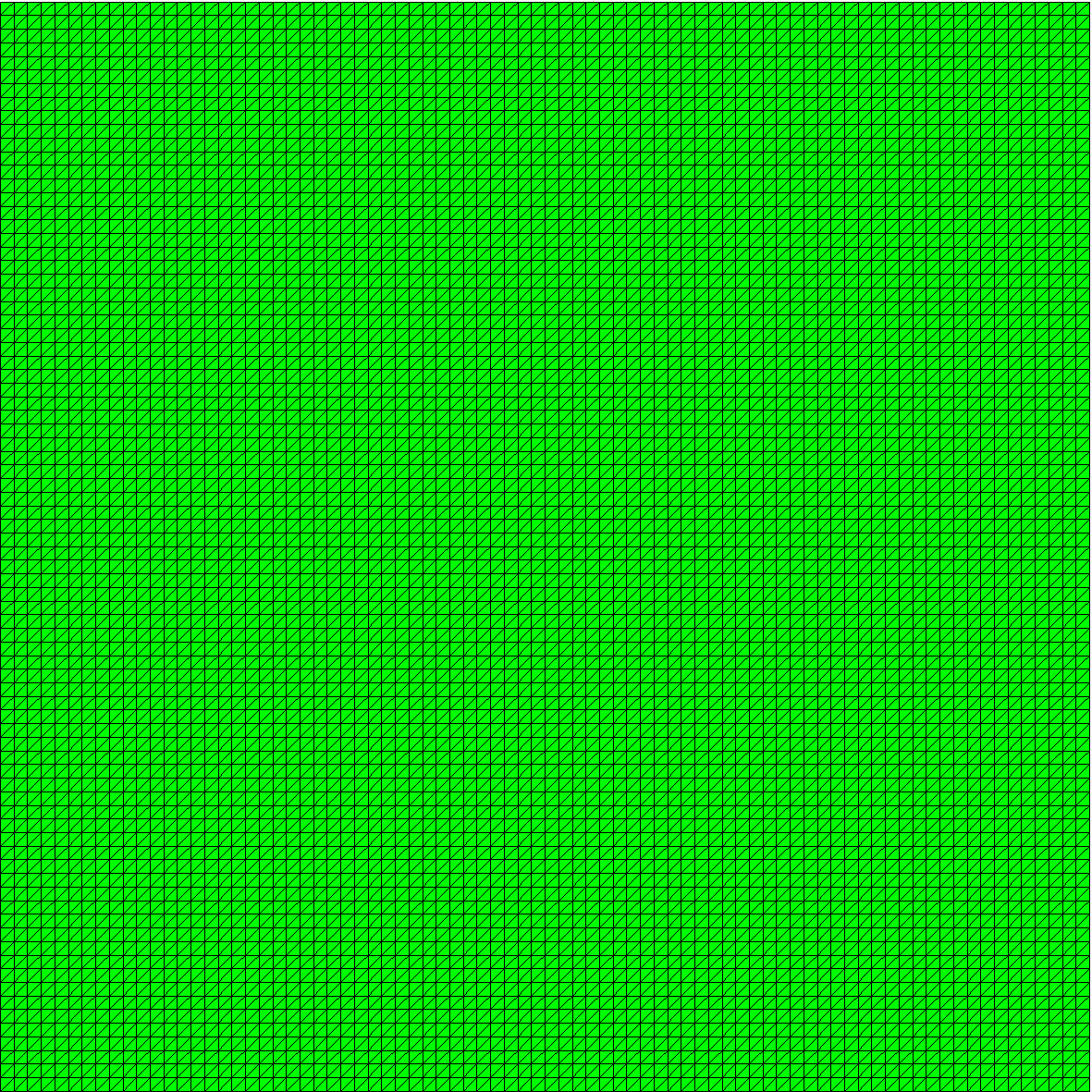}		
		\end{minipage}
	}%
	\subfigure[$\mathcal{T}_{3}^{Se}$]{\label{fig:SedovII-160x160-mesh-t0}
		\begin{minipage}[t]{0.33\linewidth}
			\centering
			\includegraphics[width=4.2cm]{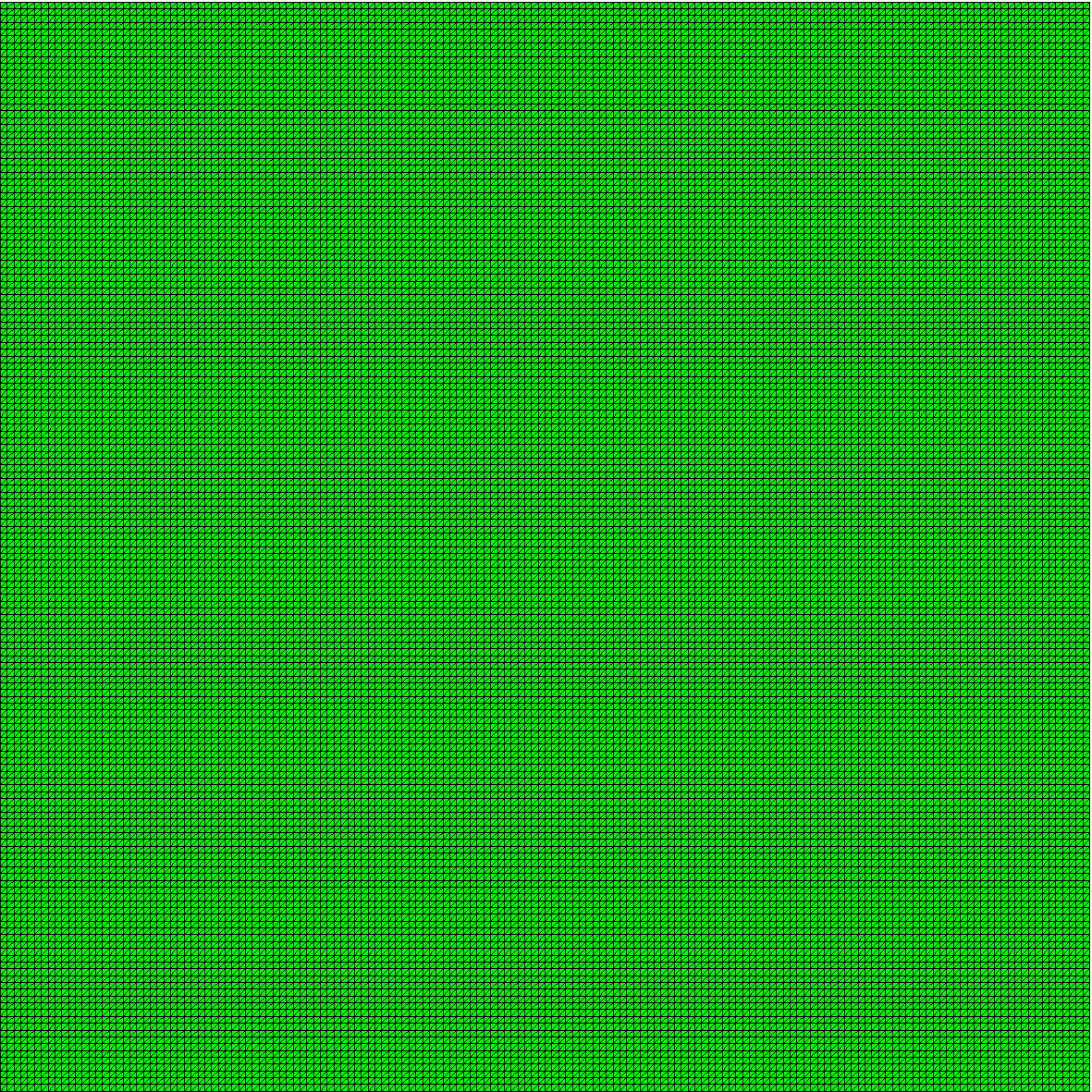}	
		\end{minipage}
	}%
	\caption{Sedov the first three sets of initial meshes of the problem.}\label{fig:SedovII-initial-mesh}
\end{figure}

\begin{example}
Consider the model problem \eqref{eq:mass-equation}-\eqref{eq:state-equation}, where domain $ \Omega = [0, 1] \times [0, 1] $ and simulation time $ t \in [0, 1]$, gas adiabatic $ \gamma = 1.4 $. We use $ 40\times40 $, $ 80\times80 $, $ 160\times160 $, $ 320\times320 $, $ 640\times640 $, $ 1000\times1000 $ uniform mesh for the domain $ \Omega $. \reftab{tab:Sedov-mesh-resolution} shows the corresponding notation, number of grid cells and number of nodes, and \reffig{fig:SedovII-initial-mesh} shows the first three sets of grids. Initial condition: initial density is 1, velocity is $ \bm{0} $, square domain in the lower left corner $[0, h_i] \times [0, h_i] $ (consists of a pair of triangles). The internal energy density is $ E_0/4h_i^2~(i=1,2,\cdots, 6)$ for the pair of cells at the down-left corner, where $ E_0 = 0.9792$, $ h_i = 0.1 \times 2^{-(i+1)}, i=1,2,\cdots,5$, $ h_6 = 10^{-3}$, and is 0 for the rest cells. Boundary condition: the left and lower boundary of the domain adopts the solid wall boundary conditions, The right and upper bounds adopt free surfaces conditions.
\end{example}

\begin{table}[!htpb]         
	\centering         
	\caption{Six kinds of initial grid information of Sedov}\label{tab:Sedov-mesh-resolution}         
	\begin{tabular}{cccc}         
		\hline         
		Notation                 & Mesh resolution  & Number of elements & Number of nodes \\
		\hline 
		$\mathcal{T}_{1}^{Se}$   &  40$ \times $40   & 3200    & 1681  \\
		\hline    
		$\mathcal{T}_{2}^{Se}$   &  80$ \times $80   & 12800   & 6561  \\ 
		\hline 
		$\mathcal{T}_{3}^{Se}$   &  160$ \times $160 & 51200   & 25921 \\ 
		\hline  
		$\mathcal{T}_{4}^{Se}$   &  320$ \times $320 & 204800  & 103041 \\
		\hline 
		$\mathcal{T}_{5}^{Se}$   &  640$ \times $640 & 819200  & 410881 \\
		\hline 
		$\mathcal{T}_{6}^{Se}$   &1000$ \times $1000 & 2000000 & 1002001 \\
		\hline           
	\end{tabular}         
\end{table} 

\begin{figure}[!htpb]
	\subfigure[Mesh, no-matterflow]{\label{fig:SedvoII-c-no-mf}
		\begin{minipage}[t]{0.33\linewidth}
			\centering
			\includegraphics[width=4.5cm]{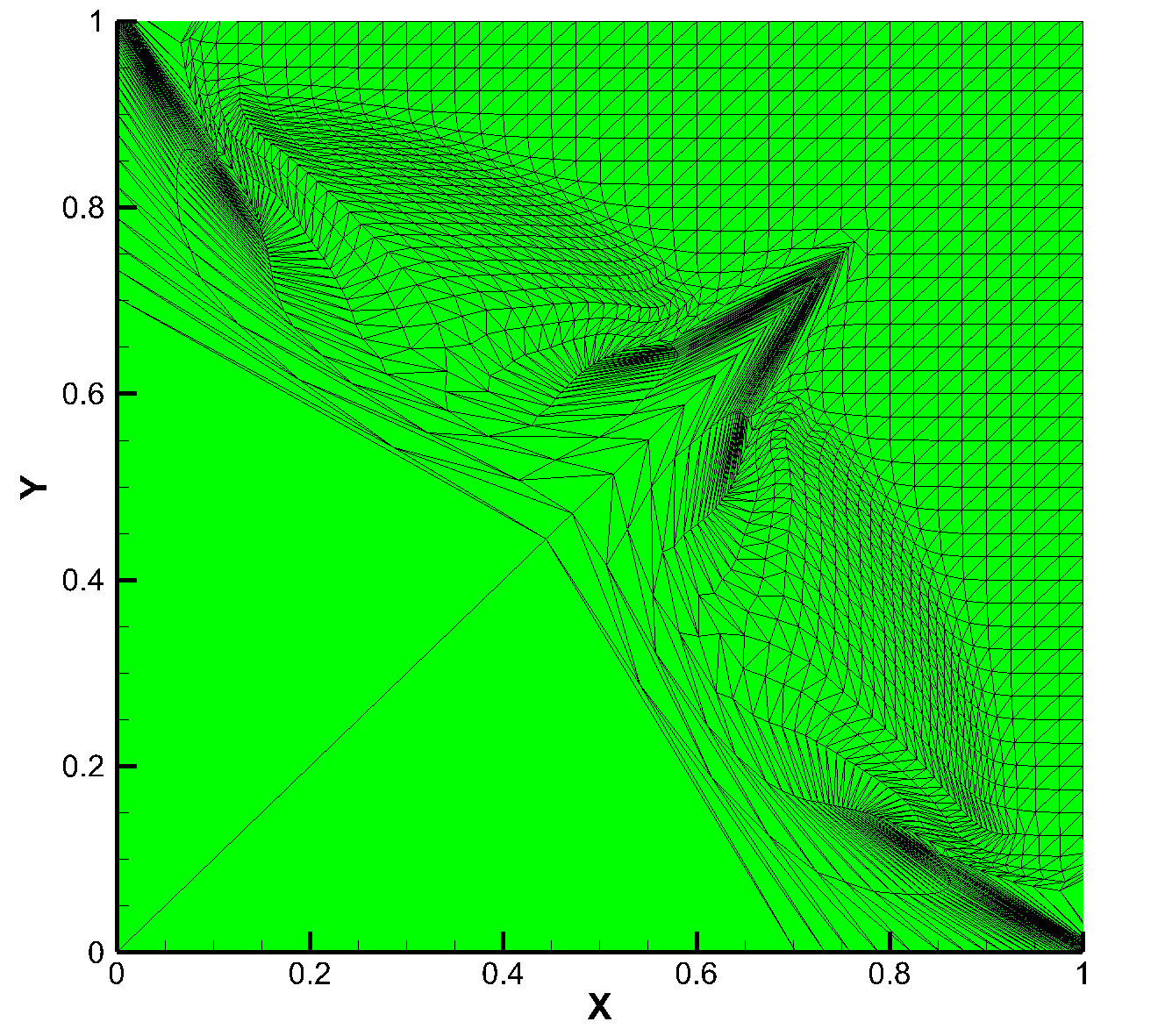}		
		\end{minipage}
	}%
	\subfigure[Density, no-matterflow]{\label{fig:SedvoII-density-no-mf}
		\begin{minipage}[t]{0.33\linewidth}
			\centering
			\includegraphics[width=4.65cm]{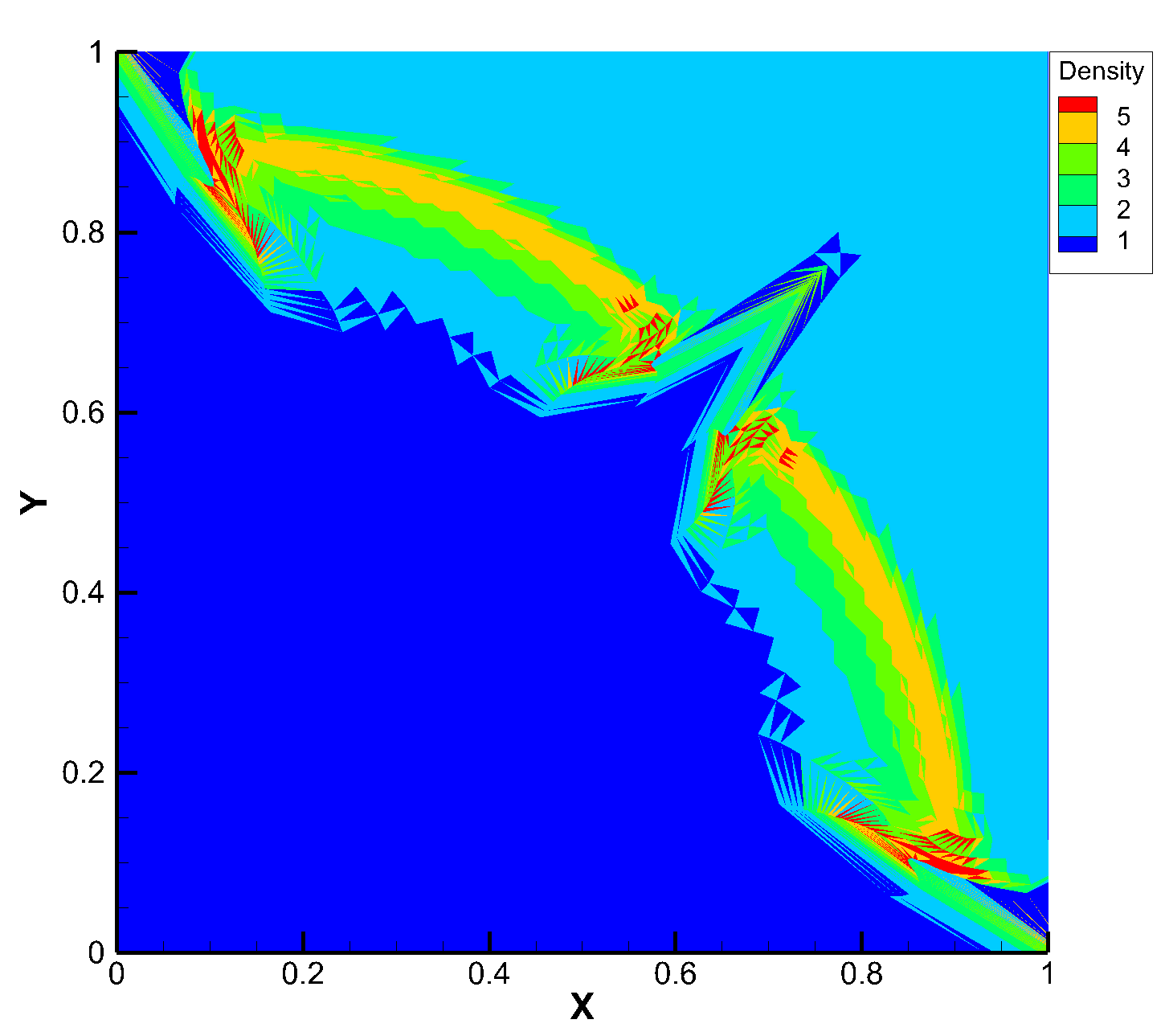}		
		\end{minipage}
	}%
	\subfigure[Pressure, no-matterflow]{\label{fig:SedvoII-pressure-no-mf}
		\begin{minipage}[t]{0.33\linewidth}
			\centering
			\includegraphics[width=4.5cm]{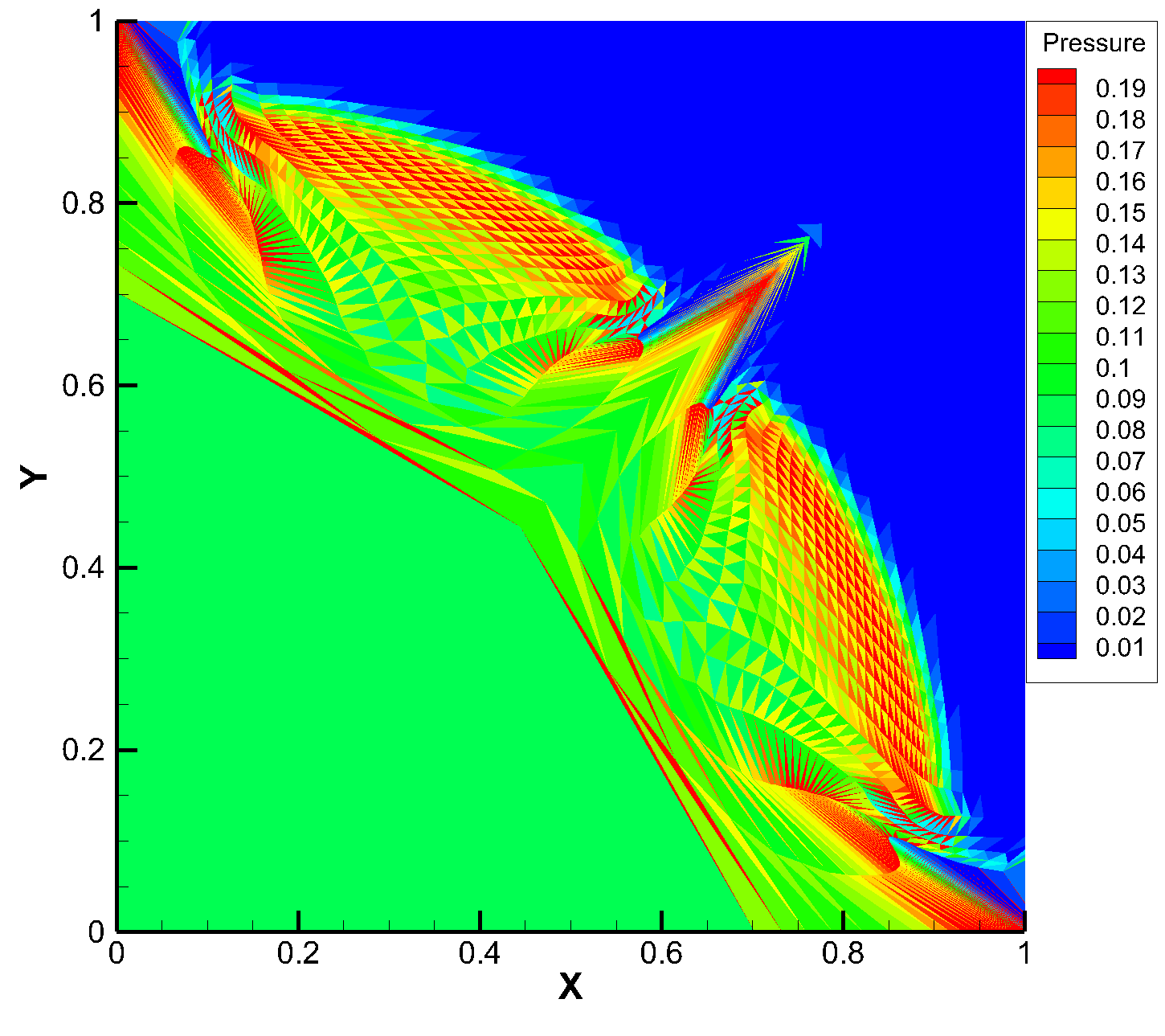}		
		\end{minipage}
	}%
	
	\subfigure[Mesh, matterflow]{\label{fig:SedvoII-c-mf}
		\begin{minipage}[t]{0.33\linewidth}
			\centering
			\includegraphics[width=4.5cm]{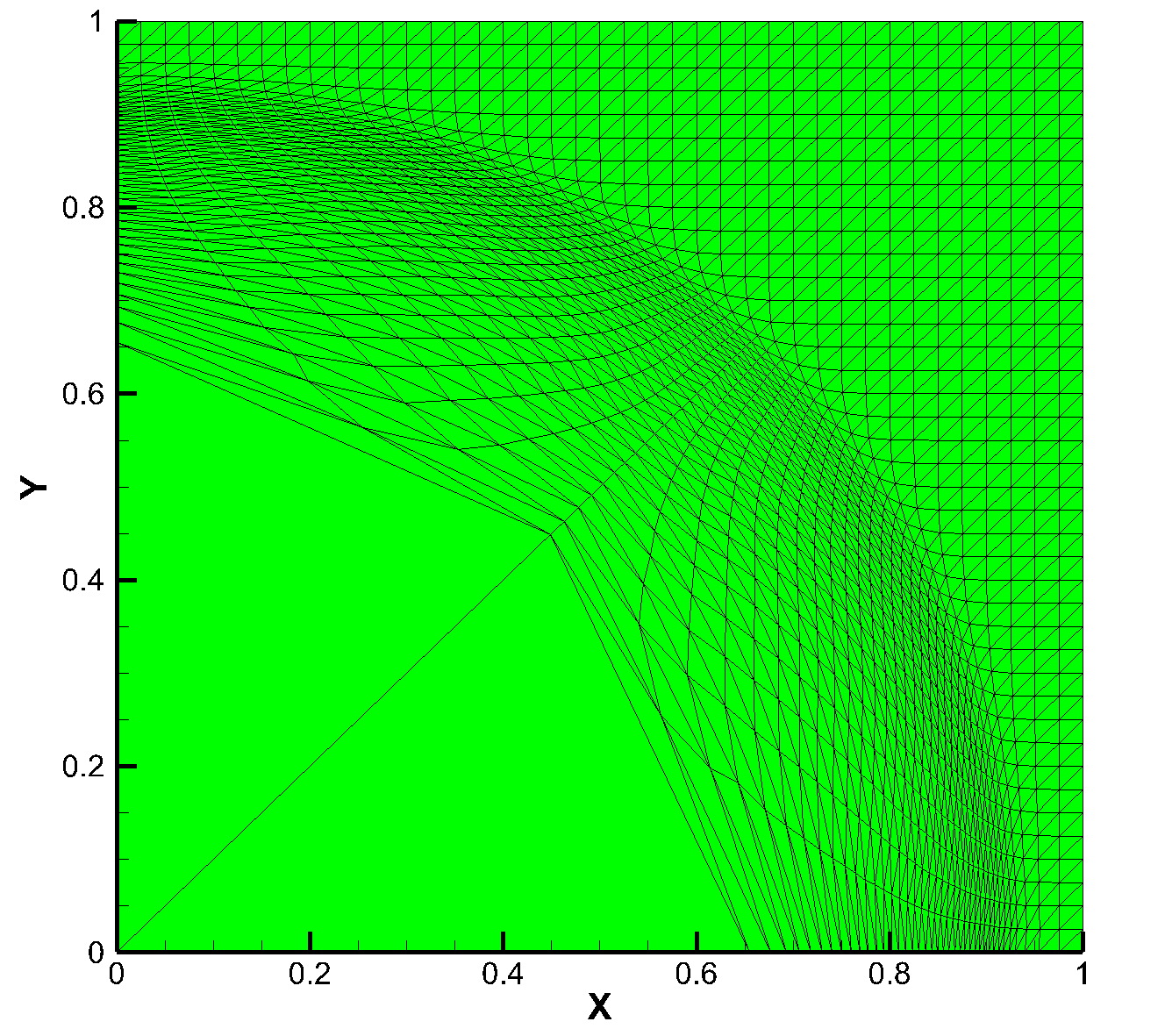}	
		\end{minipage}
	}%
	\subfigure[Density, matterflow]{\label{fig:SedvoII-density-mf}
		\begin{minipage}[t]{0.33\linewidth}
			\centering
			\includegraphics[width=4.65cm]{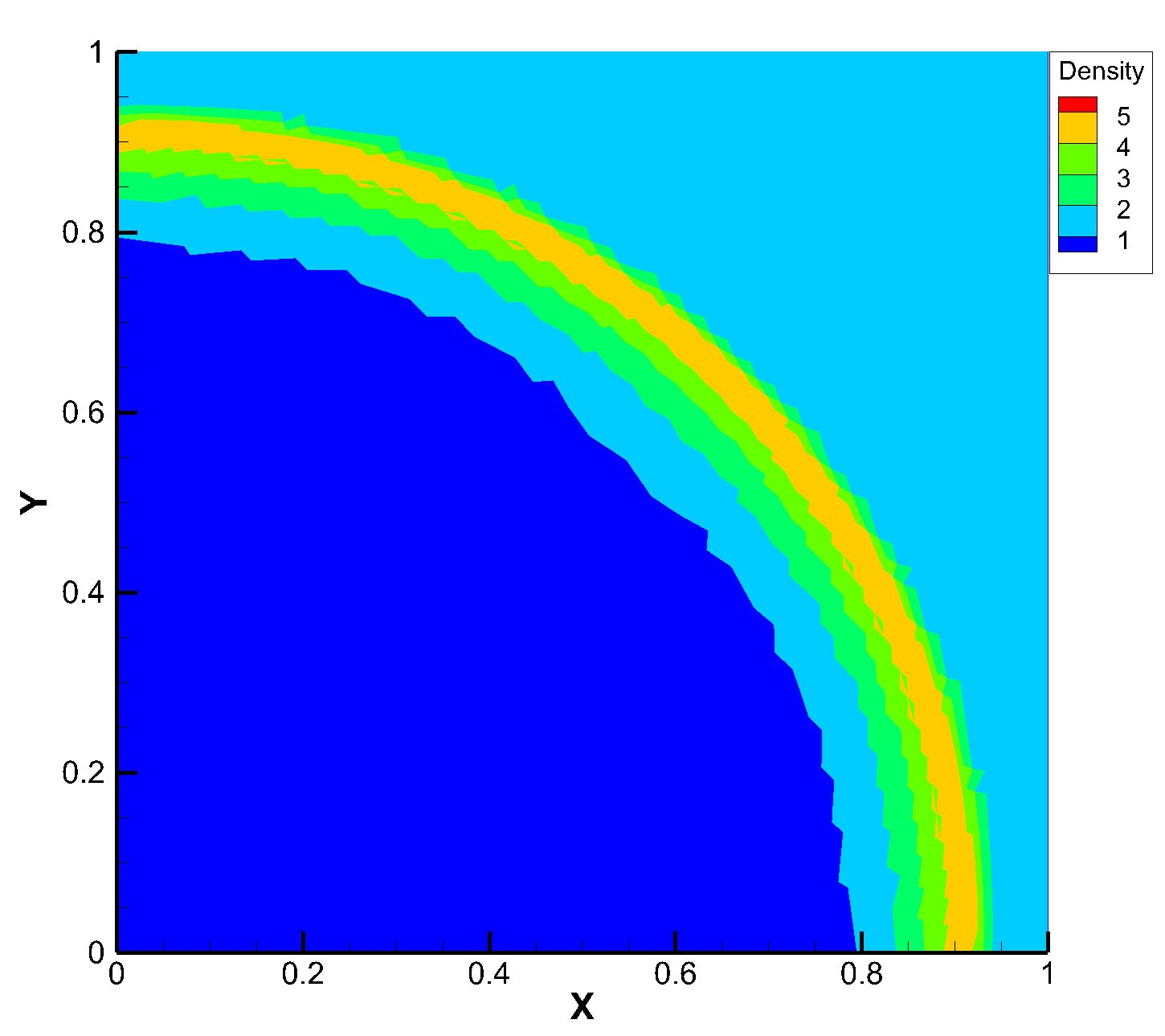}		
		\end{minipage}
	}%
	\subfigure[Pressure, matterflow]{\label{fig:SedvoII-pressure-mf}
		\begin{minipage}[t]{0.33\linewidth}
			\centering
			\includegraphics[width=4.5cm]{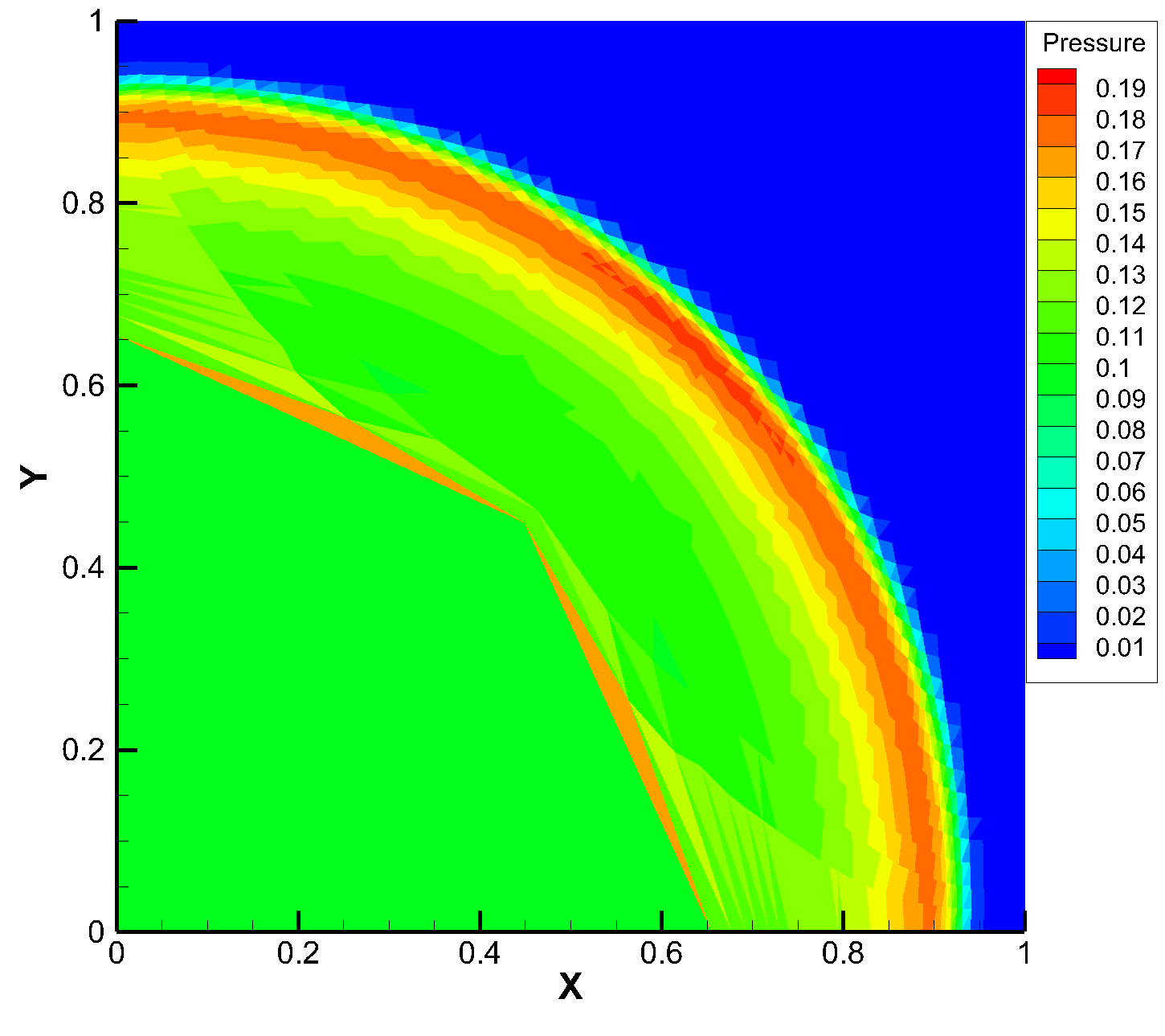}		
		\end{minipage}
	}%
	\caption{The grid, density and pressure contour diagram of $\mathcal{T}_{1}^{Se}$ at $ t = 1 $.
	}\label{fig:SedovII-compare-dist-mf}
\end{figure}

\begin{figure}[!htpb]
	\subfigure[Density]{\label{fig:SedvoII-Density-40x40-mf}
	\begin{minipage}[t]{0.33\linewidth}
		\centering
		\includegraphics[width=5.5cm]{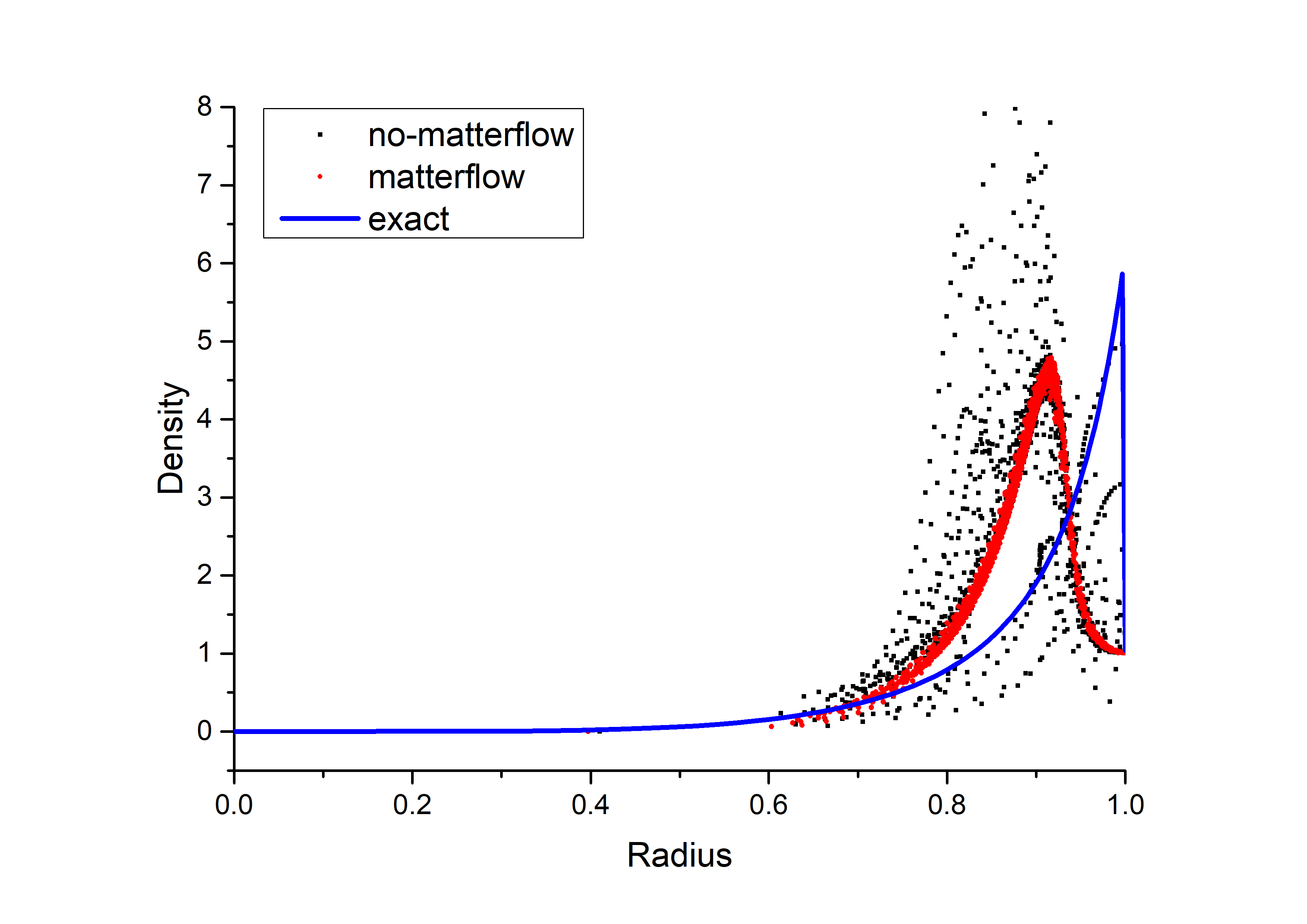}		
	\end{minipage}
	}%
	\subfigure[Pressure]{\label{fig:SedvoII-Pressure-40x40-mf}
		\begin{minipage}[t]{0.33\linewidth}
			\centering
			\includegraphics[width=5.5cm]{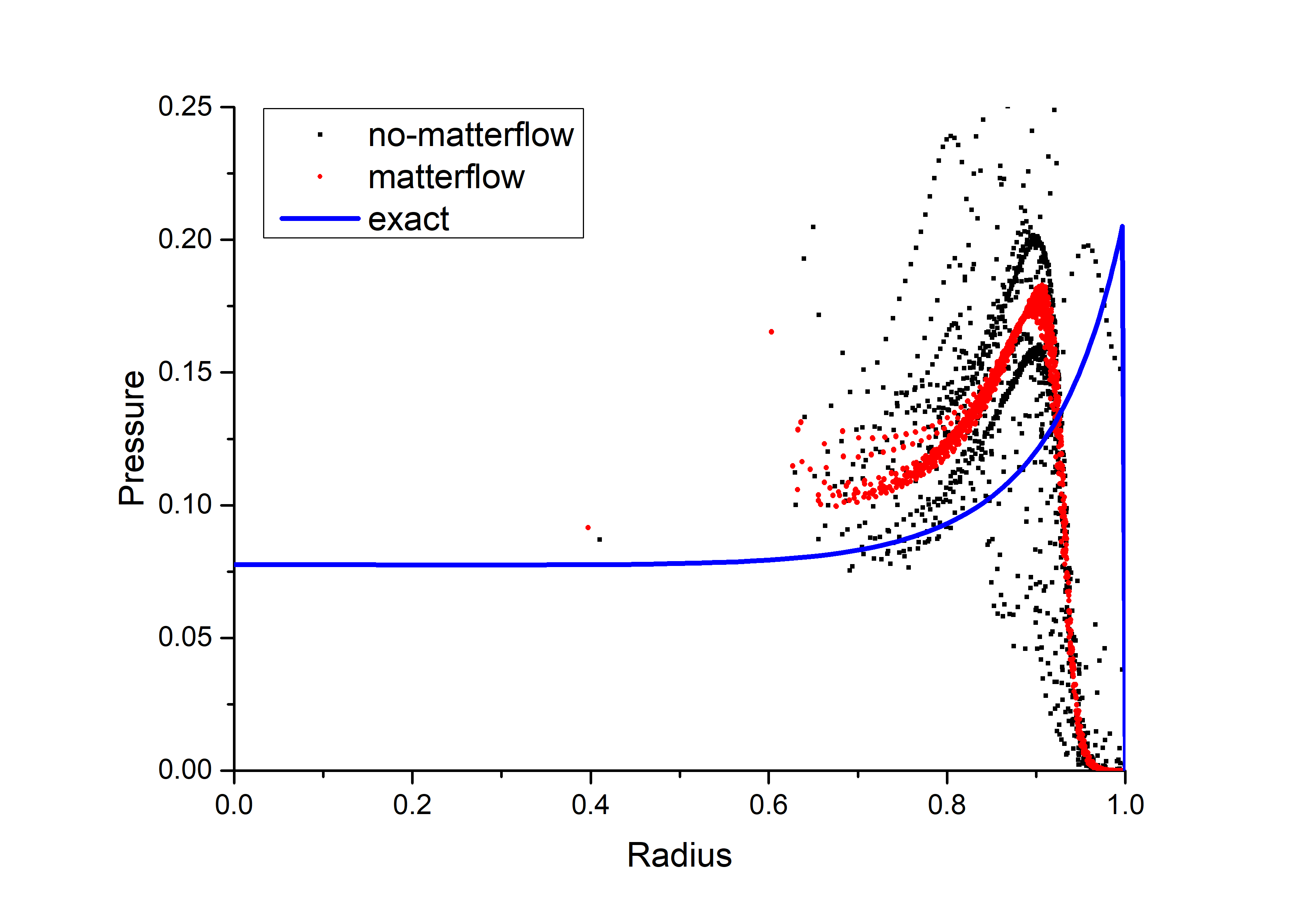}		
		\end{minipage}
	}%
	\subfigure[Velocity (r)]{\label{fig:SedvoII-Velocity-40x40-mf}
		\begin{minipage}[t]{0.33\linewidth}
			\centering
			\includegraphics[width=5.5cm]{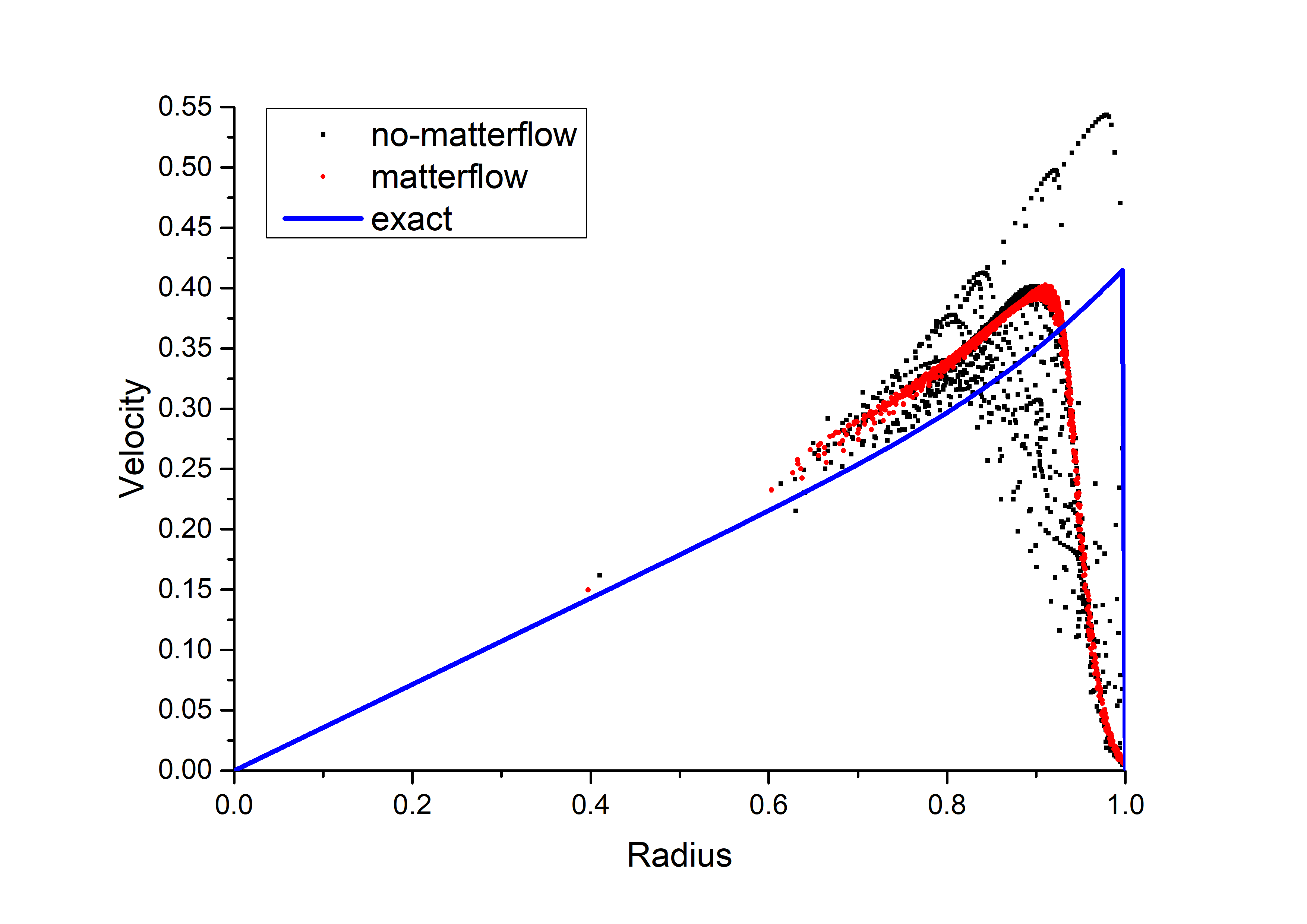}		
		\end{minipage}
	}%
	\caption{The scatter diagram of density, pressure and radial velocity of $\mathcal{T}_{1}^{Se}$ at $ t = 1 $.
	}\label{fig:SedovII-compare-exact-mf}
\end{figure}

\textbf{Firstly, the effect of material flow on numerical simulation results is discussed for initial mesh $\mathcal{T}_{1}^{Se}$.} \reffig{fig:SedovII-compare-dist-mf} shows the contour diagram of the grid, density and pressure when the $\mathcal{T}_{1}^{Se}$ grid does not use matter flow and uses matter flow at $ t = 1 $. \reffig{fig:SedovII-compare-exact-mf} shows the scatter diagram of density, pressure and radial velocity at $ t = 1 $. It can be seen from the figure that the introduction of the matter flow method can greatly alleviate the physical oscillation in the SGH Lagrangian simulation. As a side effect, the matter flow method also reduces the mesh distortion in the simulation of the Sedov problem.

\begin{figure}[!htpb]
	\subfigure[Density]{\label{fig:SedvoII-Density-multiscale-appro}
	\begin{minipage}[t]{0.33\linewidth}
		\centering
		\includegraphics[width=5.5cm]{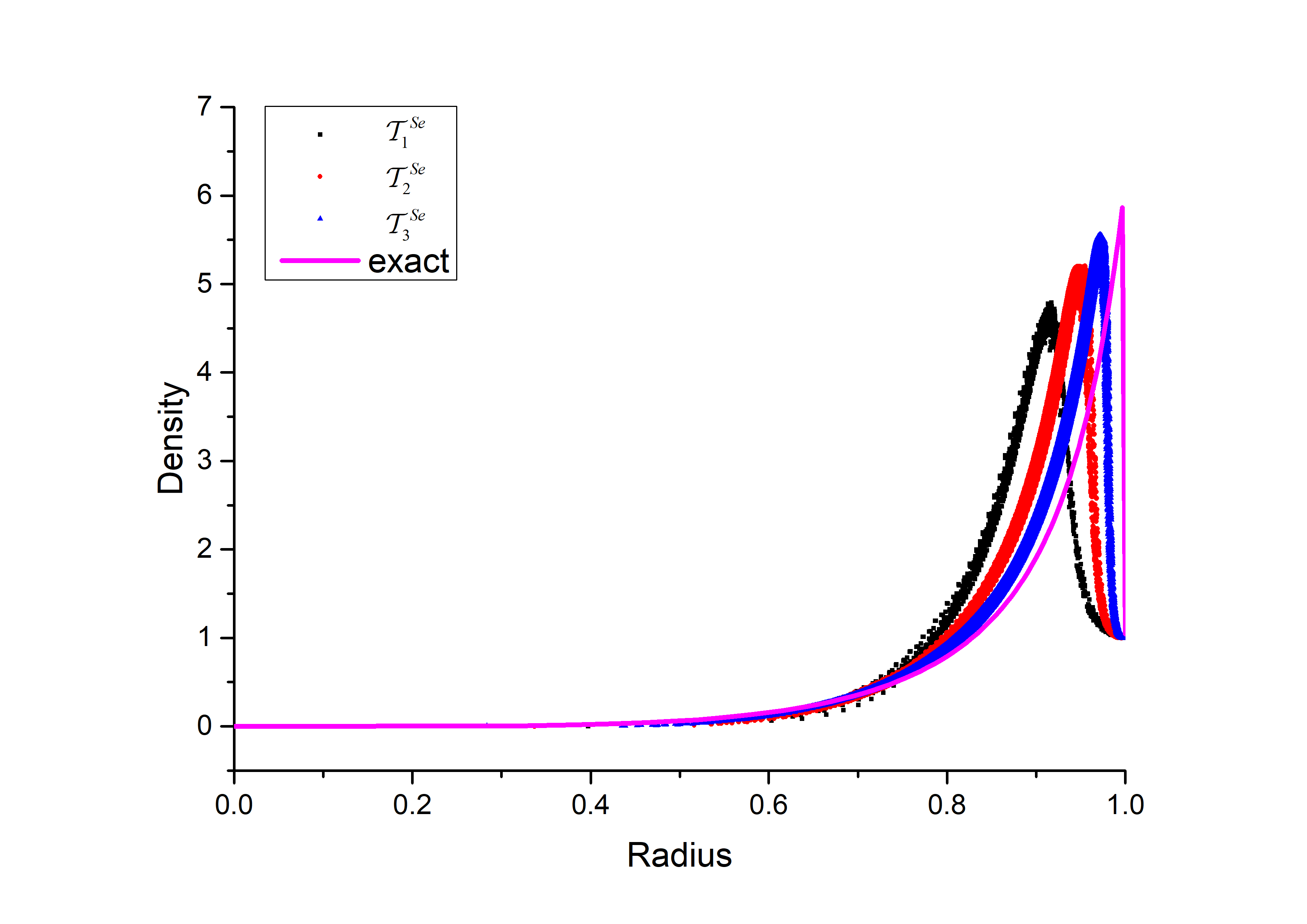}		
	\end{minipage}
	}%
	\subfigure[Pressure]{\label{fig:SedvoII-Pressure-multiscale-appro}
		\begin{minipage}[t]{0.33\linewidth}
			\centering
			\includegraphics[width=5.5cm]{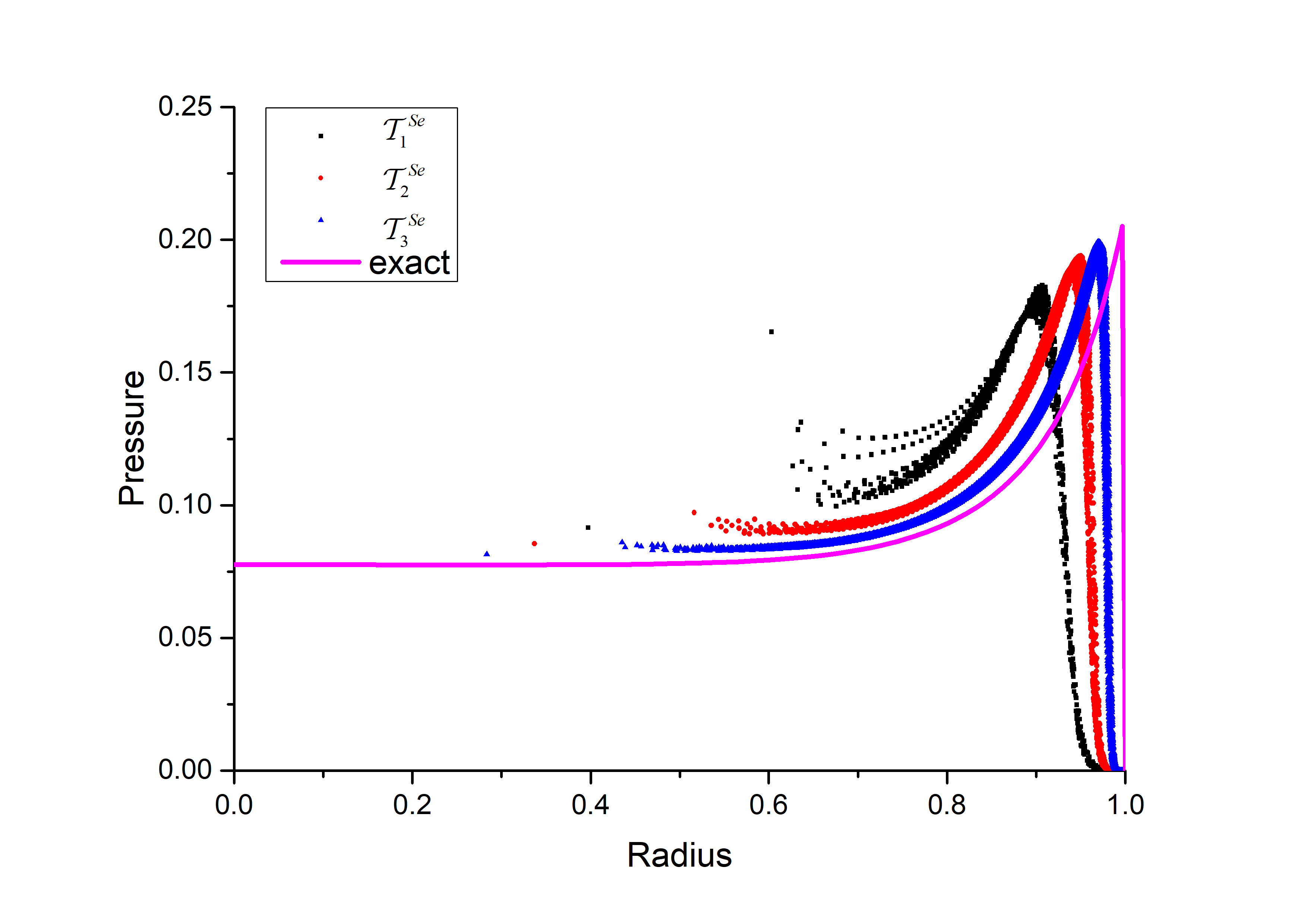}		
		\end{minipage}
	}%
	\subfigure[Velocity (r)]{\label{fig:SedvoII-Velocity-multiscale-appro}
		\begin{minipage}[t]{0.33\linewidth}
			\centering
			\includegraphics[width=5.5cm]{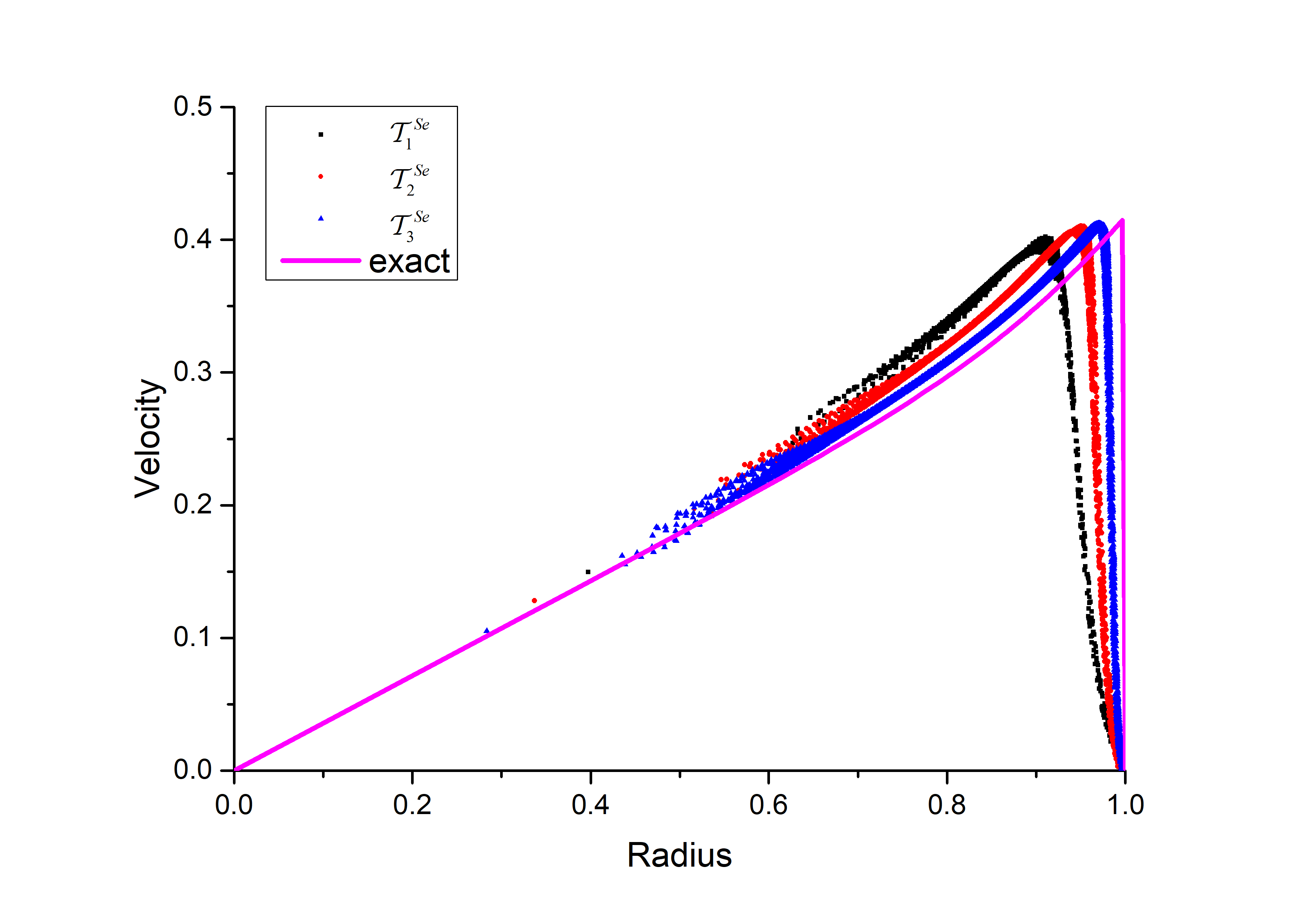}		
		\end{minipage}
	}%
	\caption{The density, pressure and radial velocity scatter diagram of $\mathcal{T}_{1}^{Se}$, $\mathcal{T}_{2}^{Se}$ and $\mathcal{T}_{3}^{Se}$ at $ t = 1 $.
	}\label{fig:SedovII-multiscale-appro}
\end{figure}

\textbf{Secondly, the effects of different grid sizes on the numerical solutions are investigated for the initial meshes $\mathcal{T}_{1}^{Se}$, $\mathcal{T}_{2}^{Se}$, $\mathcal{T}_{3}^{Se}$.} \reffig{fig:SedovII-multiscale-appro} shows the scatter diagram of the density, pressure, radial velocity of the three initial mesh sizes at $ t =1 $, respectively. The conclusions of these simulations are similar to that of Saltzman and Noh problems.

\begin{figure}[!htpb]
	\subfigure[Total mass]{\label{fig:SedvoII-Total-mass}
	\begin{minipage}[t]{0.25\linewidth}
		\centering
		\includegraphics[width=4cm]{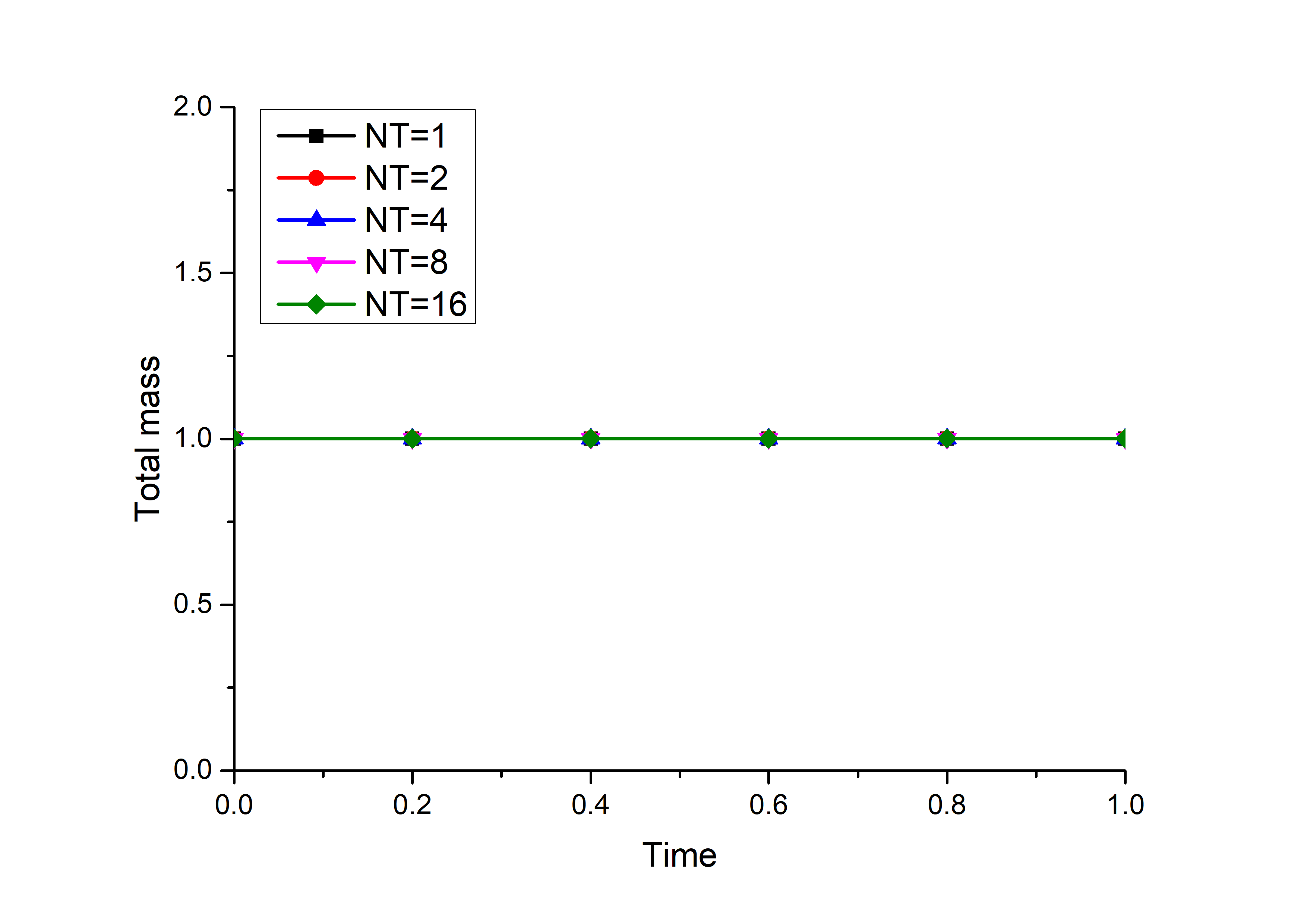}		
	\end{minipage}
	}%
	\subfigure[Total energy]{\label{fig:SedvoII-Total-energy}
		\begin{minipage}[t]{0.25\linewidth}
			\centering
			\includegraphics[width=4cm]{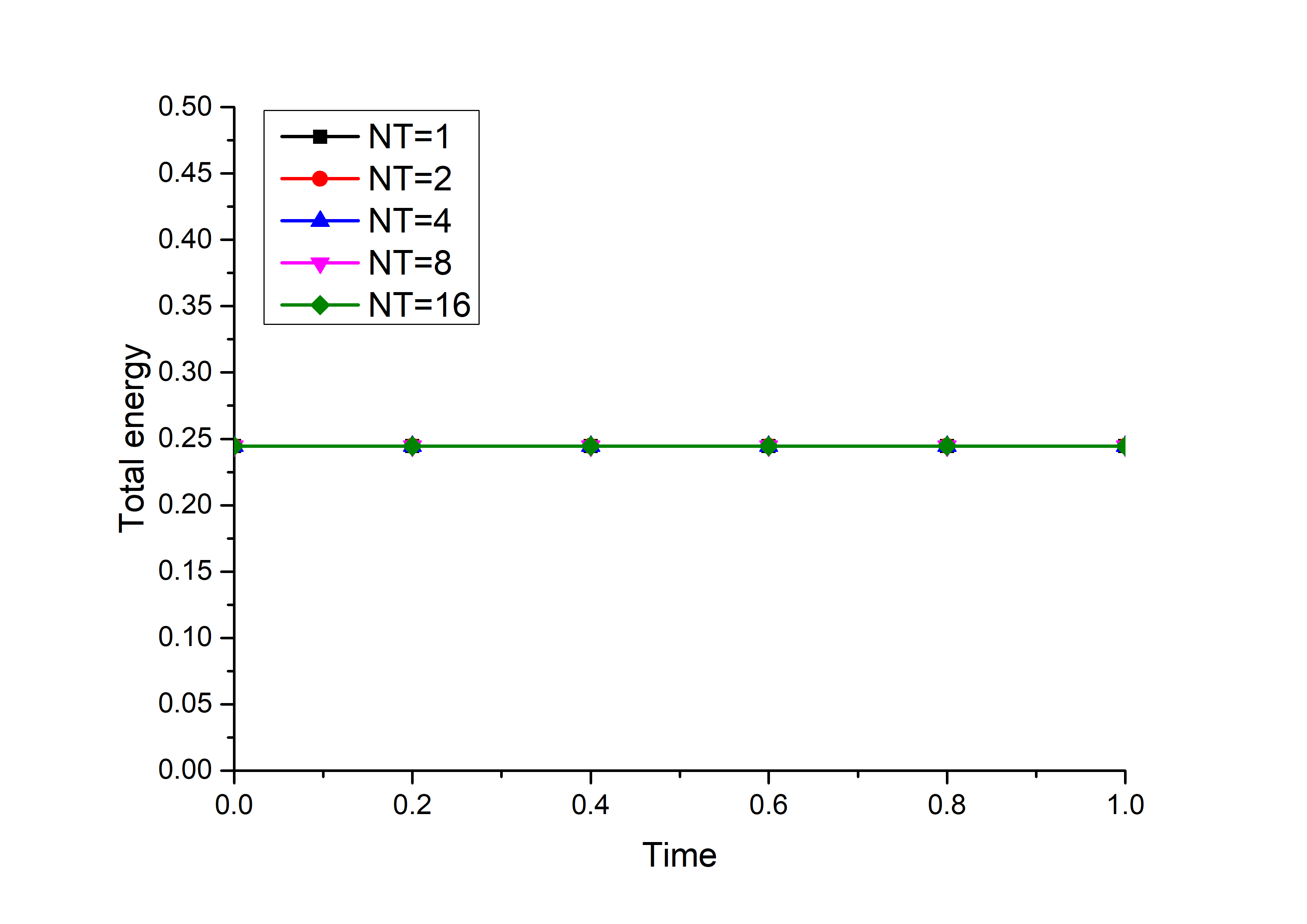}		
		\end{minipage}
	}%
	\subfigure[Total momentum(x)]{\label{fig:SedvoII-Total-momentumX}
		\begin{minipage}[t]{0.25\linewidth}
			\centering
			\includegraphics[width=4cm]{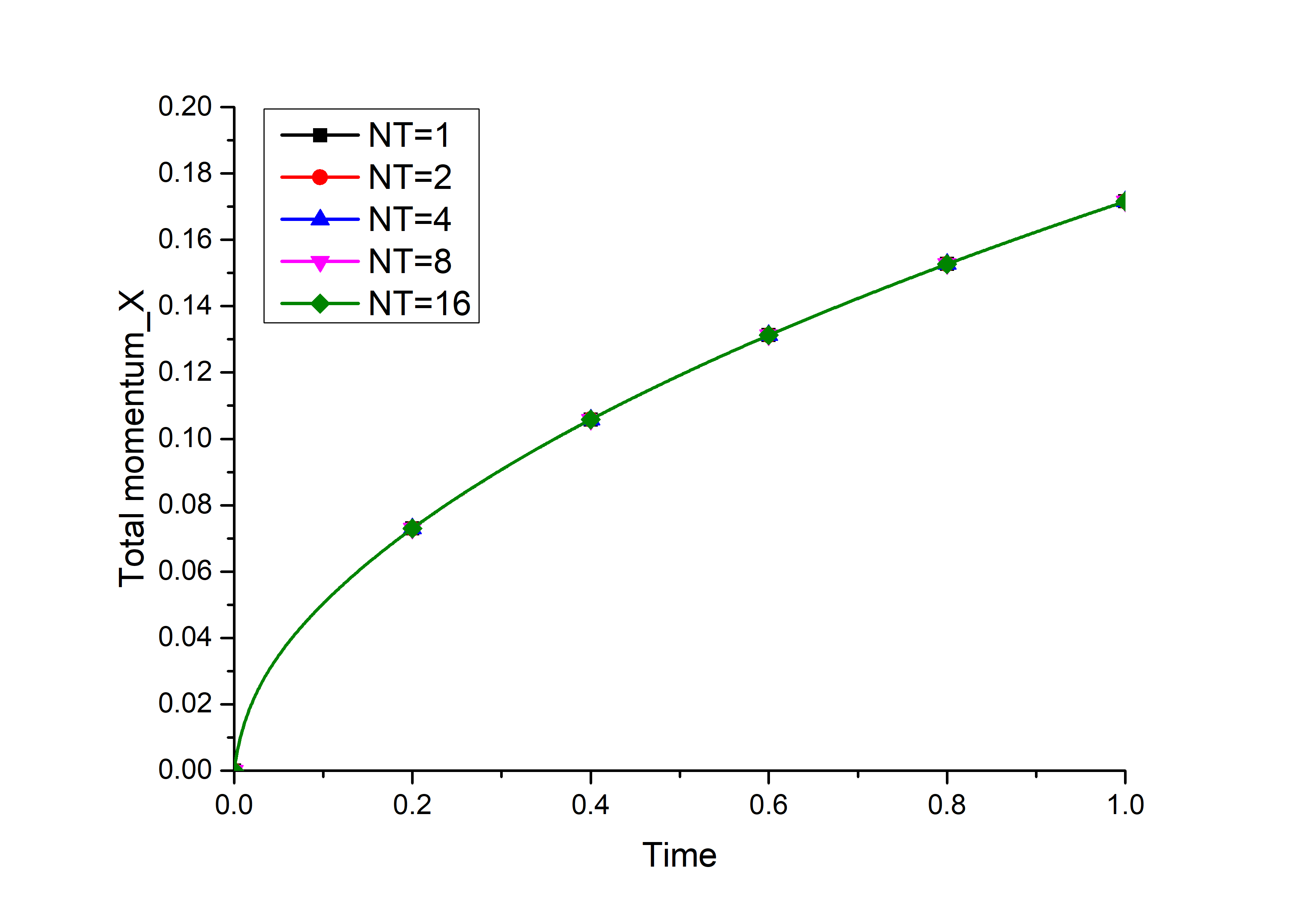}		
		\end{minipage}
	}%
	\subfigure[Total momentum(y)]{\label{fig:SedvoII-Total-momentumY}
		\begin{minipage}[t]{0.25\linewidth}
			\centering
			\includegraphics[width=4cm]{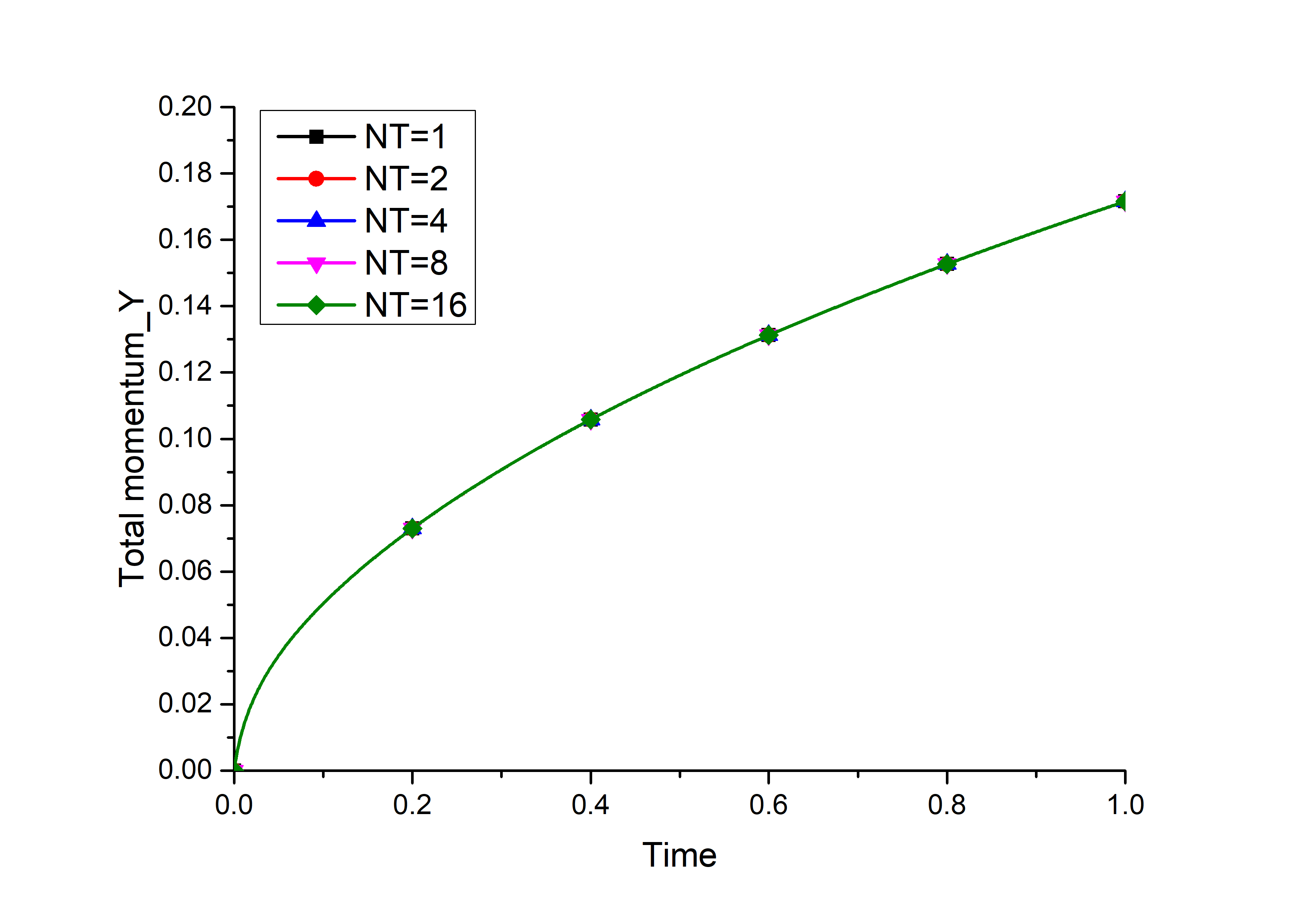}		
		\end{minipage}
	}%
	\caption{
	Total mass, total energy, total momentum variation curves of different threads. It can be seen from the diagram that the total mass and total energy are conserved, and the experimental results of multithreading are completely consistent, and the total momentum increases with the increase of simulation time, and the total momentum change curve of multithreading completely overlaps.
	}\label{fig:SedovII-80x80-parallel-total}
\end{figure}

\begin{figure}[!h]
	\subfigure[Density]{\label{fig:SedvoII-Density-80x80-parallel}
	\begin{minipage}[t]{0.33\linewidth}
		\centering
		\includegraphics[width=5.5cm]{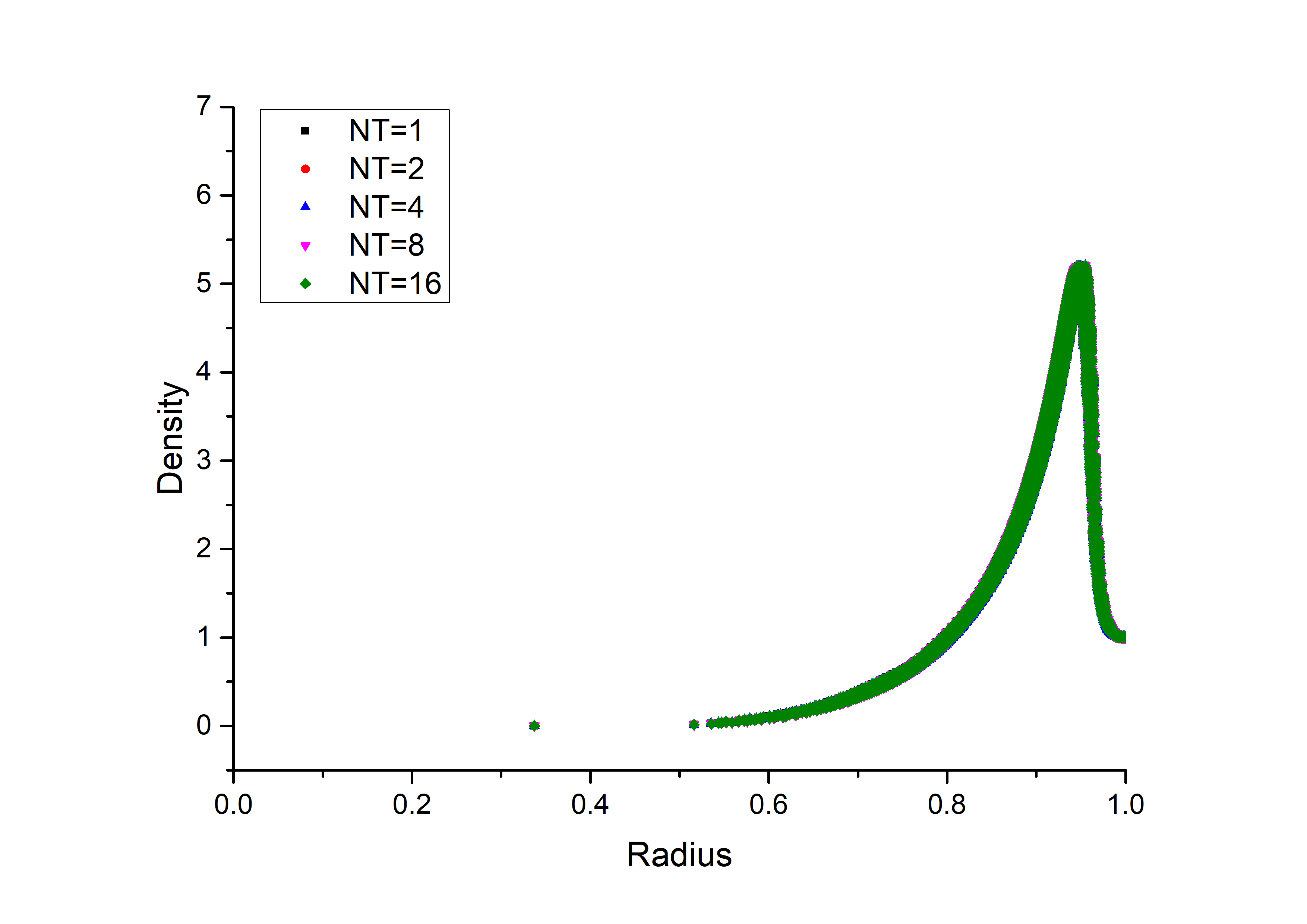}		
	\end{minipage}
	}%
	\subfigure[Pressure]{\label{fig:SedvoII-Pressure-80x80-parallel}
		\begin{minipage}[t]{0.33\linewidth}
			\centering
			\includegraphics[width=5.5cm]{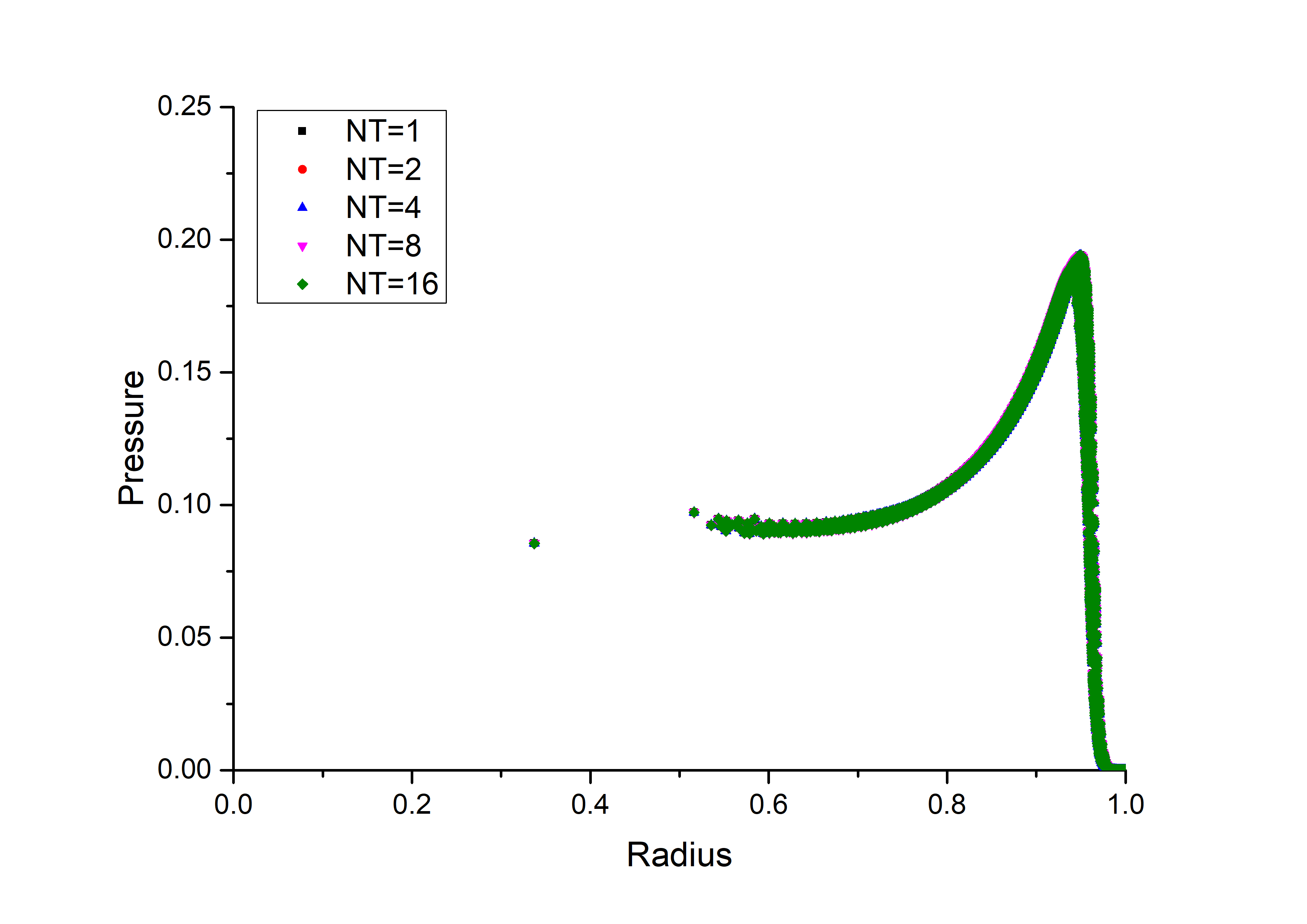}		
		\end{minipage}
	}%
	\subfigure[Radial velocity]{\label{fig:SedvoII-Velocity-80x80-parallel}
		\begin{minipage}[t]{0.33\linewidth}
			\centering
			\includegraphics[width=5.5cm]{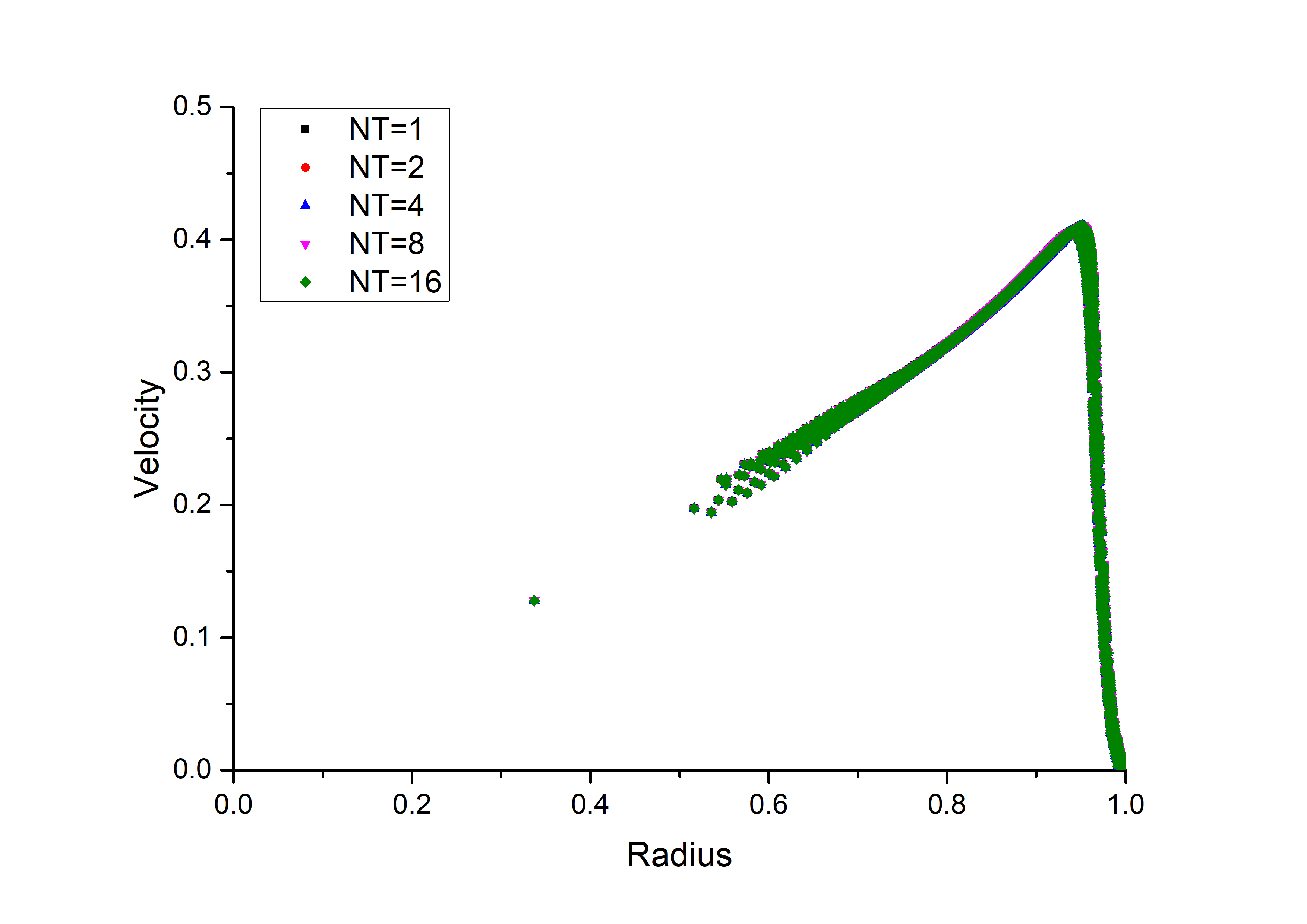}		
		\end{minipage}
	}%
	\caption{
	The scatter plot of density, pressure and radial velocity of different threads at $ t = 1 $. It can be found that the density, pressure and radial velocity calculated by different threads are exactly the same.
	 }\label{fig:SedovII-80x80-parallel}
\end{figure}

\textbf{Finally, we take the Sedov problem as an example to explore the parallel performance of the SGH Lagrangian program after adding the matter flow (the parallel test results of the other two examples are basically consistent with the examples).}
\begin{enumerate}
	\item [(1)] To validate the \textbf{correctness} of the parallel algorithm, considering the initial grid $\mathcal{T}_{2}^{Se}$, the influence of different thread number $NT = 1, 2, 4, 8, 16 $ on the numerical solution is studied.
\end{enumerate}
\reffig{fig:SedovII-80x80-parallel-total} shows the curves of total mass, total energy and total momentum of different threads with time. \reffig{fig:SedovII-80x80-parallel} shows the scatter plot of density, pressure and radial velocity of different threads at $ t =1 $.

\reffigs{fig:SedovII-80x80-parallel-total}-\ref{fig:SedovII-80x80-parallel} shows that the total mass, total energy and total momentum change curve of multithreading completely coincide with the change curve of single thread. Density, pressure and radial velocity scatter plot of multithreading at $ t =1 $ are also consistent with that of single thread, indicating that the parallel program designed is correct.

\begin{enumerate}
	\item [(2)] To validate the \textbf{scalability} of parallel programs, three groups of large-scale grids $\mathcal{T}_{4}^{Se}$, $\mathcal{T}_{5}^{Se}$, $\mathcal{T}_{6}^{Se}$, fixed iteration times of 1000 times, the number of threads are $ NT=1,2,4,8,16 $.
\end{enumerate}

\reftabs{tab:SedovII-multiscale-evolve-parallel-time}-\ref{tab:SedovII-multiscale-evolve-parallel-speedup} shows the CPU wall time and parallel speedup P-MFL different sizes in different thread numbers, respectively.

\begin{minipage}{\textwidth}
	\begin{minipage}[t]{0.5\textwidth}
		\centering
		\makeatletter\def\@captype{table}\makeatother								
		\caption{P-MFL CPU-Time~(s)}\label{tab:SedovII-multiscale-evolve-parallel-time}									
		\begin{tabular}{|c|c|c|c|c|}									
			\hline									
			NT 	&	$\mathcal{T}_{4}^{Se}$	&		$\mathcal{T}_{5}^{Se}$	&		$\mathcal{T}_{6}^{Se}$ \\
			\hline									
			1	&	3.61E+02	&	1.42E+03	&	3.35E+03	\\
			\hline											
			2	&	1.91E+02	&	7.53E+02	&	1.75E+03 	\\
			\hline											
			4	&	1.01E+02	&	3.92E+02	&	9.11E+02 	\\
			\hline											
			8	&	5.74E+01	&	2.21E+02	&	5.19E+02 	\\
			\hline											
			16	&	3.78E+01	&	1.48E+02	&	3.52E+02 	\\
			\hline						
		\end{tabular}								
	\end{minipage}
	\begin{minipage}[t]{0.5\textwidth}
		\centering
		\makeatletter\def\@captype{table}\makeatother
		\caption{P-MFL Speedup}\label{tab:SedovII-multiscale-evolve-parallel-speedup}										
		\begin{tabular}{|c|c|c|c|c|}									
			\hline									
			NT 	&	$\mathcal{T}_{4}^{Se}$	&		$\mathcal{T}_{5}^{Se}$	&		$\mathcal{T}_{6}^{Se}$ \\
			\hline									
			1	&	$\backslash$	&	$\backslash$	&	$\backslash$	\\
			\hline											
			2	&	1.89 	&	1.89	&	1.91 	\\
			\hline		 						
			4	&	3.57 	&	3.63	&	3.68 	\\
			\hline		 						
			8	&	6.29 	&	6.45	&	6.45 	\\
			\hline		 						
			16	&	9.55 	&	9.60	&	9.51 	\\
			\hline						
		\end{tabular}		
	\end{minipage}
\end{minipage}
\begin{rmk}
	The parallel speedup $ S_n = \dfrac{T_1}{T_n} $, where the $ T_1 $ is the time of execution of one processor, $ T_n $ the time of execution of $ n $ processor.
\end{rmk}

\reftab{tab:SedovII-multiscale-evolve-parallel-time} shows that when the grid amount is fixed and the number of threads increases, P-MFL CPU wall time gradually decreased. When the number of threads is fixed and the grid size increases, P-MFL CPU wall increases almost linearly with grid size.  \reftab{tab:SedovII-multiscale-evolve-parallel-speedup} shows that, when the grid is fixed,  P-MFL speedup increases with the increase of the number of threads. When the number of threads is fixed, the speedup increases gradually with the grid amount increase. Especially, If the grid is $\mathcal{T}_{6}^{Se}$ (the number of cells is 2 million) and the number of threads is 16, the speedup reached 9.51.

To sum up, the P-MFL parallel algorithm based on OpenMP is correct and has good parallel scalability.


\section{Summary and discussion}\label{sec:six}
In this paper, aiming at the checkerboard oscillation problem of triangular mesh SGH Lagrangian hydrodynamic simulation, a matter flow method is designed to alleviate the physical quantity oscillation, and parallelization is carried out. The matter flow method is similar to that of Scovazzi\cite{Scovazzi-tetrahedral-meshes} and Molgan\cite{Morgan-Godunov-like} ---- by introducing some physical quantity transport terms between elements. However, compared with these two methods, we think that the method in this paper takes into account the physical quantities that need to be transported more comprehensively. Three kinds of effects are considered in the matter flow method. Firstly, the mass, energy and momentum transport caused by matter transport. Secondly, energy transport caused by work due to the change of element density. Finally, the effect of matter flow on strain rate in the element. In contrast, Scovazzi's method only considers energy transport, while Molgan's method only considers the first kind of effect. The effectiveness of the proposed method is verified by numerical experiments.
 
Although the matter flow method in this paper has achieved some good results, there are still many limitations. Firstly, it is only suitable for scalar viscosity, and how to extend it to tensor viscosity needs further study. Secondly, when one hopes to simulate multi-material fluids, the problem of how to implement matter flow between cells with different materials needs to be solved. Finally, more work is needed in the parallel implementation algorithm of matter flow algorithm, such as designing parallel method based on MPI.

\section*{Acknowledgments}
\addcontentsline{toc}{section}{Acknowledgments}
This work was supported by the National Natural Science Foundation of China (NSFC Project number 11971414). The authors would like to thank Long Xie and Shuchao Duan for useful discussions.

\addcontentsline{toc}{section}{References}
\bibliographystyle{model1-num-names}
\bibliography{refs}

\end{document}